\documentclass[12pt]{article}
\usepackage{amsfonts}
\usepackage{amsmath, amsthm, amssymb, bbm}
\usepackage{xcolor,hyperref,multirow,graphicx}

\usepackage{lmodern}
\usepackage[scaled=0.92]{helvet}

\usepackage[T1]{fontenc}
\usepackage[latin9]{inputenc}

\usepackage{array}

\newenvironment{enumerate*}
{ \begin{enumerate}
    \setlength{\itemsep}{0em}
    \setlength{\parskip}{.25em}
    \setlength{\parsep}{0em}     }
{ \end{enumerate}                  }

\newtheorem{remark}{Remark}
\newtheorem{theorem}{Theorem}
\newtheorem{proposition}{Proposition}
\newtheorem{example}{Example}

\newtheorem{lemma}{Lemma}

\newcommand{\plim}{\operatorname*{plim}}
\newcommand{\diag}{{\rm diag}}

\newcommand{\argmax}{\operatorname*{argmax}}
\newcommand{\argmin}{\operatorname*{argmin}}

\renewcommand{\baselinestretch}{1.3}
\usepackage{hyperref}
\definecolor{darkblue}{rgb}{0.00,0.15,0.7}
\hypersetup{colorlinks,breaklinks,
            linkcolor=darkblue,urlcolor=darkblue,
            anchorcolor=darkblue,citecolor=black,
            pdfstartview= FitH,
            pdftoolbar= true,
            pdfcenterwindow=true,
           pdfpagemode=UseNone,colorlinks,breaklinks,
           }

\usepackage{verbatim}
\usepackage[authoryear]{natbib}

\begin{document}

\title{\vspace{-0cm}Bias and Consistency in Three-way Gravity Models\thanks{Thomas Zylkin is grateful for support from NUS Strategic
Research Grant WBS: R-109-000-183-646 awarded to the Global Production
Networks Centre at National University of Singapore and from the Economic and Social Research Council through ESRC grant EST013567/1. Martin Weidner acknowledges support from the Economic and Social Research Council through the ESRC Centre for Microdata Methods and Practice grant RES-589-28-0001 and from the European Research Council grants ERC-2014-CoG-646917-ROMIA
and ERC-2018-CoG-819086-PANEDA.  We also thank Treb Allen, two anonymous referees, Valentina Corradi, Riccardo D'adamo, Ivan Fern\'andez-Val, Koen Jochmans, Maia Linask, Michael Pfaffermayr, Raymond Robertson, Ben Shepherd, Amrei Stammann, and Yoto Yotov.}} 

\author{Martin Weidner\\
 Oxford\and Thomas Zylkin\thanks{\emph{Contact information}: Weidner: Nuffield College and Department of Economics, University
of Oxford, Oxford OX1 3UQ. Email: \protect\protect\href{mailto:martin.weidner@ox.ac.uk}{martin.weidner@economics.ox.ac.uk}.
Zylkin: Robins School of Business, University of Richmond, Richmond,
VA, USA 23217. E-mail: \protect\protect\href{mailto:tzylkin@richmond.edu}{tzylkin@richmond.edu}.}\\
 Richmond}

\date{\vskip .25cm
 {\Large{}{} }\today
 }

\maketitle
\vskip -.5cm \noindent
We study the incidental parameter problem for the ``three-way'' Poisson 
{Pseudo-Maximum Likelihood} (``PPML'') estimator recently recommended for identifying the effects of trade policies {and in other panel data gravity settings}. Despite the number and variety of fixed effects involved, we confirm PPML is consistent for fixed $T$ and we show it is in fact the only estimator among a wide range of PML gravity estimators that is generally consistent in this context when $T$ is fixed. At the same time, asymptotic confidence intervals in fixed-$T$ panels are not correctly centered at the true point estimates, and cluster-robust variance estimates used to construct standard errors are generally biased as well. We characterize each of these biases analytically and show both numerically and empirically that they are salient even for real-data settings with a large number of countries.  We also offer practical remedies that can be used to obtain more reliable inferences of the effects of trade policies and other time-varying gravity variables, {which we make available via an accompanying Stata package called \href{https://github.com/tomzylkin/ppml_fe_bias}{\tt{ppml\_fe\_bias}}}.

\vfill{}  \enlargethispage{\baselineskip}

  \setcounter{page}{0} \thispagestyle{empty}
\clearpage{}

\section{Introduction}

Despite intense and longstanding empirical interest, the effects of
bilateral trade agreements on trade are still considered highly difficult
to assess. %
As emphasized in a recent practitioner's guide put out  by the WTO (\citealp{yotov_advanced_2016}),
 many current estimates in the literature suffer from easily
identifiable sources of bias (or ``estimation challenges''). This
is not for a lack of awareness. Papers showing leading causes of bias
in the gravity equation are often among the most widely celebrated
and cited in the trade field, if not in all of Economics.\footnote{For some context, if we start citation counts in 2003, \citet{anderson_gravity_2003}
and \citet{santos_silva_log_2006} are, respectively, the most cited
articles in the \emph{American Economic Review} and the \emph{Review
of Economics and Statistics}. Paling only slightly in this exclusive
company, \citet{baier_free_2007} is the 4th most-cited article in the
\emph{Journal of International Economics}, having gathered ``only''
2,500 citations. Readers familiar with these other papers will also
likely be familiar with \citet{helpman_estimating_2008}'s work on
the selection process underlying zero trade flows, an issue we do
not take up here.} In particular, it is now generally accepted that trade flows
across different partners are interdependent via {the network structure of trade}
(the main contribution of \citealp{anderson_gravity_2003}), that
log-transforming the dependent variable is not innocuous (as argued
by \citealp{santos_silva_log_2006}), and\textemdash most relevant
to the context of trade agreements\textemdash that earlier, puzzlingly
small estimates of the effects of free trade agreements were almost
certainly biased downwards by treating them as exogenous (\citealp{baier_free_2007}).

As a consequence\textemdash and aided by some recent computational
developments\textemdash researchers seeking to identify the effects
of trade agreements have naturally moved towards more advanced estimation
strategies that take on board all of the above concerns.\footnote{\citet{larch2019currency}, \citet{ppmlhdfe}, and \citet{stammann2017fast} describe
algorithms that enable fast estimation of the three-way
models considered here. } In particular, a ``three-way'' fixed effects Poisson {Pseudo-Maximum Likelihood}
(``FE-PPML'') estimator with time-varying exporter and importer fixed
effects to 
account for {network dependence} and time-invariant
exporter-importer (``pair'') fixed effects to address endogeneity
has recently emerged 
as a logical workhorse method for empirical trade policy analysis.\footnote{Pair fixed effects are of course no substitute for good instruments.
However, instruments for trade policy changes which are also exogenous
to trade are understandably hard to come by. As discussed in \citet{head_gravity_2014}'s
essential handbook chapter on gravity estimation, pair fixed effects
have the advantage that the effects of trade agreements and other
trade policies are identified from time-variation in trade within
pairs. Causal interpretations follow if standard ``parallel trend''
assumptions are satisfied.} {It also has clear potential application to the study of network data more generally, such as data on urban commuting or migration (e.g., \citealp{brinkman2019freeway}; \citealp{rothenberg2020}; \citealp{allen2018border}; {\citealp{beverelli2019migration})}.

However, one reason why {some} researchers may hesitate in embracing this
estimator is the current lack of clarity regarding how the three fixed
effects in the model may bias estimation, especially in the standard
``fixed $T$'' case where the number of time periods is small. Even
though FE-PPML estimates can be shown to be asymptotically unbiased
with a single fixed effect (a well-known result) as well as in a two-way
setting where both dimensions of the panel become large (\citealp{fernandez-val_individual_2016}),
the latter result does not come strictly as a generalization of the
former one, leaving it potentially unclear whether a three-way model
with a fixed time dimension should be expected to inherit the nice
asymptotic properties of these other models.

Accordingly, the question we investigate in this paper is the extent
to which the three-way FE-PPML estimator is affected by incidental
parameter problems (IPPs). As is well known in both statistics and
econometrics (\citealp{neyman_consistent_1948,lancaster_orthogonal_2002}), IPPs arise when estimation noise from estimates of fixed effects and other ``incidental parameters''
contaminates the scores of the main parameters of interest, inducing
bias. In the worst case, this bias renders the estimates inconsistent,
making estimation inadvisable. As we will show, while inconsistency
is actually not a problem for three-way FE-PPML, both the estimated
coefficients and standard errors are affected by meaningful
biases due to IPPs that researchers should be aware of.

To state our main results more precisely, in gravity settings where the number of countries ($N$) goes to infinity and $T$ is small, we find the following: \vskip -4em
\begin{enumerate*}  
	\item Consistency of point estimates of FE-PPML: The point estimates produced by three-way FE-PPML estimator in gravity settings are asymptotically consistent.
	\item Inconsistency of other FE-PML estimators: FE-PPML is the only estimator in a set of related FE-PML estimators sometimes considered in this context that is generally consistent. FE-Gamma PML, for example, should not be used because it is only consistent under strict assumptions.
	\item Asymptotic bias: Point estimates of the three-way FE-PPML estimator are nonetheless \emph{asymptotically biased}, meaning that the asymptotic distribution of the estimates is not centered at the truth as $N \rightarrow \infty$. In other words, it approaches the truth ``at an angle'' asymptotically (see Figure  \ref{fig1} for an illustration.)  
	\item Biased standard error estimates: Estimates of cluster-robust sandwich-type standard errors are likewise asymptotically biased due to an IPP. 
	\item Bias corrections improve inferences: Simulations show that using analytical bias corrections to address each of these biases leads to improved inferences.
These corrections are available to use via the Stata package \tt{ppml\_fe\_bias}. 
\end{enumerate*}

{To first explain  our consistency results}, our basic strategy involves using the
first-order conditions of FE-PPML to ``{profile out}'' (solve
for) the pair fixed effect terms from the first-order conditions of
the other parameters. Notably, this allows us to re-express the three-way gravity model 
as a \emph{two-way} model {in which the only
remaining incidental parameters are the exporter-time and importer-time
fixed effects.} 
Three-way FE-PPML is therefore consistent in {fixed}-$T$ settings
for largely the same reasons the two-way {models} considered in \citet{fernandez-val_individual_2016}
are consistent, and we provide suitably modified versions of the regularity
conditions and consistency results established by \citet{fernandez-val_individual_2016}
for the simpler two-way case. 

At the same time, it does not also follow that \citet{fernandez-val_individual_2016}'s
earlier results for the \emph{asymptotic unbiased}-ness of the two-way
FE-PPML estimator similarly carry over to the three-way case when $T$ is fixed. {The key is that the resulting two-way estimator that is obtained after
profiling out the fixed effects has its own special properties with respect to IPPs. When $T$ is fixed, the estimation noise in the remaining exporter-time and importer-time fixed effects induces an asymptotic bias of order $1/N$ as $N$ grows large, a result that is broadly consistent with most of the two-way settings studied in \citet{fernandez-val_individual_2016}.} However, when instead both $N$ and $T$ grow large at the same rate, the estimator turns out to be unbiased asymptotically, analogous to what \citet{fernandez-val_individual_2016} found in the two-way FE-PPML case.

{The reason why asymptotic bias is a concern is that the asymptotic standard
deviation is itself of order {$1/(N\sqrt{T})$}. Thus, when $T$ is {fixed}, both the bias
in point estimates 
 will be of comparable magnitude to their
standard errors as $N \rightarrow \infty$, 
causing the asymptotic distribution of estimates to be incorrectly centered as discussed above.
In practice, this is a less severe problem
than inconsistency, but it does mean that standard hypothesis tests
for assessing statistical significance are not reliable. One of the
objectives of this paper will be to adapt some of the leading remedies
from the recent literature on ``large $T$'' IPPs (see, e.g., \citealp{arellano_understanding_2007})
in order to re-center the asymptotic distribution of estimates and
thereby restore asymptotically valid inferences.\footnote{The new literature on ``large $T$'' asymptotic bias in nonlinear
FE {models} has emerged as a recent response to the well-known ``fixed
$T$'' consistency problem first described in \citet{neyman_consistent_1948}.
Examples include \citet{phillips_linear_1999}, \citet{hahn_asymptotically_2002},
\citet{lancaster_orthogonal_2002}, \citet{woutersen_robustness_2002},
\citet{alvarez_time_2003}, \citet{carro_estimating_2007}, \citet{arellano_robust_2009},
\citet{fernandez-val_bias_2011}, and \citet{kato_asymptotics_2012}.
Unlike in most other settings explored in this literature, the panel
estimator we consider is consistent regardless of $T$.}}

The bias in the estimated standard errors
is similar to one that has been found in two-way
gravity {settings} by several recent studies (\citealp{egger_glm_2015,jochmans_two-way_2016,pfaffermayr2019gravity,pfaffermayr2021confidence}). 
Intuitively, because the origin-time and destination-time
fixed effects in the {model} each converge to their true values at a
rate of only $1/\sqrt{N}$ (not $1/N$), the cluster-robust sandwich
estimator for the variance has a leading bias of order $1/N$ (not
$1/N^{2}$), and standard errors in turn have a bias of order $1/\sqrt{N}$.
This latter type of bias is related to the general result that standard
``heteroskedasticity-robust'' variance estimators are downward-biased
in small samples (see, e.g., \citealp{mackinnon1985some,imbens2016robust}),
including for PML estimators (\citealp{kauermann2001note}), but is more
severe in this setting due to an IPP. We should therefore be concerned
that estimated confidence intervals may be too narrow in addition
to being off-center.

For
the bias in point estimates, we construct two-way analytical and jackknife
bias corrections inspired by the corrections proposed in Fern\'andez-Val
and Weidner \citeyearpar{fernandez-val_individual_2016,ARE}. For
the bias in standard errors, we show how \citet{kauermann2001note}'s
method for correcting the PML sandwich estimator may be adapted to
the case of a conditional estimator with multi-way fixed effects and
cluster-robust standard errors. Our simulations confirm that these
methods are usually effective at improving inferences. The jackknife
correction reduces more of the bias in point estimates than
the analytical correction in smaller samples, but the analytical correction does a better
job at improving coverage, especially when also paired with corrected
standard errors.

\enlargethispage{1em}
For our empirical applications, we {first} estimate the average effects of
a free trade agreement (FTA) on trade for a range of different industries
using what would typically be considered a large trade data set, with
167 countries and 5 time periods. The biases we uncover vary in size
across the different industries, but are generally large enough to
indicate that our bias corrections should be worthwhile in most three-way
gravity settings. For aggregate trade data (which yields results that
are fairly representative), the estimated coefficient for FTA has an implied downward bias about 15\%-22\% of the estimated
standard error, and the implied downward bias in the standard error
itself is about 11\% of the original standard error.  As a means of further demonstration, we also apply our corrections to replication data from several recent papers that have used three-way gravity {models}. This latter exercise reveals several instances in which our methods make a material difference for assessing statistical significance. It also highlights the possibility that the bias in standard errors can sometimes be severe, as much as 40\% or more in some cases.

Aside from \citet{fernandez-val_individual_2016}'s
work on 
two-way nonlinear {models}, \citet{pesaran_estimation_2006},
\citet{bai_panel_2009}, \citet{hahn_reducing_2006}, and \citet{moon_dynamic_2017}
have each conducted similar analyses for two-way linear
{models} with interacted individual and time fixed effects. Turning
to three-way models, \citet{hinzetal} have recently
developed bias corrections for dynamic three-way  probit and logit {models} based on asymptotics suggested by \citet{ARE} where all three panel dimensions grow at the same rate. 
Though widely applicable, this approach is not appropriate for our setting because of the different role played by the time dimension when the estimator is FE-PPML.\footnote{Also related are the GMM-based differencing strategies for two-way
FE {models} proposed by \citet{charbonneau_multiple_2012} and \citet{jochmans_two-way_2016}.
These strategies rely on differencing the data in such as way that
the resulting GMM moments do not depend on any of the incidental parameters.
In principle, these methods could be extended to allow for differencing
across a time dimension as well in a three-way panel.} 
In the network context, \citet{graham2017econometric}, \citet{dzemski2018empirical},
and \citet{chen2014nonlinear} have studied asymptotic bias in network
{models} with node-specific (possibly sender- and receiver-specific) fixed effects.
\citet{chen2014nonlinear}'s analysis is espeically notable in that they allow these node-specific effects to be vectors
rather than scalars, similar to the exporter-time and importer-time
fixed effects that feature in gravity {models}. Our bias expansions
substantially differ from those of \citet{chen2014nonlinear} because the
equivalent outcome variable in our setting (trade flows observed over
time for a given pair) is also a vector rather than a scalar and because
we work with a conditional moment {model} where the distribution of
the outcome may be misspecified. 

In what follows, Section \ref{sec2} first provides a discussion of why IPPs are a concern for gravity {models} and of the no-IPP properties of FE-PPML. Section \ref{sec3} then establishes bias and consistency
results for the three-way gravity {model} specifically and discusses
how to implement bias corrections. Sections \ref{sim_results} and
\ref{emp_app} respectively present simulation evidence and empirical
applications. Section \ref{conclusion} concludes, and an Appendix
adds further simulation results and technical details, including proofs. 

\section{Gravity Models and IPPs} \label{sec2}

Gravity models are now routinely estimated using FE-PPML with multiple
sets of fixed effects. As we discuss in this section, these practices
follow naturally from the gravity model's theoretical microfoundations
but are not without need for further scrutiny. In particular, because
the underlying model is nonlinear, it is important to clarify that,
while PPML is known to be free from incidental parameter bias in some
special cases, it is by no means immune to IPPs in general. It will
also be useful for us to provide some general discussion of IPPs and
the different ways in which they may manifest.

\subsection{Fixed Effects and Gravity Models}

As documented in \citet{head_gravity_2014}, the emergence of rich
and varied theoretical foundations for the gravity equation has fueled
a ``fixed effects revolution'' in the gravity literature over the
last two decades. As such, we find it useful to briefly describe a
simple trade model and discuss how it may be used to motivate an estimating
equation with either two-way or three-way fixed effects.

To establish some notation we will use throughout the paper, we will
consider a world with $N$ countries and we will let $i$ and $j$
respectively be indices for exporter and importer. For now, we will
focus on deriving a two-way gravity model where the two fixed effects
account for each country's multilateral resistance. Later, we will
add a time dimension and a third fixed effect that absorbs all time-invariant
components of trade costs.

To add some theoretical structure, suppose that trade flows are given
by the following gravity equation: 
\begin{align}
y_{ij} & =\frac{y_{i}}{\Pi_{i}^{-\theta}}\frac{y_{j}}{P_{j}^{-\theta}}\tau_{ij}^{-\theta}.\label{eq:theory-1}
\end{align}
Here, $y_{i}:=\sum_{j}y_{ij}$ and $y_{j}:=\sum_{i}y_{ij}$ are the
market sizes of the two countries, $\tau_{ij}\ge1$ is a bilateral
trade cost, $\theta>0$ is the trade elasticity, and $\Pi_{i}$ and
$P_{j}$ respectively are the outward and inward multilateral resistances
from \citet{anderson_gravity_2003}, which capture how bilateral trade
flows depend on each country's opportunities for trade with third
countries. More formally, these latter terms are derived from the
following two relationships that are inherent to all general equilibrium
gravity models: 
\begin{align}
\Pi_{i}^{-\theta} & =\sum_{j=1}^{N}\frac{y_{j}}{P_{j}^{-\theta}}\tau_{ij}^{-\theta}, & P_{j}^{-\theta} & =\sum_{i=1}^{N}\frac{y_{i}}{\Pi_{i}^{-\theta}}\tau_{ij}^{-\theta}.\label{eq:MRs}
\end{align}
As shown, these terms respectively aggregate the exporter's ability
to export goods to more desirable import markets and the importer's
ability to import from more capable exporters.\footnote{This presentation of the gravity model readily conforms to the trade
models used in \citet{eaton_technology_2002} or \citet{anderson_gravity_2003},\textbf{
}though the interpretation of $\theta$ differs across the two models.
With some minor modifications, this setup can also be made compatible
with any of the theoretical gravity models considered in \citet{head_gravity_2014}
or \citet{costinot_trade_2014}. To be clear, our econometric results
do not require any particular microfoundation for the gravity equation.}

For estimation, it is typical to parameterize the trade cost $\tau_{ij}$
as depending exponentially on some variables of interest, i.e., 
\begin{align}
\tau_{ij}^{-\theta} & =\exp(x_{ij}\beta)\omega_{ij},\label{eq:tau}
\end{align}
where $x_{ij}$ are the components of trade costs whose effects we
wish to estimate. Because not all trade costs are reflected in $x_{ij}$,
we also allow for ``unobserved'' trade costs via the {idiosyncratic}
trade cost term $\omega_{ij}$. Combining \eqref{eq:theory-1} with
\eqref{eq:tau} then delivers the following estimating equation: 
\begin{align}
y_{ij} & =\exp\left(\alpha_{i}+\gamma_{j}+x_{ij}\beta\right)\omega_{ij},\label{eq:twoway}
\end{align}
where $\alpha_{i}=\ln(y_{i}/\Pi_{i}^{-\theta})$ and $\gamma_{j}=\ln(y_{j}/P_{j}^{-\theta})$
are origin and destination fixed effects that absorb market sizes
and multilateral resistances and $\omega_{ij}$ now provides a multiplicative
error term.\footnote{Alternatively, it is sometimes common to write trade costs as a log-linear
function, i.e, $\ln\tau_{ij}=x_{ij}\beta+e_{ij}$, with $e_{ij}$
now reflecting unobserved (log) trade costs. Interestingly, these
two ways of specifying the error term do not necessarily have equivalent
implications for estimation. In the log-linear formulation, if the
log-error term $e_{ij}$ is assumed to be heteroskedastic with mean
zero, Jensen's inequality implies that estimation in levels will be
biased and log-OLS will be consistent.} When we introduce the three-way model, all of the terms shown in
\eqref{eq:twoway} will have a further subscript for time, and the
the unobserved trade cost will have a time-invariant component that
will motivate the use of an added $ij$ fixed effect.

To motivate the arc of the rest of the paper, several points stand
out from the estimation suggested by \eqref{eq:twoway-moment}. First,
the implied moment condition for estimation is 
\begin{align}
\mathbb{E}(y_{ij}|x_{ij},\alpha_{i},\gamma_{j}) & =\lambda_{ij}:=\exp\left(\alpha_{i}+\gamma_{j}+x_{ij}\beta\right),\label{eq:twoway-moment}
\end{align}
{which follows after imposing $\mathbb{E}(\omega_{ij}|x_{ij},\!\alpha_i,\!\gamma_j)=1.$}\footnote{Note that consistent estimation of $\beta$ actually does not require
$\mathbb{E}(\omega_{ij}|\cdot)=1$ in this case but rather $\mathbb{E}(\omega_{ij}|\cdot)=\widetilde{\omega}_{i}\widetilde{\omega}_{j}$,
where $\widetilde{\omega}_{i}$ and $\widetilde{\omega}_{j}$ could
be country-specific components of unobserved trade costs that would
be absorbed by the fixed effects. For the three-way model, one requires $\mathbb{E}(\omega_{ijt}|\cdot)=\widetilde{\omega}_{it}\widetilde{\omega}_{jt}\widetilde{\omega}_{ij}$.} 
As discussed in \citet{santos_silva_log_2006}, consistent estimation
of the trade cost parameters in $\beta$ therefore generally requires
a nonlinear model. {Second, because unobserved trade costs enter
the country-specific terms $\alpha_{i}$ and $\gamma_{j}$ through the system
of multilateral resistances, we treat $\alpha_{i}$ and $\gamma_{j}$ as unknown parameters that will be noisily estimated,
raising concerns about a possible IPP}.\footnote{%
{As demonstrated in \citet{pfaffermayr2021confidence}, if we assume (i) there are no unobserved trade costs (such that $\omega_{ij}$ does not enter the system  in \eqref{eq:MRs}) and (ii) the aggregate quantities $y_i$ and $y_j$ are perfectly observed,
then it is better to regard $\alpha_i$ and $\gamma_j$ as reflecting constraints rather than as incidental parameters. Since \citet{fally_structural_2015} shows that constrained PPML and FE-PPML produce the same estimates in this context, there is no concern about IPPs if these assumptions are met.}}
As we go on to discuss, the FE-PPML estimator that is most often used
in this context has some special robustness against IPPs, but this
robustness does not hold for FE-PPML in general, especially
once we deviate from the two-way gravity setting implied by \eqref{eq:twoway-moment}.

\subsection{The Incidental Parameter Problem} \label{theIPP}

In the context of fixed effects models, IPPs occur when the estimation
noise in the fixed effects contaminates the scores of the other parameters
being estimated, inducing a bias. This bias can manifest in a variety
of different ways; thus it is useful to provide a generic characterization
that can illustrate the different possibilities that may arise. To
that end, let $n$ be the total number of observations and let $p$
be the total number of parameters being estimated, inclusive of any
fixed effects. As described in \citet{ARE}, what we need to be concerned
with the number of observations that are available to estimate each
fixed effect, i.e., $\ensuremath{n/p}$. More precisely, when appropriate
regularity conditions are satisfied, the estimated $\widehat{\beta}$
may generally be thought of as having the following bias and standard
deviation: 
\begin{align}
{\rm bias}(\widehat{\beta}) & =\frac{b\,p}{n}+o(p/n), & {\rm std}(\widehat{\beta}) & =\frac{c}{\sqrt{n}}+o(n^{-1/2}),\label{BiasGeneral}
\end{align}
where $b\in\mathbb{R}$ and $c>0$ are constants that depend on the
model being estimated. {As this presentation emphasizes, {all} estimators in nonlinear settings are generally biased in small samples, but this is not the same thing as saying that the bias always poses a problem for inferring statistical significance. As we will discuss, in larger samples, what matters is whether the bias disappears faster than the standard error as $n\rightarrow \infty.$\footnote{In the panel data literature, these results for the bias and standard
deviation are usually derived not for $\widehat{\beta}$ directly,
but for the asymptotic distribution of $\widehat{\beta}$ (because
there are cases where $\widehat{\beta}$ may not have a first or second
moment, but nevertheless has a well-defined limiting distribution
with finite moments). We ignore this distinction for our heuristic
discussion here.}

To provide a simple taxonomy of the cases that can arise, consider first the standard textbook
treatment of maximum likelihood estimation, where we usually have $p$ fixed while $n\rightarrow\infty$.
In this case,  the bias in $\widehat{\beta}$ becomes
asymptotically negligible as compared to the standard deviation, which crucially means that estimated confidence intervals can be expected to be centered at the truth
when the data becomes sufficiently large. By contrast, in the classical
IPP of \citet{neyman_consistent_1948}, the number of parameters grows
at the same rate as the number of observations, implying that the
bias does not converge to zero asymptotically. In that case, the fixed
effect estimator is inconsistent.}%

The gravity model with two-way fixed effects then serves to illustrate
a \emph{third} possibility that will also be applicable to the results that
follow for three-way gravity models. In the two-way gravity setting,
$p$ is on the order of $2N$, where $N$ is the number of countries,
and $n$ is on the order of $N^{2}$. Consequently, the bias and standard
deviation of $\widehat{\beta}$ are given by 
\begin{align}
{\rm bias}(\widehat{\beta}) & =\frac{2\,b}{N}+o(1/N), & {\rm std}(\widehat{\beta}) & =\frac{c}{N}+o(1/N),\label{BiasTwoWay}
\end{align}
In this case, as $N\rightarrow\infty$, the estimated $\widehat{\beta}$
is consistent (both the standard deviation and bias converge to zero),
but we also have 
\[
\lim_{N\rightarrow\infty}\;\frac{{\rm bias}(\widehat{\beta})}{{\rm std}(\widehat{\beta})}=\frac{2\,b}{c}.
\]
{As we discuss below, the two-way PPML gravity {estimator} is a special case
where we actually have that $b=0$. However, for other two-way gravity
{estimators} (such as two-way Gamma PML for example), we generally
have that $b\neq0$, meaning the bias will \emph{not} disappear
relative to the standard error as $N\rightarrow\infty$.} Compared to \citet{neyman_consistent_1948},
the IPP these estimators suffer from is not an inconsistency problem but
rather an \emph{asymptotic bias} 
problem, whereby the slow convergence of the fixed effects causes
the {asymptotic distribution} for $\widehat{\beta}$ to be incorrectly
centered as it converges to the truth.\footnote{Consistency here follows from how the number of fixed effects grows
only with the square root of the sample size, as discussed in \citet{egger_trade_2011}.
The bias in the asymptotic distribution for two-way {models} was proven
by \citet{fernandez-val_individual_2016}, discussed below.} This version of the IPP is more benign, but ignoring the bias will
nonetheless result in invalid inferences and test results. 
The ``large
$T$'' panel data literature therefore discusses various methods
for bias correction of $\widehat{\beta}$ that restore asymptotically
valid inference. Importantly, the degree to which inferences are biased
depends on the bias constant $b$, which {cannot be} known beforehand
without applying such a correction.

{To provide a more visual illustration of these ideas,
Figure \ref{fig1} presents simulation results for the three cases
we have just discussed: inconsistency (top-left), no asymptotic bias
(top-right), and asymptotic bias (bottom-left). Since asymptotic bias
will ultimately be our focus, it is worth noting from the figure how the estimates
are consistent in this case\textemdash the distribution will collapse
to the true value as $N\rightarrow\infty$\textemdash but confidence
bounds based on these estimates will clearly be inappropriate.}

\renewcommand{\baselinestretch}{1.05}
\begin{figure}
\centering{}\includegraphics[scale=0.8]{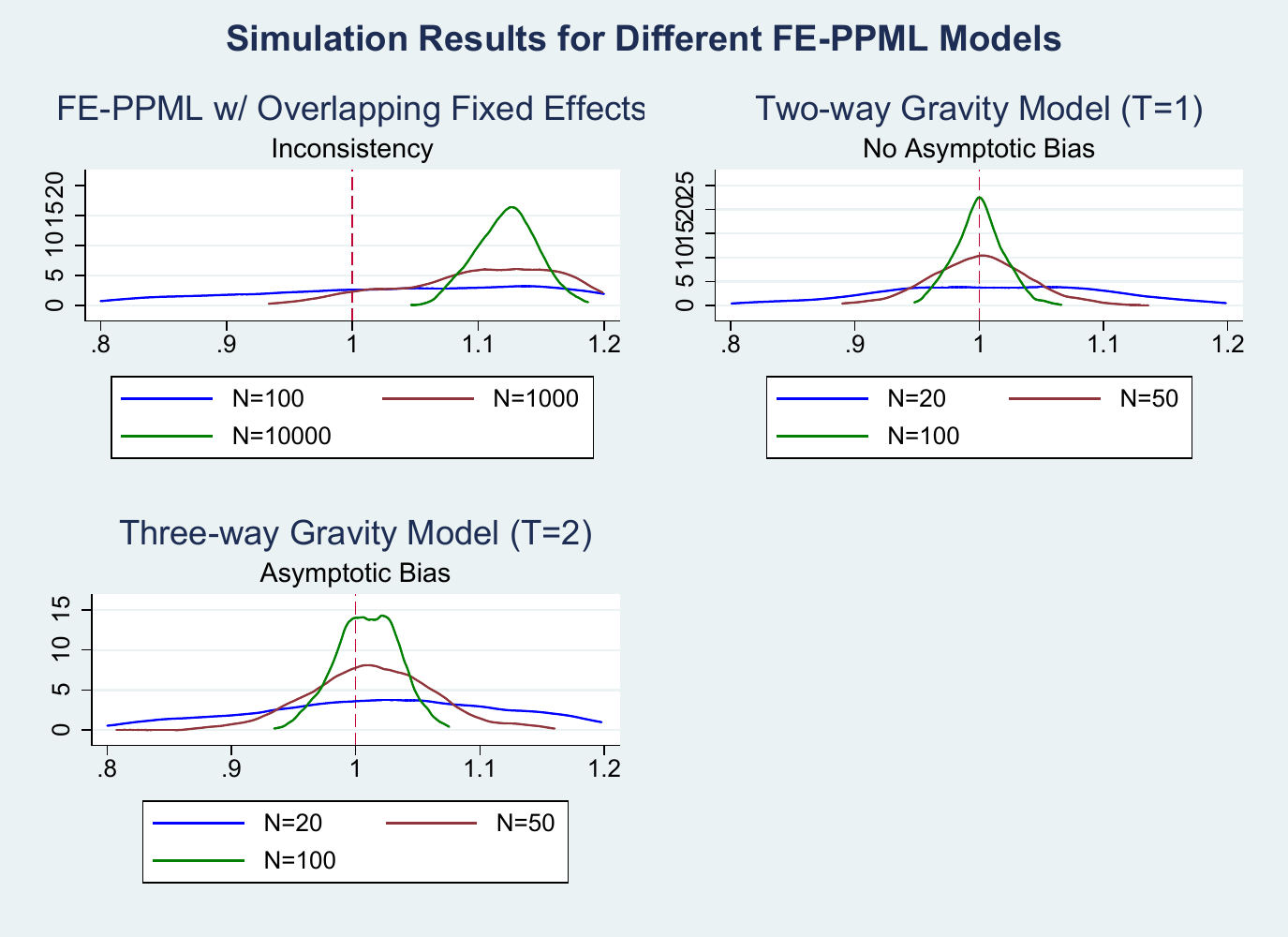}\caption{{\footnotesize Kernel density plots of FE PPML estimates for 3 different models, using 500 replications. Clockwise from top left, the 3 models are:
$y_{it}=\exp[\alpha_{i}\times1(t\le2)+\gamma_{i}\times1(t\ge2)+x_{it}\beta]\omega_{it}$, with the $t$ dimension of the panel fixed at $T=3$; 
a two-way gravity model with $y_{ij}=\exp[\alpha_{i}+\gamma_{j}+x_{ij}\beta]\omega_{ij}$; 
a three-way gravity model with $y_{ijt}=\exp[\alpha_{it}+\gamma_{jt}+\eta_{ij}+x_{ijt}\beta]\omega_{ijt}$ and $T=2$. {Respectively, these models exemplify inconsistency, no asymptotic bias, and asymptotic bias.} The $i$ and $j$ dimensions of the panel both have size $N$ in the latter two models. The true value of $\beta$ is 1 (indicated
by the vertical dotted lines) and the data is generated using ${\rm Var}(y|\cdot)=\mathbb{E}(y|\cdot)$.
}\label{fig1}}
\end{figure}
\renewcommand{\baselinestretch}{1.3}

\subsection{How FE-PPML is Different}

Our discussion of IPPs thus far has been for the generic estimation of a nonlinear model. However, the PPML estimator that is most commonly used to estimate
gravity models actually behaves very differently than other estimators
in this context. In the classic panel data setting with ``one way'' fixed
effects, for example, FE-PPML has the very special property that the
IPP bias constant $b$ turns out to be zero, meaning that it is {asymptotically
unbiased} in situations where other estimators tend to be inconsistent.
As discussed in \citet{wooldridge_distribution-free_1999}, the reason
behind this result is that {the same estimator can be obtained from a multinomial model} that does not depend on the fixed effects.\footnote{The earliest references to present versions of this result include
\citet{andersen1970asymptotic}, \citet{palmgren_fisher_1981}, and
\citet{hausman_econometric_1984}. \citet{wooldridge_distribution-free_1999}'s
contribution is to show that FE-PPML is consistent even when the assumed
distribution of the data is misspecified. Our Lemma \ref{lemma1}
in the Appendix clarifies that FE-PPML is relatively unique in this
regard versus similar models.}

This special property of FE-PPML has important implications for estimating
gravity models as well. For two-way gravity settings, the asymptotic
bias of $\widehat{\beta}$ when $N\rightarrow\infty$ was worked out
in \citet{fernandez-val_individual_2016}. They show that the two
IPP contributions from $\alpha_{i}$ and $\gamma_{j}$ ``decouple''
asymptotically, such that the overall bias can be decomposed as the
sum of two bias terms that would be expected in a one-way setting,
i.e., $b_{(\alpha)}/N+b_{(\gamma)}/N$, {where the $N$'s come from
the number of observations associated with each fixed effect.} Because
$b=0$ for the one-way FE-PPML case, we also have $b_{(\alpha)}=b_{(\gamma)}=0$
in the two-way setting; that is, two-way FE-PPML gravity estimates
for $\widehat{\beta}$ are asymptotically unbiased just as one-way
FE-PPML estimates are.\footnote{Note that Theorem~4.1 in \citet{fernandez-val_individual_2016} is
written for the correctly specified case, where $y_{ij}$ is actually
Poisson distributed. However, Remark 3 in the paper gives the extension
to conditional moment models, where for the FE-PPML case only the
moment condition in \eqref{eq:twoway-moment} needs to hold. Their
paper considers standard panel models, as opposed to trade models,
but the only technical difference is that $y_{ij}$ is often not observed
for the trade model when $i\neq j$. This missing diagonal has no
meaningful effect on any of the results we discuss.}

Taken together, these results might create the impression that FE-PPML
is generally immune to IPPs, regardless of what fixed effects are
included in the model. Thus, it is important to clarify that a key
feature of the two-way gravity model is that both fixed effect dimensions
grow only with the square root of the panel size, such that the estimation
noise in the estimated $\widehat{\alpha}_{i}$'s and $\widehat{\gamma}_{j}$'s
disappears asymptotically. {As we discuss in the Appendix,
if we instead consider a model where both fixed effects grow
with $n$ rather than with its square root, the IPPs associated with
each fixed effect do not decouple from one another, and FE-PPML in
this case is actually inconsistent.}

To synthesize these points, FE-PPML has a very special property\textemdash one
can condition out one of the fixed effects\textemdash but this property
has an important limitation\textemdash the resulting multinomial {model}
does not inherit the same no-bias properties as the original PPML
estimator with respect to any further fixed effects. Both of these
results will be fleshed out in more detail in the following section
when we recast the three-way gravity model as a {two-way multinomial
model} in order to obtain an appropriate expression for the bias. Doing so
will also allow us to highlight another reason why FE-PPML is
not immune to IPPs: even for the two-way gravity model, while the
$\alpha_{i}$ and $\gamma_{j}$ parameters do not induce an IPP bias
in $\widehat{\beta}$, they nonetheless have implications for the
estimated variance that are not innocuous; we thus will devote
attention to this issue as well. 

\section{Results for the Three-way Gravity Model} \label{sec3}

To recap the sequence of results just described, we know that FE-PPML
estimates with one fixed effect do not suffer from an IPP. We also
know that
FE-PPML may have an IPP in models with more than one fixed effect, but it is both consistent and asymptotically 
unbiased in two-way gravity settings where neither fixed effect dimension grows at the same rate as the size of the panel.
As we will now show,
each of these earlier results will be useful for understanding the
more complex case of a three-way gravity model that adds a time
dimension and a third set of fixed effects to the above two-way model.
We also describe a series of bias corrections for the three-way model, 
including for the possible downward bias
of the estimated standard errors.

\subsection{Consistency} \label{sec3_1}

{To formally introduce the three-way model, we add an explicit
time subscript $t\in\{1,\ldots,T\}$ to $y_{ij}$, $x_{ij}$, and $\omega_{ij}$ from the prior model and add a ``country-pair''-specific fixed effect $\eta_{ij}$, such that trade costs are now given by $\tau_{ijt}^{-\theta}=e^{x_{ijt}'\beta+\eta_{ij}}\omega_{ijt}$. All other elements in the original trade model likewise acquire a time subscript, meaning that $\alpha_{it}=\ln{y_{it} / \Pi_{it}^{-\theta}}$ and $\gamma_{jt}=\ln{y_{jt}/P_{jt}^{-\theta}}$ also must be indexed by $t$.} The model now reads  
\begin{align}
\mathbb{E}(y_{ijt}|x_{ijt},\alpha_{it},\gamma_{jt},\eta_{ij}) & =\lambda_{ijt}:=\exp(x_{ijt}'\beta+\alpha_{it}+\gamma_{jt}+\eta_{ij}),\label{MeanY}
\end{align}
where the three fixed effects now respectively index exporter-time,
importer-time, and country-pair.\footnote{{Note that a multiplicative error term is not necessary to deliver the moment condition in \eqref{MeanY}. {We could instead have an additive error term $\varepsilon_{ijt}=y_{ijt}-\lambda_{ijt}$. In this case}, it would be more natural to think of it as coming from measurement error. In addition, note that we assume the true model is as written in \eqref{MeanY} and assume away, e.g., any unobserved heterogeneity in $\beta$. Allowing for this type of heterogeneity is an important extension for future work to address.}} The unobserved trade cost $\omega_{ijt}\ge0$ continues to serve as an {error term}, such that $y_{ijt}=\lambda_{ijt}\omega_{ijt}\ge0$. {We thus allow for zero trade flows}.
For the asymptotics using the three-way model, we consider $T$ fixed,
while $N\rightarrow\infty$. The FE-PPML estimator maximizes 
\begin{align*}
     \mathcal{L}(\beta, \alpha, \gamma, \eta)
     &:=  \sum_{i=1}^N \sum_{\begin{minipage}[c]{0.55cm} $\scriptstyle j=1$ \\[-10pt] $\scriptstyle j \neq i$ \end{minipage}}^N \sum_{t=1}^T
   \left( y_{ijt} \log \lambda_{ijt} - \lambda_{ijt} \right) 
\end{align*}
over $\beta$, $\alpha$, $\gamma$ and $\eta$.

{Our strategy for showing the consistency of this estimator will capitalize on the special properties of FE-PPML discussed in the previous section. In particular, we will exploit the fact} that not all of the fixed effect dimensions grow at the same rate as $N$ increases. The numbers of exporter-time and importer-time fixed effects each increase with $N$ (as before), but the dimension of {the pair fixed effect} $\eta$ increases with $N^2$, since adding another country to the data adds another $N-1$ {pairs} to the estimation. It therefore makes sense to first ``profile out'' (i.e., solve for)  $\eta$  {so that we may deal with the remaining two fixed effects in turn}. For given values of $\beta$, $\alpha$, $\gamma$, {maximizing over $\eta$ gives us}
\begin{align}
      \exp\left[ \widehat \eta_{ij}(\beta, \alpha, \gamma) \right]
      &=
      \frac{ \sum_{t=1}^T y_{ijt}  }
            { \sum_{t=1}^T   \mu_{ijt}  } ,
       &
       \mu_{ijt} & := \exp( x_{ijt}' \beta + \alpha_{it} + \gamma_{jt})     \label{eta} .
\end{align}
We therefore have
\begin{align}
     \mathcal{L}(\beta, \alpha, \gamma)
     & := 
     \max_{\eta} \mathcal{L}(\beta, \alpha, \gamma, \eta)
    =   \sum_{i=1}^N \sum_{\begin{minipage}[c]{0.55cm} $\scriptstyle j=1$ \\[-10pt] $\scriptstyle j \neq i$ \end{minipage}}^N 
        \ell_{ij}(\beta, \alpha_{it}, \gamma_{jt})  \label{profile},
\end{align}
with {
\begin{align}
       \ell_{ij}(\beta, \alpha_{it}, \gamma_{jt}) 
       &:= 
       \sum_{t=1}^T
    y_{ijt} \log\left(  \frac{ \mu_{ijt} } {\sum_{s=1}^T \mu_{ijs}} \right)  
    + \text{terms not depending on parameters},  
    \label{DefLikelihood}
\end{align}
thus leaving us with the likelihood
of a multinomial model where the only incidental parameters are $\alpha_{it}$ and $\gamma_{jt}$. } 
{Using \eqref{MeanY}, one can easily verify that
there is no bias in the score of the profile log-likelihood 
$  \ell_{ij}(\beta, \alpha_{it}, \gamma_{jt}) $ when evaluated at the true parameters $\beta^0$, $\alpha_{it}^0$, and $\gamma_{jt}^0$.} {The reason for this is exactly the same as for the classic panel data setting discussed above.} {Furthermore, the remaining fixed effects $\alpha_{it}$ and $\gamma_{jt}$ grow only with the square root of the sample size as $N\rightarrow\infty$, implying that they are consistently estimated.} {This in turn leads us to  the following result}:

\begin{proposition} 
So long as the set of non-fixed effect regressors $x_{ijt}$ is exogenous to the disturbance $\omega_{ijt}$ after conditioning on the fixed effects $\alpha_{it}$, $\gamma_{jt}$, and $\eta_{ij}$, FE-PPML estimates of $\beta$ from the three-way gravity model are consistent for $N\rightarrow\infty$.\footnote{%
This consistency result can be seen as a corollary of the asymptotic normality result in Proposition~\ref{bias-results} below,
for which formal regularity conditions are stated in Assumption~\ref{ass:MAIN} of the Appendix.
}
\label{prop1} 
\end{proposition} 

{{{Intuitively, this result follows because of how the special properties of FE-PPML allow us to rewrite the three-way gravity model as a two-way model without introducing a $1/T$ bias. The form of the bias in $\widehat{\beta}$ is therefore the same as in \eqref{BiasTwoWay}, such that three-way FE-PPML is consistent as $N\rightarrow\infty$ largely for the same reason two-way FE-PPML and other two-way PML gravity estimators are generally consistent. However, in the context of \emph{three-way} estimators, we can also state a stronger result that applies more narrowly to FE-PPML in particular:

\begin{proposition} Assume the conditional mean is given by $\lambda_{ijt}=\exp(x_{ijt}'\beta+\alpha_{it}+\gamma_{jt}+\eta_{ij})$ and consider the class of ``three-way'' FE-PML gravity estimators with FOC's given by 
\begin{align*}
\widehat{\beta}\!\!:\,\sum_{i=1}^{N}\sum_{\begin{minipage}[c]{0.55cm} $\scriptstyle j=1$ \\[-10pt] $\scriptstyle j \neq i$ \end{minipage}}^{N}\sum_{t=1}^{T}\,\!x_{ijt}\!\left(y_{ijt}-\widehat{\lambda}_{ijt}\right)\!g(\widehat{\lambda}_{ijt}) & =0, & \widehat{\alpha}_{it}\!\!:\,\sum_{j=1}^{N}\left(y_{ijt}-\widehat{\lambda}_{ijt}\right)\!g(\widehat{\lambda}_{ijt}) & =0,\\
\widehat{\gamma}_{jt}\!\!:\,\sum_{i=1}^{N}\left(y_{ijt}-\widehat{\lambda}_{ijt}\right)\!g(\widehat{\lambda}_{ijt}) & =0, & \widehat{\eta}_{ij}\!\!:\,\sum_{t=1}^{T}\left(y_{ijt}-\widehat{\lambda}_{ijt}\right)\!g(\widehat{\lambda}_{ijt}) & =0,
\end{align*}
where $i,j=1,\ldots,N$, $t=1,...,T,$ and $g(\widehat{\lambda}_{ijt})$
is an arbitrary function of $\widehat{\lambda}_{ijt}$ {that can be specialized to construct various PML estimators. For example, $g(\widehat{\lambda}_{ijt})=1$ delivers PPML,  $g(\widehat{\lambda}_{ijt})=\widehat{\lambda}_{ijt}^{-1}$ delivers Gamma PML, etc.} If $T$ is
{fixed}, then for $\widehat{\beta}$ to be consistent under 
general assumptions about ${\rm Var}(y|x,\alpha,\gamma,\eta)$, 
we must have that $g(\lambda_{ijt})$ is constant over the range of $\lambda$'s that are realized 
in the data-generating process. That is, the estimator must be equivalent to 
FE-PPML.\label{prop2}\end{proposition}

{In other words, three-way FE-PPML is unique among three-way PML estimators} in that its consistency does not require strong assumptions about the conditional variance of $y_{ijt}$. To draw an appropriate contrast, it is possible to obtain a closed form solution for the pair fixed effect $\widehat{\eta}_{ij}$ so long as $g(\widehat{\lambda}_{ijt})$ is of the form $g(\widehat{\lambda}_{ijt})=\widehat{\lambda}_{ijt}^q$, where $q$ can be any real number. Notably, this latter class of estimators not only includes FE-PPML (for which $q=0$), but also includes other popular gravity estimators such as Gamma PML ($q=-1$) and Gaussian PML ($q=1$). However, as we discuss in the Appendix, these other estimators are only consistent if the conditional variance is proportional to ${\lambda}_{ijt}^{1-q}$, in which case they inherit the properties of their associated MLE estimators.}

\subsection{Asymptotic Bias \label{sec3_2}}
 
Because three-way FE-PPML inherits the consistency properties
of the two-way estimator, one might expect that it also inherits its
{``no asymptotic bias'' properties} as well. However, this is where the
limitations of PPML's no-IPP properties become apparent. While the
profile log-likelihood in \eqref{profile} is now of a similar form to
the two-way {models} considered in \citet{fernandez-val_individual_2016},
notice that it no longer resembles the original FE-PPML log-likelihood.
Their no-bias result for two-way FE-PPML 
therefore does not carry over to {the three-way model}, and it
is possible to show that FE-PPML {estimates have} an asymptotic bias in this setting.

{Before proceeding, it is helpful to first revisit the intuition established
in Section \ref{theIPP} that shapes how we expect the bias to behave. As we have discussed, in a model with $p$
parameters and $n$ observations, the bias should be proportional
to $p/n$, whereas the standard error should vary with $1/\sqrt{n}$.
After profiling out $\eta$, the resulting two-way model has $\sim2NT$
parameters vs. $\sim N^{2}T$ observations. If $T$ is held fixed, we would expect
the bias and the standard error to decrease at the same rate ($1/N$), raising concerns about
a possible asymptotic bias problem and guiding us as to its form. Readers should keep this intuition in mind 
in reading through the technical details that follow.}

{To illustrate more precisely where the bias comes from}, {it is necessary to examine how the estimated
fixed effects enter the score for ${\beta}$ using a Taylor
expansion. 
To that end, first let $\phi:={\rm vec}(\alpha,\gamma)$
be a vector that collects all of the exporter-time and importer-time fixed effects,
such that we can rewrite $\ell_{ij}$ slightly as $\ell_{ij}=\ell_{ij}(\beta,\phi)$.
We can then similarly define the function $\widehat{\phi}(\beta)$ as collecting the estimated fixed effects $\widehat{\alpha}$ and $\widehat{\gamma}$ as functions of $\beta$.
Next, we construct a second-order expansion of the 
score for ${\beta}$ around the true {set of fixed effects}
$\phi^{0}$ and evaluated at the true parameter $\beta^{0}$: 
\begin{align}
\mathbb E \left[ \frac{\partial\ell_{ij}(\beta^{0},\widehat{\phi}(\beta^{0}))}{\partial\beta} \right] & \approx\underbrace{\mathbb E \left[\frac{\partial\ell_{ij}(\beta^{0},\phi^{0})}{\partial\beta} \right]}_{=0}+\underbrace{\mathbb E \left[ \frac{ \partial^{2}\ell_{ij}(\beta^{0},\phi^{0})}{\partial\beta\partial\phi^{\prime}}\left(\widehat{\phi}(\beta^{0})-\phi^{0}\right) \right] }_{\neq 0 \text{ (reflects estimation error in FEs)} }\nonumber \\
 & +\underbrace{\frac{1}{2}\sum_{f,g}^{\dim\phi} \mathbb E \left[  \frac{\partial^{3}\ell_{ij}(\beta^{0},\phi^{0})}{\partial\beta\partial\phi_{f}\partial\phi_{g}}\left(\widehat{\phi}_{f}(\beta^{0})-\phi_{f}^{0}\right)\left(\widehat{\phi}_{g}(\beta^{0})-\phi_{g}^{0}\right) \right]}_{\neq 0  \text{ (reflects variances and covariances of FEs)}}.\label{eq:expansion}
\end{align}
}This expression is near-identical to a similar expansion that appears
in \citet{fernandez-val_individual_2016}\textemdash differing mainly
in that $\ell_{ij}$ is a vector rather than a scalar\textemdash and
communicates the same essential insights: because the latter two terms
in \eqref{eq:expansion} are generally not equal to zero, the score
for ${\beta}$ is biased, with the bias depending on the interaction
between the higher-order partial derivatives of $\ell_{ij}$ and the
estimation errors in $\widehat{\alpha}_{i}$ and $\widehat{\gamma}_{j}$
as well as their variances and covariances.

Demonstrating the bias in {this particular setting} then requires that we  introduce some additional notation, mainly to provide some shorthand for the higher-order partial derivatives of $\ell_{ij}$ that appear in \eqref{eq:expansion}. To do so, we first find it convenient to let $\vartheta_{ijt}:=\lambda_{ijt}/\sum_{\tau}\lambda_{ij\tau}$. 
{We then define
the $T\times1$ ``score'' vector $S_{ij}$, the $T \times T$ ``Hessian'' matrix $H_{ij}$ and the $T\times T\times T$ cubic tensor  $G_{ij}$ (the ``third partial''), with their respective elements given by
\begin{align*}
     S_{ij,t} &= \frac{\partial \ell_{ij} } {\partial \alpha_{it}} = y_{ijt}-\vartheta_{ijt}\sum_{\tau}y_{ij\tau} ,
    \\
     H_{ij,ts} &= - \frac{\partial^2 \ell_{ij}} {\partial \alpha_{it} \, \partial \alpha_{is}} =
     \left\{ \begin{array}{ll}
                           \vartheta_{ijt}\left(1-\vartheta_{ijt}\right)\sum_{\tau}y_{ij\tau} & \text{if $t=s$,}
                           \\
                        -\vartheta_{ijs}\vartheta_{ijt}\sum_{\tau}y_{ij\tau} & \text{if $t\neq s$,}
              \end{array} 
              \right.
 \\
        G_{ij,tsr} &=  \frac{\partial^3  \ell_{ij} } {\partial \alpha_{it} \, \partial \alpha_{is} \, \partial \alpha_{ir}},
\end{align*}
where it should be understood that all of these terms are evaluated at the true values for all parameters.} 
Explicit formulas for $G_{ij,tsr}$ are provided in the Appendix. 

{The value of defining these objects is that they allow us to easily form terms identified by \eqref{eq:expansion} as being important for the bias of the score. For example,
$S_{ij}$ 
allows
us to obtain $\partial\ell_{ij}/\partial\beta^{k}={x_{ij,k}^{\prime}S_{ij}}$.
Likewise, we also have that 
$\partial^{2}\ell_{ij}/\partial\alpha_{i}\partial\beta^{k}=\partial^{2}\ell_{ij}/\partial\gamma_{j}\partial\beta^{k}=-H_{ij}x_{ij,k}$
and that
\[
\frac{\partial^{3}\ell_{ij}}{\partial\alpha_{i}\partial\alpha_{i}^{\prime}\partial\beta^{k}}=\frac{\partial^{3}\ell_{ij}}{\partial\alpha_{i}\partial\gamma_{j}^{\prime}\partial\beta^{k}}=\frac{\partial^{3}\ell_{ij}}{\partial\gamma_{j}\partial\alpha_{i}^{\prime}\partial\beta^{k}}=\frac{\partial^{3}\ell_{ij}}{\partial\gamma_{j}\partial\gamma_{j}^{\prime}\partial\beta^{k}}=G_{ij}x_{ij,k},
\]
where we use the convention that  $G_{ij}x_{ij,k}$ is
a $T\times T$ matrix with elements $[G_{ij}x_{ij,k}]_{st}=\sum_{r}G_{ijrst}x_{ijr,k}$.} We also find it useful to define the expected
Hessian $\bar{H}_{ij}=\mathbb{E} (H_{ij} \, |\,x_{ij} )$ and, similarly,
the expected third partial $\bar{G}_{ij}=\mathbb{E} (G_{ij} \,|\,x_{ij} )$.\footnote{%
{Because $\bar{H}_{ij}$
is only positive semi-definite (not positive definite),
we use a Moore-Penrose pseudoinverse whenever the analysis requires we work with an inverse of $\bar{H}_{ij}$. Specifically, we have that $\bar{H}_{ij}\,\iota_{T}=0$, where $\iota_{T}=(1,\ldots,1)^{\prime}$
is a T-vector of ones. Thus, $\bar{H}_{ij}$ is only of rank $T-1$
rather than of rank $T$.} }
Finally,  we define the $K$-vector  $\widetilde{x}_{ij}$ 
as an appropriate two-way within-transformation of $x_{ij}$ {that purges it of the fixed effects}; see the Appendix for details.}

{Obtaining a tractable expression for the bias then involves following the logic of \eqref{eq:expansion} and plugging in the just-defined objects $S_{ij}$, $H_{ij}$, $G_{ij}$, and $\widetilde{x}_{ij}$ where appropriate. Before doing so, we invoke the assumption that observations are serially correlated within pairs but independent across pairs, as is commonly assumed in the literature (see \citealp{yotov_advanced_2016}.) This assumption turns out to cause the IPPs associated with $\alpha_i$ and $\gamma_j$ to ``decouple'',\footnote{%
In particular, all elements of the cross-partial objects $\mathbb{E}[\partial^{2}\ell_{ij}/\partial\alpha_i \partial\gamma_{j}]$, $\mathbb{E}[\partial^{3}\ell_{ij}/\partial\alpha_i \alpha_i^{\prime} \partial\gamma_{j}]$, etc. can be shown to be asymptotically small for $N\rightarrow \infty$. Thus, in what follows, $B_N$ reflects the contribution of the $\alpha_i$ parameters to the bias and $D_N$ reflects the contribution of the $\gamma_j$ parameters. {As we discuss in the Appendix, relaxing this assumption can change the expression of the bias.}}  leading to the following proposition:}

\begin{proposition} Under appropriate regularity conditions
 {(Assumption~\ref{ass:MAIN} in the Appendix)}, for $T$ fixed and $N\rightarrow\infty$ we have 
\begin{align*}
\sqrt{N\,(N-1)}\;\left(\widehat{\beta}-\beta^{0}-\frac{W_{N}^{-1}(B_{N}+D_{N})}{N-1}\right)\,\rightarrow_{d}\,{\cal N}\left(0,W_{N}^{-1}\,\Omega_{N}\,W_{N}^{-1}\right),
\end{align*}
where $W_{N}$ and $\Omega_{N}$ are $K\times K$ matrices given by
\begin{align*}
W_{N} & =\frac{1}{N\,(N-1)}\sum_{i=1}^{N} \sum_{j\in\mathfrak{N}\setminus\{i\}} \widetilde{x}_{ij}^{\prime}\,\bar{H}_{ij}\,\widetilde{x}_{ij},\\
\Omega_{N} & =\frac{1}{N\,(N-1)}\sum_{i=1}^{N} \sum_{j\in\mathfrak{N}\setminus\{i\}} \widetilde{x}_{ij}^{\prime}\,\left[{\rm Var}\left(S_{ij}\,\big|\,x_{ij}\right)\right]\,\widetilde{x}_{ij},
\end{align*}
and $B_{N}$ and $D_{N}$ are $K$-vectors with elements given by
\begin{align*}
B_{N}^{k} & =-\frac{1}{N}\sum_{i=1}^{N}\mathrm{Tr}\left[\left(\sum_{j\in\mathfrak{N}\setminus\{i\}}\bar{H}_{ij}\right)^{\dagger}\sum_{j\in\mathfrak{N}\setminus\{i\}}\mathbb{E}\left(H_{ij} \, \widetilde x_{ij,k} \, S_{ij}'\big|x_{ij,k}\right)\right]\\
 & \hspace{-0.5cm}+\frac{1}{2\,N}\sum_{i=1}^{N}\mathrm{Tr}\left[\left(\sum_{j\in\mathfrak{N}\setminus\{i\}}\bar{G}_{ij}\,\widetilde{x}_{ij,k}\right)\left(\sum_{j\in\mathfrak{N}\setminus\{i\}}\bar{H}_{ij}\right)^{\dagger}\left[\sum_{j\in\mathfrak{N}\setminus\{i\}}\mathbb{E}\left(S_{ij}\,S{}_{ij}^{\prime}\big|x_{ij,k}\right)\right]\left(\sum_{j\in\mathfrak{N}\setminus\{i\}}\bar{H}_{ij}\right)^{\dagger}\right],\\
D_{N}^{k} & =-\frac{1}{N}\sum_{j=1}^{N}\mathrm{Tr}\left[\left(\sum_{i\in\mathfrak{N}\setminus\{j\}}\bar{H}_{ij}\right)^{\dagger} \sum_{i\in\mathfrak{N}\setminus\{j\}}\mathbb{E}\left(H_{ij} \, \widetilde x_{ij,k} \, S_{ij}'\big|x_{ij,k}\right)\right]\\
 & \hspace{-0.5cm}+\frac{1}{2\,N}\sum_{j=1}^{N}\mathrm{Tr}\left[\left(\sum_{i\in\mathfrak{N}\setminus\{j\}}\bar{G}_{ij}\,\widetilde{x}_{ij,k}\right)\left(\sum_{i\in\mathfrak{N}\setminus\{j\}}\bar{H}_{ij}\right)^{\dagger}\left[\sum_{i\in\mathfrak{N}\setminus\{j\}}\mathbb{E}\left(S_{ij}\,S'_{ij}\big|x_{ij,k}\right)\right]\left(\sum_{i\in\mathfrak{N}\setminus\{j\}}\bar{H}_{ij}\right)^{\dagger}\right],
\end{align*}
where a $\dagger$ denotes a Moore-Penrose pseudoinverse.
\label{bias-results}\end{proposition}
The above proposition establishes the asymptotic
distribution of the three-way gravity estimator as $N\rightarrow \infty$, including the asymptotic
bias $(N-1)^{-1}W_{N}^{-1}(B_{N}+D_{N})$.
{Intuitively, this bias
can be decomposed as the product of the {inverse expected Hessian with respect to $\beta$} (i.e.\ $W_{N}^{-1}$), the rate of asymptotic convergence (essentially $1/N$), and the bias of the score from \eqref{eq:expansion}, which here is given by the combined term $B_{N}+D_{N}$. $B_N$ reflects the contribution to the bias from the noise in $\widehat \alpha_i$, whereas $D_N$ reflects the contribution from the noise in $\widehat \gamma_j$. {The first terms in both $B_N$ and $D_N$ come from the second term in \eqref{eq:expansion}, reflecting the estimation error in the estimated fixed effects, and the second terms in $B_N$ and $D_N$ echo the third term in \eqref{eq:expansion}, reflecting their variance.}} 

Thus, in the end, the bias reduces to essentially the same simple formula we gave in Section \ref{theIPP} for the  bias of the two-way gravity model, i.e.,
\[
\frac{1}{N-1}\,b_{(\alpha)}+\frac{1}{N-1}\,b_{(\gamma)},
\] where $b_{(\alpha)}=W_{N}^{-1} B_{N}$ and $b_{(\gamma)}=W_{N}^{-1} D_{N}$ are constants that do not vary with $N$.
Importantly, and unlike in the two-way FE-PPML setting,
the three-way model does not give us the no-bias result that $B_{N}=D_{N}=0$, as the following discussion helps to illustrate.

\subsubsection*{Illustrating the Bias using the $T=2$ Case}

Admittedly, the complexity of the objects that appear in Proposition
\ref{bias-results} may make it difficult to appreciate the general
point that the three-way estimator is not {asymptotically unbiased}. One way to make
these details more transparent is to focus our attention on the simplest
possible panel model where $T=2$. The convenient thing about this
simplified setting is that the likelihood function $\ell_{ij}$ can be
reduced to just a scalar: $\ell_{ij}=y_{ij1}\log\vartheta_{ij1}+y_{ij2}\log\left(1-\vartheta_{ij1}\right)$,
where now
{
\begin{align*}
\vartheta_{ij1} & =\frac{\exp\left(\Delta x_{ij}\beta+ \Delta \alpha_{i} + \Delta \gamma_{j} \right)}{\exp\left(\Delta x_{ij}\beta+ \Delta \alpha_{i} + \Delta \gamma_{j} \right)+1} ,
\end{align*}
with $\Delta x_{ij}=x_{ij1}-x_{ij2}$, $\Delta \alpha_{i}= \alpha_{i1}- \alpha_{i2}$, and $\Delta \gamma_{j}= \gamma_{j1}- \gamma_{j2}$. {These normalizations allow us to express all of the objects that appear in Proposition \ref{bias-results} as also just scalars, and we can therefore easily derive the following result:}
}

\begin{remark}For $T=2$, we calculate $S_{ij}=\vartheta_{ij2}y_{ij1}-\vartheta_{ij1}y_{ij2},$
$H_{ij}=\vartheta_{ij1}\vartheta_{ij2}(y_{ij1}+y_{ij2})$, $\bar{H}_{ij}=\vartheta_{ij1}\lambda_{ij2}$,
$G_{ij}=\vartheta_{ij1}\vartheta_{ij2}(\vartheta_{ij1}-\vartheta_{ij2})(y_{ij1}+y_{ij2})$,
$\bar{G}_{ij}=\vartheta_{ij1}(\vartheta_{ij1}-\vartheta_{ij2})\lambda_{ij2}$,
and $\Delta\widetilde{x}_{ij}=\widetilde{x}_{ij1}-\widetilde{x}_{ij2}$.
The bias term $B_{N}^{k}$ in Proposition \ref{bias-results} can
then be written as
\begin{align*}
B_{N}^{k} & =\plim_{N\rightarrow\infty}\left[-\frac{1}{N}\sum_{i=1}^{N}\frac{\sum_{j\neq i}\Delta\widetilde{x}_{ij}\vartheta_{ij1}\vartheta_{ij2}\!\left[\vartheta_{ij2}\mathbb{E}(y_{ij1}^{2})-\vartheta_{ij1}\mathbb{E}(y_{ij2}^{2})+(\vartheta_{ij2}-\vartheta_{ij1})\mathbb{E}(y_{ij1}y_{ij2})\right]}{\sum_{j\neq i} \vartheta_{ij1}\lambda_{ij2}}\right.\\
 & \hspace{-0.5cm}\left.\!\!+\frac{1}{2\!N}\!\sum_{i=1}^{N}\!\frac{\left\{ \sum_{j\neq i}\!\Delta\widetilde{x}_{ij}\vartheta_{ij1}\hspace{-1bp}(\vartheta_{ij1}\!-\!\vartheta_{ij2})\lambda_{ij2}\!\right\} \!\!\left\{ \sum_{j=1}^{N}\!\vartheta_{ij2}^{2}\mathbb{E}(y_{ij1}^{2})\!+\!\vartheta_{ij1}^{2}\mathbb{E}(y_{ij2}^{2})\!-\!2\vartheta_{ij1}\vartheta_{ij2}\mathbb{E}(y_{ij1}y_{ij2})\!\right\} }{\left[\sum_{j\neq i} \vartheta_{ij1}\lambda_{ij2}\right]^{2}}\!\right]\!\!,
\end{align*}
with an analogous expression also following for $D_{N}^{k}$.
\label{T-equals-2}
\end{remark} 

{Two points then stand out based on the above expression: (i) all bias terms in $B_{N}^{k}$ and $D_{N}^{k}$ generally depend on the distribution of $y_{ij}$ and do not depend on it in the same way; (ii) none of these terms generally equals 0. The first of these two observations can be seen from how the bias depends on the expected second moments of $y_{ij}$ (e.g., $\mathbb{E}(y_{ij1}^{2})$, $\mathbb{E}(y_{ij1}y_{ij2})$, etc.), marking an important difference from the models that were considered in \citet{fernandez-val_individual_2016}.\footnote{The specific examples they use are the Poisson model, which is unbiased, and the probit model, which
requires the distribution of $y_{ij}$ to be correctly specified. They also provide a
bias expansion for ``conditional moment'' models that allow the
distribution of $y_{ij}$ to be misspecified. Beyond this theoretical discussion, bias corrections for
misspecified models have yet to receive much attention, however.} Among other things, the difficulty associated with estimating these second moments means that analytical bias corrections may not necessarily offer superior performance to distribution-free method such as the jackknife. The second observation mainly follows from the first. It can also be shown that $\sum_{j\neq i}\bar{G}_{ij}\Delta\widetilde{x}_{ij}\neq0$, which ensures that the second term can never be zero.}

\subsubsection*{What if $T$ is Large?} \label{sec3_3}

While Proposition \ref{bias-results} only focuses on asymptotics where $N\rightarrow\infty$, 
the three-way gravity panel also features a time dimension ($T$), and it is interesting to wonder how the above results may depend on changes in $T$. 
{As we show in the Appendix}, {
large $T$ only makes a difference for the asymptotic order of the bias of $\widehat \beta$
if there is only weak time dependence between observations belonging to the same country pair, in the sense described by \citealp{hansen2007asymptotic}.\footnote{By ``weak'' time dependence, we mean that any such dependence dissipates as the temporal distance between observations increases. Alternatively, if observations are correlated regardless of how far apart they are in time, the standard error is always of order $1/N$ (see \citealp{hansen2007asymptotic}), and the same will also be true for the asymptotic bias. The latter is arguably a less natural assumption in this context, however.} We will henceforth assume any time dependence is weak.
The following remark then describes some additional asymptotic results for when $T$ is large.
}

\begin{remark}
   \label{Remark:LargeT}
Under asymptotics where $T \rightarrow \infty$, we have the following:
\item[(i)] If $N$ is fixed and $T\rightarrow\infty$, then $\widehat{\beta}$
is generally inconsistent.
\item[(ii)] As $N,T\rightarrow \infty$, the combined bias term $(N-1)^{-1} W_{N}^{-1}(B_{N}+D_{N})$ goes to zero at a rate of $1/(NT)$. Therefore, because the standard error is of order $1/(N\sqrt{T})$, there is no bias in the asymptotic distribution of $\widehat{\beta}$ as $N$ and $T$ both $\rightarrow \infty$.\label{remark-T}
\end{remark}

To elaborate further, letting $T\rightarrow\infty$ is obviously not sufficient for either
$\alpha$ or $\gamma$ to be consistently estimated and does not solve
the IPP, as stated in part (i). However, as part (ii) tells us, $T$ still plays an interesting role in conditioning the bias 
when both $N$ and $T$ jointly become large. Intuitively, because $W_{N}^{-1}$ is of order $1/T$ as $T\rightarrow\infty$, {whereas $B_{N}$ and
 $D_{N}$  
are both of order 1}, 
the bias in $\widehat{\beta}$ effectively vanishes at a rate of $1/(NT)$ as both $N,T\rightarrow \infty$, such that it disappears asymptotically in relation to the order-$1/(N\sqrt{T})$ standard error. However, since $T$ is usually small relative to $N$ in this context, it remains to be seen whether these asymptotic results carry over to practical settings. 

\subsection{Downward Bias in Robust Standard Errors \label{sec3_3}} 

Of course, even if the point estimates are correctly centered, inferences
will still be unreliable if the estimates of the variance used to
construct confidence intervals are not themselves unbiased. For PPML, 
confidence intervals are typically obtained using a ``sandwich''
estimator for the variance that accounts for the possible misspecification
of the model. However, as shown by \citet{kauermann2001note}, the
PPML sandwich estimator is generally downward-biased in finite samples.
Furthermore, for gravity models (both two-way and three-way), the
bias in the sandwich estimator can itself be formalized as a kind
of IPP.\footnote{This type of IPP has similar origins to the one described
in \citet{verdier2018}, who considers a dyadic linear model with
two-way FEs and sparse matching between the two panel dimensions.}

To illustrate the bias of the sandwich estimator in our three-way
setting, recall that we can express the variance of $\widehat{\beta}$
as ${\rm Var}(\widehat{\beta}-\beta)=N^{-1}(N-1)^{-1}W_{N}^{-1}\Omega_{N}W_{N}^{-1}$. 
As is also true for the linear model (cf., \citealp{mackinnon1985some,imbens2016robust}),
{the bias arises because plugin estimates for the ``meat'' of the sandwich $\Omega_{N}$ depend
on the \emph{estimated} score variance $\mathbb{E}(\widehat{S}_{ij}\widehat{S}_{ij}^{\prime})$
rather than on the true variance $\mathbb{E}(S_{ij}S_{ij}^{\prime})$.}
Even though $\mathbb{E}(\widehat{S}_{ij}\widehat{S}_{ij}^{\prime})$
is a consistent estimate for $\mathbb{E}(S_{ij}S_{ij}^{\prime})$,
it will generally be downward-biased in finite samples. Notably, this
bias may be especially slow to vanish for models with gravity-like
fixed effects. 

To see this, we follow the same approach as \citet{kauermann2001note}. Specifically, we
use the special case where $\mathbb{E}(S_{ij}S_{ij}^{\prime})=\kappa\bar{H}_{ij}$
(such that $\Omega_{N}=\kappa W_{N}$, meaning PPML is correctly
specified) to demonstrate that $\mathbb{E}(\widehat{S}_{ij}\widehat{S}_{ij}^{\prime})$
generally has a downward bias. Under this assumption, it is possible to show that the expected outer product of the fitted score $\mathbb{E}(\widehat{S}_{ij}\widehat{S}_{ij}^{\prime})$
has a first-order bias of} 
\begin{align}
\mathbb{E}(\widehat{S}_{ij}\widehat{S}_{ij}^{\prime}-S_{ij}S_{ij}^{\prime}) & \approx
-\underbrace{\frac{\kappa}{N \! \left(N \! - \! 1\right)}\bar{H}_{ij}\widetilde{x}_{ij}W_{N}^{-1}\widetilde{x}_{ij}^{\prime}\bar{H}_{ij}}_{\text{order } 1/N^2}
-\underbrace{\frac{\kappa}{N  \! \left(N  \! - \! 1\right)}\bar{H}_{ij}d_{ij}W_{N}^{(\phi)-1}d_{ij}^{\prime}\bar{H}_{ij}}_{\text{order } 1/N \text{ (due to IPP)}}
\label{eq:SS-bias}
\end{align}
where $W_{N}^{(\phi)}:=\mathbb{E}_{N}[-\partial^{2}\ell_{ij}/\partial\phi\partial\phi^{\prime}]$
captures the expected Hessian of the concentrated likelihood with
respect to $\alpha$ and $\gamma$ and where $d_{ij}$ is a  $T\times dim(\phi)$ matrix of dummies such that each row satisfies $d_{ijt}\phi=\alpha_{it}+\gamma_{jt}$.\footnote{A detailed derivation of \eqref{eq:SS-bias} is provided in the Appendix.}

The two terms on the right-hand side of \eqref{eq:SS-bias} are both
negative definite, implying that  the sandwich estimator is generally downward-biased\textemdash and
definitively so if the model is correctly specified. {Since we work with cluster-robust standard errors,
a relevant comparison to draw here is with \citet{cameron2008bootstrap}, who have previously shown 
that the cluster-robust sandwich estimator has a downward bias that depends on the number of clusters. In our setting,
the standard \citet{cameron2008bootstrap} bias is reflected in the first term on the righthand-side of \eqref{eq:SS-bias},
which captures how the the bias depends on the variance of $\widehat{\beta}$. The second term, 
which arises because of an IPP, captures how much of the bias is due to the variance in the estimated origin-time and destination-time fixed effects in
$\widehat{\phi}$. The former term decreases with $1/N^{2}$---i.e., with the number of pairs/clusters---but the latter term only decreases with $1/N$, since increasing $N$
by $1$ only adds $1$ additional observation of each origin-time and destination-time fixed effect.\footnote{%
The $[N (N-1)]^{-1} W_{N}^{(\phi)-1}$ matrix that appears in the second term is the inverse Hessian with respect 
to the fixed effects and thus reflects their variance.
Because adding a new country only adds one new observation of each fixed effect, the diagonal elements of this matrix decrease with only $1/N$ as $N$ increases, despite how the formula is written. 
\citet{pfaffermayr2019gravity} makes a similar point about the order of the bias of the standard errors for the two-way FE-PPML estimator, albeit using a slightly different analysis.}}



All together, this analysis implies that the estimated standard error
for $\widehat{\beta}$ will exhibit a bias that only disappears at
the relatively slow rate of $1/\sqrt{N}$. We should therefore be
concerned that asymptotic confidence intervals for $\widehat{\beta}$
may exhibit inadequate coverage even in moderately large samples,
similar to what has been found for the two-way FE-PPML estimator in recent
simulation studies by \citet{egger_glm_2015}, \citet{jochmans_two-way_2016},
and \citet{pfaffermayr2019gravity}. Indeed, the bias approximation
we have derived in \eqref{eq:SS-bias} can be readily adapted to the
two-way setting or even to more general settings with $k$-way fixed effects.

\subsection{Bias Corrections for the Three-way Gravity Model \label{sec3_4}} 

We now present two methods for correcting the bias in estimates: a jackknife method based on the split-panel jackknife of \cite{dhaene2015split} and an analytical correction based on the expansion shown in Proposition \ref{bias-results}. We also provide an analytical correction for the downward bias in standard errors.

\subsubsection*{Jackknife Bias Correction}

The advantage of the jackknife correction is that it does not require
explicit estimation of the bias yet still has a simple and powerful
applicability. To see this, note first that the asymptotic bias we characterize can
be written as 
\[
\frac{1}{N}B^{\beta}+ o_p(N^{-1}), \label{eq:bias}
\]
where $B^{\beta}$ is a combined term that captures any suspected asymptotic bias contributions of order $1/N$. 
The specific jacknife we will apply for our current purposes is a
split-panel jackknife based on \cite{dhaene2015split}. As in \cite{dhaene2015split},
we want to divide the overall data set into subpanels of roughly even size
and then estimate $\widehat{\beta}_{(p)}$ for each subpanel $p$. Given
the gravity structure of the model, we first divide the set of countries
into evenly-sized groups $a$ and $b$. We then consider 4 subpanels
of the form ``$(a,b)$'', where ``$(a,b)$'' denotes a subpanel
where 
exporters from group $a$ are matched with importers from group $b$. 
The other three subpanels are  $(a,a)$, $(b,a)$, and $(b,b)$.
For randomly-generated data, we can define $a$ and $b$ based on
their ordering in the data (i.e., $a:={i:i\le N/2}$; $b:={i:i> N/2}$).
For actual data, it would be more sensible to draw these subpanels
randomly and repeatedly.\footnote{This is just one possible way to construct a jackknife correction for two-way panels. We have also experimented with splitting the panel one dimension at a time as in \citet{fernandez-val_individual_2016}, but we find the present method performs significantly better at reducing the bias.}

The split-panel jackknife estimator for $\beta$, $\widetilde{\beta}_{N}^J$,
is then defined as 
\begin{align}
\widetilde{\beta}_{N}^{J} & :=2\widehat{\beta}-\sum_{p}\frac{\widehat{\beta}_{(p)}}{4}.\label{eq:jackknife}
\end{align}
This correction works to reduce the bias because,  {so long as the distribution of $y_{ij}$ and $x_{ij}$ is homogeneous across both the $i$ and $j$ dimensions of the panel},\footnote{%
{
By ``homogeneity'' we mean that the vector $(y_{ij}, x_{ij}, \alpha_i, \gamma_j)$ is identically distributed across both $i$ and $j$,
which is the appropriate translation of Assumption 4.3 in  \citet{fernandez-val_individual_2016} to our setting. This does allow $(y_{ij}, x_{ij})$ to be heterogeneously distributed {conditional on the fixed effects} in the sense that the fixed effects themselves introduce heterogeneity into the model. Nonetheless, this is a strong assumption. One of the main advantages of the analytical bias correction is that it does not require such assumptions}.
}
each $\widehat{\beta}_{(p)}$ has a leading bias term equal to
$2B^{\beta}/N$.
The average $\widehat{\beta}_{(p)}$ across these four subpanels thus
also has a leading bias of $2B^{\beta}/N$ and any terms depending
on $B^{\beta}/N$ cancel out of \eqref{eq:jackknife}.
 Thus, the bias-corrected
estimate $\widetilde{\beta}_{N}^{J}$ only has a bias of order $o_p(N^{-1})$,
which is obtained by combining the second-order bias from $\widehat{\beta}$
with that of the average subpanel estimate. This latter bias  can be shown to be larger
than the original second-order bias in \eqref{eq:bias}, but the overall bias should still be smaller
because of the elimination of the leading bias term.

\subsubsection*{Analytical Bias Correction}

{
Our analytical correction for the bias is based on the bias expression in Proposition \ref{bias-results}.
In the Appendix, we show how appropriate sample analogs $\widehat{W}_{N}$, $\widehat{B}_{N}$, $\widehat{D}_{N}$
of the expressions for $W_N$, $B_N$, $D_N$ can be formed.
The resulting bias-corrected estimate is 
then given by  
$$\widehat \beta - 
(N-1)^{-1}\widehat{W}_{N}^{-1}(\widehat{B}_{N}+\widehat{D}_{N}) .
$$
It is possible to show that these plug-in corrections lead to estimates that are asymptotically unbiased as $N\rightarrow\infty$.}
 Still, for finite samples, it is evident that the bias in some of these plug-in objects
could cause the analytical bias correction to itself exhibit some bias. For this reason, it is not obvious \emph{a priori} whether the analytical correction will outperform the jackknife at reducing the bias in $\widehat{\beta}$. One clear advantage the analytical correction has over the jackknife is that {it does not require any homogeneity restrictions on the distribution of $y_{ij}$ and $x_{ij}$}  in order to be valid.

\subsubsection*{Bias-corrected Standard Errors}

Under the assumption of clustered errors within pairs, a natural correction
for the variance estimate is available based on \eqref{eq:SS-bias}.
Specifically, let
\[
\widehat{\Omega}^{U}\!:=\!\frac{1}{N\!\left(N\!-\!1\right)}\!\sum_{i,j}\widehat{\widetilde{x}}_{ij}\!\!\left[\mathbf{I}_{T}-\frac{1}{N\!\left(N\!-\!1\right)}\bar{H}_{ij}\widehat{\widetilde{x}}_{ij}\widehat{W}_{N}^{-1}\widehat{\widetilde{x}}^{\prime}\!-\frac{1}{N\!\left(N\!-\!1\right)}\bar{H}_{ij}d_{ij}\widehat{W}_{N}^{(\phi)-1}d_{ij}^{\prime}\right]^{-1}\!\!\!\widehat{S}_{ij}\widehat{S}_{ij}^{\prime}\widehat{\widetilde{x}}_{ij},
\]
where $\mathbf{I}_{T}$ is a $T\times T$ identity matrix and $\widehat{W}_{N}^{(\phi)}$ 
is a plugin estimate for $W_{N}^{(\phi)}$. The corrected variance
estimate is then given by
\begin{align*}
\widehat{V}^{U} & =\frac{1}{N\!\left(N\!-\!1\right)\!-\!1}\widehat{W}^{-1}\widehat{\Omega}^{U}\widehat{W}^{-1}.
\end{align*}
The logic of this adjusted variance estimate follows directly from
Kauermann and Carroll (2001): if the PPML estimator is correctly specified
(such that $E(S_{ij}S_{ij}^{\prime})=\kappa\bar{H}_{ij}$), then $\widehat{V}^{U}$
can be shown to eliminate the first-order bias in $\widehat{V}(\widehat{\beta}-\beta^{0})$
shown in \eqref{eq:SS-bias}. It is not generally unbiased otherwise,
but it is plausible that it should eliminate a significant
portion of any downward bias under other variance assumptions as
well.

{
\subsubsection*{Other Practicalities}

As we have noted, implementations of our analytical corrections are available via our Stata command {\tt ppml\_fe\_bias}. Since this command can be applied to data sets that do not conform exactly to our theoretical framework, we provide here some brief comments on its applicability in general cases that may not be explicitly covered in the notation and formulas used above. For example, it is worth pointing out that our corrections may be used with data sets that have missing values. They also continue to apply if   the data includes the diagonal ($i=j$) terms versus treating them as missing or inapplicable. Similarly, we can allow for the data to have unequal numbers of exporters and importers. In the latter case, we need the numbers of exporters and importers to grow at the same rate asymptotically for our results to apply.

One practicality that is not covered in our current framework  is the case of ``four way'' gravity models that have an added index for industry. Though a full analytical characterization of this type of model is left for future work, our Appendix describes how a modified version of the heuristic from \citet{ARE} may be used to assess the order of the bias and derive a suitable jackknife correction. Interestingly, this discussion reveals that the asymptotic bias problem may be more severe for four-way models than for three-way models. The heuristic we propose may be used to assess IPPs in other, non-trade settings as well.}

\section{Simulation Evidence \label{sim_results}}

For our simulation analysis, we assume the following: (i) the data
generating process (DGP) for the dependent variable is of the form
$y_{ijt}=\lambda_{ijt}\omega_{ijt}$, where $\omega_{ijt}$ is a log-normal
disturbance with mean $1$ and variance $\sigma_{ijt}^{2}$. (ii)
$\beta=1$. (iii) The model-relevant fixed effects $\alpha$, $\gamma$,
and $\eta$ are each $\sim{\cal N}(0,1/16)$. (iv) $x_{ijt}=x_{ijt-1}/2+\alpha+\gamma+\nu_{ijt}$,
where $\nu_{ijt}\sim{\cal N}(0,1/16)$.\footnote{These assumptions on $\alpha$, $\gamma$, $\eta$, $x_{ijt}$, and
$\nu_{ijt}$ are taken from \citet{fernandez-val_individual_2016}.
Notice that $x_{ijt}$ is strictly exogenous with respect to $\omega_{ijt}$
conditional on $\alpha$, $\gamma$, and $\eta$.} (v) Taking our cue from \citet{santos_silva_log_2006}, we consider
4 different assumptions about the {disturbance} $\omega_{ijt}$:
\begin{align*}
\text{\textbf{DGP\,I}:\,\,}\, & \sigma_{ijt}^{2}=\lambda_{ijt}^{-2}; &  & {\rm Var}(y_{ijt}|x_{it},\alpha,\gamma,\eta)=1.\\
\text{\textbf{DGP\,II}:\,\,\,} & \sigma_{ijt}^{2}=\lambda_{ijt}^{-1}; &  & {\rm Var}(y_{ijt}|x_{it},\alpha,\gamma,\eta)=\lambda_{ijt}.\\
\text{\textbf{DGP\,III}:\,\,\,} & \sigma_{ijt}^{2}=1; &  & {\rm Var}(y_{ijt}|x_{it},\alpha,\gamma,\eta)=\lambda_{ijt}^{2}.\\
\text{\textbf{DGP\,IV}:\,\,\,} & \sigma_{ijt}^{2}=0.5\lambda_{ijt}^{-1}+0.5e^{2x_{ijt}}; &  & {\rm Var}(y_{ijt}|x_{it},\alpha,\gamma,\eta)=0.5\lambda_{ijt}+0.5e^{2x}\lambda_{ijt}^{2},
\end{align*}
where we also allow for serial correlation within pairs by imposing
\begin{align*}
{\rm Cov}[\omega_{ijs},\omega_{ijt}] & =\exp\left[{0.3^{\left|s-t\right|}\times\sqrt{\ln(1+\sigma_{ijs}^{2})}\sqrt{\ln(1+\sigma_{ijt}^{2})}}\right]-1,
\end{align*}
such that the degree of correlation weakens for observations further
apart in time.\footnote{The $0.3$ that appears here serves as a quasi-correlation parameter.
Replacing $0.3$ with $1$ would be analogous to assuming disturbances
are perfectly correlated within pairs. Replacing it with $0$ removes
any serial correlation. Choosing other values for this parameter produces
similar results.}

The relevance of these various assumptions to commonly used error
distributions is best described by considering the conditional variance
${\rm Var}(y_{ijt}|x_{it},\alpha,\gamma,\eta)$. For example, DGP
I assumes that the conditional variance is constant, as in a Gaussian
process with i.i.d disturbances. In DGP II, the conditional variance
equals the conditional mean, as in a Poisson distribution. DGP III\textemdash which
we will also refer to as ``log-homoskedastic''\textemdash is
the unique case highlighted in \citet{santos_silva_log_2006} where
the assumption that the conditional variance is proportional to the
square of the conditional mean leads to a homoskedastic error when
the model is estimated in logs using a linear model. Finally, DGP
IV provides a ``quadratic'' error distribution that mixes DGP II
and DGP III and also allows for overdispersion that depends on $x_{ijt}$.
As shown by \citet{santos_silva_log_2006}, this type of DGP tends
to induce a relatively large degree of bias.

Tables \ref{table1} and \ref{table2} present simulation evidence
comparing the uncorrected three-way FE-PPML estimator with results
computed using the analytical and jackknife corrections described
in Section \ref{sec3_4}. As in the prior simulations, we again compute
results for a variety of different panel sizes\textemdash in this
case for $N=20,50,100$ and $T=2,5,10$.\footnote{Note that the trade literature currently recommends using wide intervals
of 4-5 years between time periods so as to allow trade flows time
to adjust to changes in trade costs (see \citealp{cheng_controlling_2005}.)
Thus, for practical purposes, $T=10$ may be thought of as a relatively
``long'' panel in this context that might span 40+ years. IPPs do
not necessarily vanish for larger values of $T$, as discussed further
below.} In order to validate our analytical predictions regarding these estimates,
we compute the average bias of each estimator, the ratios of the average
bias to the average standard error and of the average standard error
to the standard deviation of the simulated estimates, and the probability
that the estimated $95\%$ confidence interval covers the true estimate
of $\beta=1$. In particular, we expect that the bias in $\widehat{\beta}$
should be decreasing in either $N$ or $T$ but should remain large
relative to the estimated standard error and induce inadequate coverage
for small $T$. We are also interested in whether the usual cluster-robust
standard errors accurately reflect the true dispersion of estimates.
Results for DGPs I and II are shown in Table \ref{table1}, whereas
Table \ref{table2} shows results for DGPs III and IV.

The results in both tables collectively confirm the presence of bias
and the viability of the analytical and jackknife bias corrections.
The average bias is generally larger for DGPs I and IV than II and
III. As expected, it generally falls with both $N$ and $T$ across
all the different DGPs, though only weakly so for DGP III (the log-homoskedastic
case), which generally only has a small bias.\footnote{Numerically, we have found that the two terms that appear
in both $B_N$ and $D_N$ in Proposition \ref{bias-results} tend to have
opposite signs when the DGP is log-homoskedastic, thereby mitigating one another.} To use DGP II\textemdash the Poisson case, where PPML should otherwise
be an optimal estimator\textemdash as a representative example, we
see that the average bias falls from $3.840\%$ for the smallest sample
where $N=20$, $T=2$ to a low of $0.249\%$ at the other extreme
where $N=100,$ $T=10$. For DGP IV, the least favorable of these
cases, the average bias ranges from $-6.1364\%$ down to $-1.878\%$.
On the whole, these results support our main theoretical findings
that $\beta$ should be consistently estimated even for {fixed} $T$
but has an asymptotic bias that depends on the number of countries
and on the number of time periods.

Interestingly, while the average bias almost always decreases with
$T$, the ratio of the bias to standard error usually does not, seemingly
contrary to the expectations laid out in Remark \ref{remark-T}. {Evidently, increasing $T$ does not 
automatically reduce the bias at a rate of $1/T$. As we discuss in more detail below, researchers should thus be careful
to note that the implications of Remark \ref{remark-T} do not necessarily
apply to settings with small $T$ or even moderately large $T$.} 
Furthermore, the estimated cluster-robust standard errors themselves
clearly exhibit a bias in all cases as well. Even when $N=100$, SE/SD
ratios are uniformly below 1; generally they are closer to $0.9$
or $0.95$, and for DGPs I and IV, they are often closer to $0.85$
or even $0.8$. Because of these biases, the simulated FE-PPML coverage
ratios are unsurprisingly below the $0.95$ we would expect for an
unbiased estimator.

Bias corrections to the point estimates do help with addressing some,
but not all, of these issues. The jackknife generally performs more
reliably than the analytical correction at reducing the average bias
when compared across all values of $N$ and $T$; notice
how, for the Poisson case, for example, the average bias left by the
jackknife correction is never greater than $0.25\%$ in absolute magnitude,
whereas the analytical-corrected estimates still have average biases
ranging between $0.01\%$ and $1.12\%$. However, when $N=100$, the
analytical correction begins to closely match the jackknife, especially
when $T=10$. All the same, both corrections generally have a positive
effect, and the better across-the-board bias-reduction performance
of the jackknife comes at the important cost of a relatively large
increase in the variance. Thus, the analytical correction generally
performs as well as or better than the jackknife in terms of improving
coverage even in the smaller samples. Neither correction is sufficient
to bring coverage ratios to $0.95$, however, {though
corrected Gaussian-DGP estimates and Poisson-DGP estimates both exceed
$0.94$ using the analytical correction when $N=100$ and $T=10$,
with the latter reaching $0.948$.}

Table \ref{table3} then evaluates the efficacy of our bias correction
for the estimated variance. Keeping in mind that this correction is
calibrated for the case of a correctly specified variance (which corresponds
to DGP II), we would naturally expect that the effect of this correction
should vary depending on the conditional distribution of the data.
In that light, it is encouraging that we observe positive effects
across all cases. The best results by far are for the DGPs I, II, and III, where
combining the analytical bias correction for the point estimates with
the correction for the variance yields coverage ratios that range
between $0.925$ and $0.952$ when $N$ is either 50 or 100 and generally
get closer to the the target value of $0.95$ as either $N$ or $T$
increases. These corrections lead to dramatic improvements in coverage
for DGP IV as well, but there the remaining biases in both the point
estimate and the standard error remain large even for $N=100$ and
$T=10$.

To summarize, these simulations suggest that combining an analytical
bias correction for $\widehat{\beta}$ with a further correction for
the variance based on \eqref{eq:SS-bias} should be a reliable way
of reducing bias and improving coverage. At the same time, it should
be noted that neither should be expected to offer a complete bias removal. For smaller
samples, if reducing bias on average is heavily favored, {and
if the distribution of $y_{ij}$ and $x_{ij}$ can be reasonably assumed
to be homogeneous across $i$ and $j$}, then the split-panel jackknife method might be
preferable to the analytical correction method. 

\renewcommand{\baselinestretch}{1.05}
\begin{figure}[h]
\centering{}\includegraphics[scale=0.8]{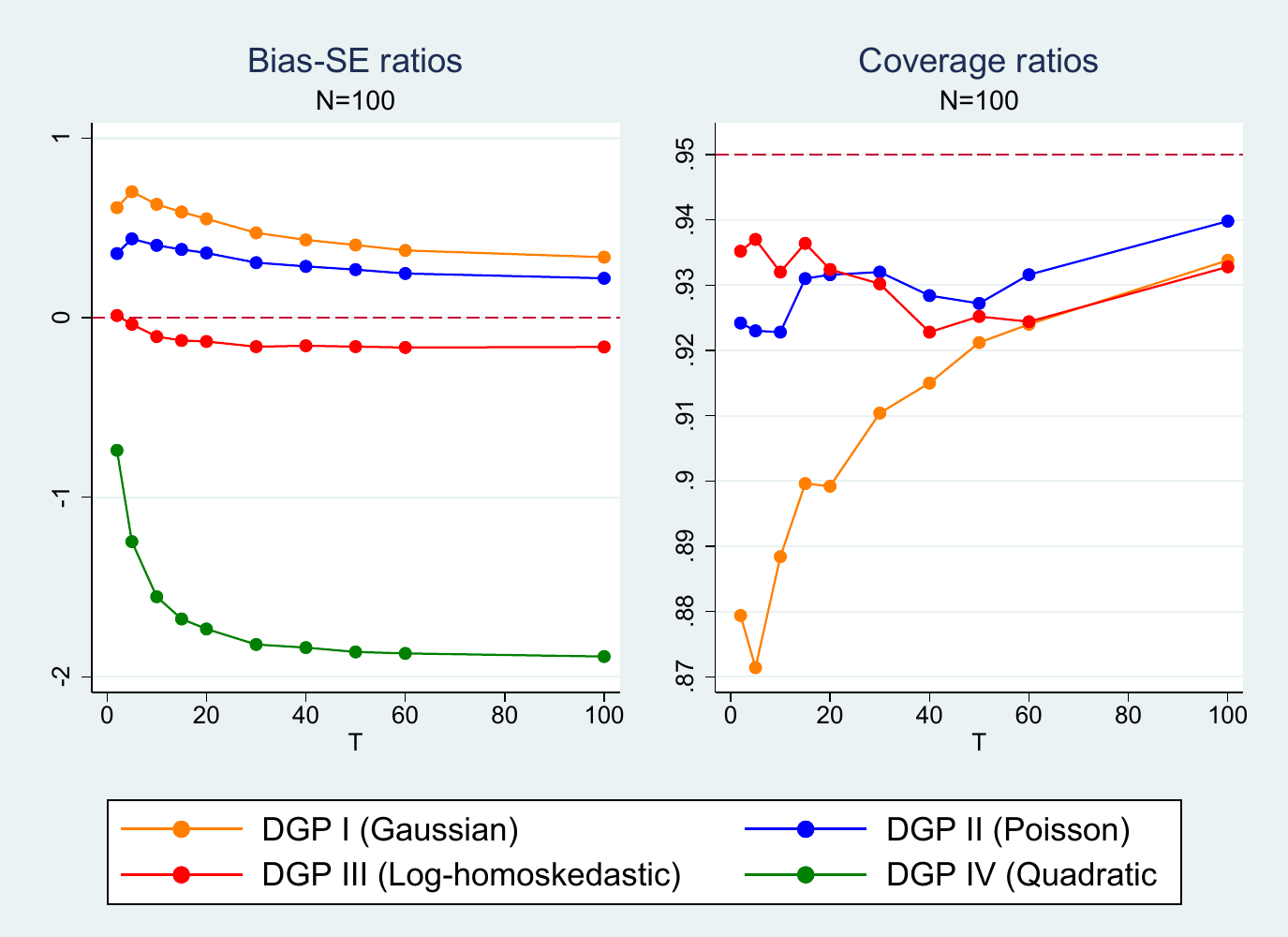}\caption{{\footnotesize Simulated bias/SE and coverage ratios for three-way FE-PPML estimates as the time dimension ($T$) becomes increasingly large. The data is generated in the same way as in Tables \ref{table1}-\ref{table3}. The $i$ and $j$ dimensions are fixed at $N=100$. The values for $T$ are 2, 5, 10, 15, 20, 30, 40, 50, 60, and 100.  Coverage for DGP IV (not shown) ranges between 0.52 and 0.84 and generally gets progressively worse at higher values of $T$.}
\label{fig2}} 
\end{figure}
\renewcommand{\baselinestretch}{1.3}

\subsubsection*{What happens for larger values of $T$? }

{Based on our Remark \ref{Remark:LargeT}, one might expect that increasing the size
of the time dimension should reduce the asymptotic bias in relation
to the standard error. However,
in the range of $T$ values we used in Tables \ref{table1}-\ref{table3}, this is not
what we observe. The question thus arises: can we say if there exists a ``large enough''
value of $T$ beyond which researchers may feel relatively secure about IPPs?

Figure \ref{fig2} addresses this question by presenting simulated
bias/SE and coverage ratios for a wider range of $T$ values, spanning
from $2$ to $100$. $N$ is fixed at 100, and the data is otherwise generated the same way as before. If we focus just on the
first two DGPs, the Gaussian and Poisson cases, we do indeed observe
steady improvements in both ratios as $T$ increases, though coverage
fails to hit $0.95$ in either case. However, for both DGP III (log-homoskedastic)
and DGP IV (quadratic), we actually observe bias/SE ratios getting
worse as $T$ approaches 100. In the case of DGP IV, coverage actually
gets worse as well.\footnote{%
{For scale reasons, coverage results for DGP IV are not shown. For $T=100$, we find that coverage is only $0.52$ in this case. As shown in the left-hand panel of Figure \ref{fig2}, the reason is because the bias tends to decrease more slowly than the standard error as $T$ increases while $N$ is fixed under this DGP.}} The main takeaway is that gravity panels with
seemingly large time spans are not necessarily immune to IPPs. To
reconcile these findings with our theory, note that Remark \ref{Remark:LargeT} only says that three-way PPML estimates become
asymptotically unbiased as both $N$ \emph{and} $T$ become large
simultaneously. In further simulations, we have confirmed that both
the bias/SE ratio and coverage improve across all DGPs when
we compare, e.g., $N=T=200$ with $N=T=100.$} 

\section{Empirical Applications\label{emp_app}}

For our {main} empirical application, we estimate the average effects of
an FTA for a variety of different industries using a panel with a relatively large number of countries. The value of this exercise is that we expect that
trade flows could be distributed very differently across different industries. Based on our results so far,
this should lead to a range of differerent bias behaviors in the data.

Our trade data is from the BACI database of \citet{gaulier_baci:_2010}, from which
we extract data on trade flows between 167 countries for the years
1995, 2000, 2005, 2010, and 2015. Countries are chosen so that the
same 167 countries always appear as both exporters and importers in
every period; hence, the data readily maps to the setting just described
with $N=167$ and $T=5$. We combine this trade data with data on
FTAs from the NSF-Kellogg database maintained
by Scott Baier and Jeff Bergstrand, which we crosscheck against data
from the WTO in order to incorporate agreements from more recent years.\footnote{This database is available for download on Jeff Bergstrand's website:
\url{https://www3.nd.edu/~jbergstr/}. The most recent version runs
from 1950-2012. The additional data from the WTO is needed to capture
agreements that entered into force between 2012 and 2015. } The specification we estimate is 
\begin{align}
y_{ijt} & =\exp[\alpha_{it}+\gamma_{jt}+\eta_{ij}+\beta FTA_{ijt}]\omega_{ijt},\label{eq:empirical}
\end{align}
where $y_{ijt}$ is trade flows (measured in current USD), $FTA_{ijt}$
is a $0/1$ dummy for whether or not $i$ and $j$ have an FTA at
time $t$, and $\omega_{ijt}$ is an error term. As we have noted, estimation of specifications such
as \eqref{eq:empirical} via PPML has become an increasingly standard
method for estimating the effects of FTAs and other trade
policies and is currently recommended as such by the WTO (see \citealp{yotov_advanced_2016}.)

Table \ref{table4} presents results from FE-PPML estimation of
\eqref{eq:empirical}, including results obtained using our bias corrections. The estimation is applied separately to 28 2 digit ISIC (rev. 3) industries as well as to aggregate trade. While we do indeed see a range of different biases in the industry-level estimates, the results for aggregate
trade flows, shown in the bottom row of Table \ref{table4}, are
fairly representative. To provide some basic interpretation, the coefficient on $FTA_{ijt}$ for aggregate trade is initially estimated to be $0.082$, which equates to an $e^{0.082}-1=8.5\%$
average ``partial'' effect of an FTA on trade.\footnote{The term ``partial effect'' is conventionally used to distinguish this
type of estimate from the ``general equilibrium'' effects of an
FTA, which would typically be calculated by solving a general equilibrium
trade model where prices, incomes, and output levels (which are otherwise
absorbed by the $\alpha_{it}$ and $\gamma_{jt}$ fixed effects) are
allowed to evolve endogenously in response to the FTA. In the context
of such models, $\beta$ can usually be interpreted as capturing the
average effect of an FTA on bilateral trade frictions specifically,
holding fixed all other determinants of trade.} The estimated standard error is $0.027$, implying that this effect
is statistically different from zero at the $p<0.01$ significance
level. Our bias-corrected estimates do not paint an altogether different
picture, but do highlight the potential for meaningful refinement.
Both the analytical and jackknife bias corrections for $\beta$ suggest
a downward bias of $0.04$-$0.06$, or about $15\%$-$22\%$ of the estimated
standard error. As our bias-corrected standard errors show (in the
last column of Table \ref{table4}), the initially estimated standard
error itself has an implied downward bias of $11\%$ (i.e., $0.027$
versus $0.030$).

Turning to the industry-level estimates, the analytical bias correction
more often than not indicates a downward bias ranging between $5\%$-$20\%$
of the estimated standard error, though exceptions are present on both sides
of this range. Estimates for the Chemical and Furniture industries
appear to be unbiased, for example, and some (such as Tobacco) are
associated with an upward bias. On the other end of the spectrum,
implied downward biases can also be larger than $20\%$ of the standard
error, as is seen for Petroleum ($47\%$), Fabricated Metal Products
($31\%$), Electrical Equipment ($27\%$), and Agriculture ($22\%$). The biases implied
by the jackknife are often even larger (see Fabricated Metal Products,
for example), consistent with what we found in our simulations for smaller panel sizes. One possible interpretation is that the jackknife-corrected estimates
are giving us a less conservative alternative to the analytical corrections in these cases. Indeed, the general correspondence between 
the two sets of results adds validity to both methods.
However, as we have noted, these jackknife estimates could also be reflecting
{non-homogeneity in the data} and/or the higher variance
introduced by the jackknife. Implied biases in the standard error,
meanwhile, tend to range between $10\%$-$20\%$ of the original standard
error, again with some exceptions. 

{To further illustrate the types of results that can occur, we also obtain
replication data for several recent articles that have used three-way
gravity models and re-examine their findings using our bias corrections.
The results, reported in Table \ref{table5}, help to demonstrate how these corrections
can matter for assessing statistical significance. Instances where
conventional significance levels are affected include the coefficients
for $\text{EIA}\times\text{CONTIG}$ and $\text{EIA}\times\text{LEGAL}$
from \citet{baier2018heterogeneous} and the coefficient for log approval
rating from \citet{rose2019soft}.\footnote{Note that \citet{baier2018heterogeneous}
and \citet{baier_economic_2014} use first-differenced OLS with added pair time trends.
We estimate the three-way FE-PPML equivalents.} The implied bias to standard error ratios are sometimes as large
as 40\%-45\%, as occurs for the $\text{EIA}\times\text{LANG}$ coefficient
from \citet{baier2018heterogeneous} and the Total EIA Effect from
\citet{bergstrand_economic_2015}. However, the largest effects are
actually for the standard error, which is downward-biased by more
than 40\% in several cases (the Total FTA effect from \citet{baier2019widely},
for example). Notably, there are meaningful differences found even
for the data used in \citet{larch2019currency}, a large data set
with 213 countries and 66 time periods. These results reinforce our
earlier finding that bias corrections may be useful even in settings
with seemingly large $N$ and $T$}. %

{
\section{Conclusion}\label{conclusion}

Thanks to recent methodological and computational advances, nonlinear
models with three-way fixed effects have become increasingly popular
for investigating the effects of trade policies on trade flows. However,
the asymptotic and finite-sample properties of three-way fixed effects estimators have
not been rigorously studied, especially with regards to potential
IPPs. The performance of the FE-PPML estimator in particular is of
natural interest in this context, both because FE-PPML is known to
be relatively robust to IPPs as well as because it is likely to be
a researcher's first choice for estimating three-way gravity models.
Our results regarding the consistency of PPML in this setting reflect
these unique properties of PPML and support its current status as
a workhorse estimator for estimating the effects of trade polices.

Given the consistency of PPML in this setting, and given the nice
IPP-robustness properties of PPML in general, it may come as a surprise
that three-way PPML estimates nonetheless suffer from an asymptotic
bias that affects the validity of inferences. In theory, the bias
should become less of a problem when the country and time dimensions
are both large, but our experiments with the time dimension indicate
the bias can be of comparable magnitude to the standard error even in ostensibly large trade data
sets. Typical cluster-robust estimates of the standard error are also
biased, implying estimated confidence intervals not only off-center
but also too narrow.

These issues are not so severe that they leave researchers in the
wilderness, but we do recommend taking advantage of the corrective measures we have described.
In particular, we find that analytical bias corrections based
on Taylor expansions to both the point estimates and standard errors
generally lead to improved inferences when applied simultaneously.
We caution that we have not found these corrections to be a panacea,
however, and several avenues remain open for future work. For example,
confidence interval estimates could be adjusted further to account
for the uncertainty in the estimated variance\textemdash \citet{kauermann2001note}
describe such a correction for the PPML case. A quasi-differencing
approach similar to \citet{jochmans_two-way_2016} could provide another angle of attack, 
and a recent contribution by \citet{pfaffermayr2021confidence} 
suggests that jackknife and bootstrap confidence interval methods hold promise as well. 
Turning to broader applications, the essential
dyadic structure of our bias corrections could be easily {adapted}
to network models that study changes in network behavior over time,
including settings that involve studying the number of interactions
between network members.
}

\setlength{\bibsep}{.67em}
\bibliographystyle{econometrica}
\bibliography{references}

\clearpage

\begin{table}
\centering{}\caption{Finite-sample Properties of Three-way FE-PPML Estimates}
\scalebox{.70}{%
\begin{tabular}{clllcccccccccccc}
\hline 
 &  &  &  & N=20 &  &  &  & N=50 &  &  &  & N=100 &  &  & \tabularnewline
\cline{9-11} \cline{13-16} 
 &  &  & \hspace{0.75em} & T=2 & T=5 & T=10 & \hspace{0.75em} & T=2 & T=5 & T=10 & \hspace{0.75em} & T=2 & T=5 & T=10 & \tabularnewline
\hline 
\multicolumn{16}{l}{\vspace{-0.6cm}
}\tabularnewline
\multicolumn{14}{l}{\textbf{A. Gaussian DGP (``DGP I'')}} &  & \tabularnewline
\multicolumn{14}{l}{\emph{Average bias ($\times100$)}} &  & \tabularnewline
 & \multicolumn{2}{l}{FE-PPML} &  & 6.647  & 3.764  & 2.364  &  & 2.854  & 1.497  & 0.913  &  & 1.463  & 0.755  & 0.428  & \tabularnewline
 & \multicolumn{2}{l}{Analytical} &  & 2.261  & 1.005  & 0.543  &  & 0.529  & 0.165  & 0.104  &  & 0.152  & 0.043  & 0.011  & \tabularnewline
 & \multicolumn{2}{l}{Jackknife} &  & 0.558  & 0.024  & -0.304  &  & 0.247  & 0.014  & -0.035  &  & 0.074  & 0.006  & -0.025  & \tabularnewline
\multicolumn{16}{l}{\vspace{-0.5cm}
}\tabularnewline
\multicolumn{3}{l}{\emph{Bias / SE ratio}} &  &  & \tabularnewline
 & \multicolumn{2}{l}{FE-PPML} &  & 0.690  & 0.796  & 0.753  &  & 0.668  & 0.732  & 0.687  &  & 0.644  & 0.708  & 0.626  & \tabularnewline
 & \multicolumn{2}{l}{Analytical} &  & 0.235  & 0.213  & 0.173  &  & 0.124  & 0.081  & 0.078  &  & 0.067  & 0.040  & 0.016  & \tabularnewline
 & \multicolumn{2}{l}{Jackknife} &  & 0.058  & 0.005  & -0.097  &  & 0.058  & 0.007  & -0.026  &  & 0.033  & 0.006  & -0.037  & \tabularnewline
\multicolumn{16}{l}{\vspace{-0.5cm}
}\tabularnewline
\multicolumn{3}{l}{\emph{SE / SD ratio}} &  &  &  &  & \tabularnewline
 & \multicolumn{2}{l}{FE-PPML} &  & 0.836  & 0.846  & 0.868  &  & 0.901  & 0.916  & 0.936  &  & 0.934  & 0.949  & 0.968  & \tabularnewline
 & \multicolumn{2}{l}{Analytical} &  & 0.792  & 0.813  & 0.845  &  & 0.859  & 0.887  & 0.921  &  & 0.899  & 0.929  & 0.958  & \tabularnewline
 & \multicolumn{2}{l}{Jackknife} &  & 0.718  & 0.752  & 0.783  &  & 0.831  & 0.863  & 0.900  &  & 0.887  & 0.918  & 0.949  & \tabularnewline
\multicolumn{16}{l}{\vspace{-0.5cm}
}\tabularnewline
\multicolumn{14}{l}{\emph{Coverage probability (should be $0.95$ for an unbiased estimator)}} &  & \tabularnewline
 & \multicolumn{2}{l}{FE-PPML} &  & 0.836  & 0.804  & 0.823  &  & 0.856  & 0.846  & 0.870  &  & 0.874  & 0.870  & 0.889  & \tabularnewline
 & \multicolumn{2}{l}{Analytical} &  & 0.871  & 0.883  & 0.895  &  & 0.907  & 0.921  & 0.930  &  & 0.922  & 0.933  & 0.941  & \tabularnewline
 & \multicolumn{2}{l}{Jackknife} &  & 0.841  & 0.866  & 0.876  &  & 0.901  & 0.911  & 0.926  &  & 0.916  & 0.929  & 0.939  & \tabularnewline
\multicolumn{16}{l}{\vspace{-0.45cm}
}\tabularnewline
\multicolumn{16}{l}{\textbf{B. Poisson DGP (``DGP II'') }}\tabularnewline
\multicolumn{16}{l}{\emph{Average bias ($\times100$)}}\tabularnewline
 & \multicolumn{2}{l}{FE-PPML} &  & 3.840  & 2.161  & 1.358  &  & 1.621  & 0.857  & 0.541  &  & 0.825  & 0.435  & 0.249  & \tabularnewline
 & \multicolumn{2}{l}{Analytical} &  & 1.193  & 0.621  & 0.421  &  & 0.237  & 0.095  & 0.094  &  & 0.063  & 0.024  & 0.010  & \tabularnewline
 & \multicolumn{2}{l}{Jackknife} &  & 0.229  & 0.002  & -0.153  &  & 0.079  & 0.007  & 0.001  &  & 0.018  & 0.004  & -0.013  & \tabularnewline
\multicolumn{16}{l}{\vspace{-0.5cm}
}\tabularnewline
\multicolumn{3}{l}{\emph{Bias / SE ratio}} &  &  & \tabularnewline
 & \multicolumn{2}{l}{FE-PPML} &  & 0.403  & 0.466  & 0.446  &  & 0.394  & 0.439  & 0.430  &  & 0.386  & 0.436  & 0.390  & \tabularnewline
 & \multicolumn{2}{l}{Analytical} &  & 0.125  & 0.134  & 0.138  &  & 0.058  & 0.049  & 0.075  &  & 0.029  & 0.024  & 0.016  & \tabularnewline
 & \multicolumn{2}{l}{Jackknife} &  & 0.024  & 0.000  & -0.050  &  & 0.019  & 0.004  & 0.001  &  & 0.008  & 0.004  & -0.020  & \tabularnewline
\multicolumn{16}{l}{\vspace{-0.5cm}
}\tabularnewline
\multicolumn{3}{l}{\emph{SE / SD ratio}} &  &  & \tabularnewline
 & \multicolumn{2}{l}{FE-PPML} &  & 0.865  & 0.869  & 0.891  &  & 0.931  & 0.940  & 0.957  &  & 0.957  & 0.965  & 0.982  & \tabularnewline
 & \multicolumn{2}{l}{Analytical} &  & 0.824  & 0.838  & 0.869  &  & 0.903  & 0.924  & 0.947  &  & 0.939  & 0.957  & 0.977  & \tabularnewline
 & \multicolumn{2}{l}{Jackknife} &  & 0.748  & 0.774  & 0.804  &  & 0.875  & 0.898  & 0.924  &  & 0.926  & 0.942  & 0.966  & \tabularnewline
\multicolumn{16}{l}{\vspace{-0.5cm}
}\tabularnewline
\multicolumn{14}{l}{\emph{Coverage probability (should be $0.95$ for an unbiased estimator)}} &  & \tabularnewline
 & \multicolumn{2}{l}{FE-PPML} &  & 0.887  & 0.880  & 0.892  &  & 0.912  & 0.905  & 0.919  &  & 0.918  & 0.919  & 0.925  & \tabularnewline
 & \multicolumn{2}{l}{Analytical} &  & 0.888  & 0.897  & 0.902  &  & 0.920  & 0.931  & 0.938  &  & 0.934  & 0.939  & 0.948  & \tabularnewline
 & \multicolumn{2}{l}{Jackknife} &  & 0.857  & 0.870  & 0.884  &  & 0.916  & 0.922  & 0.934  &  & 0.928  & 0.936  & 0.945  & \tabularnewline
\multicolumn{16}{l}{\vspace{-0.6cm}
}\tabularnewline
\hline 
\hline 
\multicolumn{16}{>{\centering}p{19cm}}{\raggedright{}\textbf{\footnotesize{}Notes}{\footnotesize{}: Results
computed using 5,000 repetitions. The model being estimated is $y_{ijt}=\lambda_{ijt}\omega_{ijt}$,
where $\lambda_{ijt}=\exp(\alpha_{it}+\gamma_{jt}+\eta_{ij}+\beta x_{ijt})$.
The data is generated using $\alpha_{it}\sim\mathcal{N}(0,1/16)$,
$\gamma_{jt}\sim\mathcal{N}(0,1/16)$, $\eta_{ij}\sim\mathcal{N}(0,1/16)$
and $\beta=1$. $x_{ijt}=x_{ijt-1}/2+\alpha_{it}+\gamma_{jt}+\eta_{ij}+\nu_{ijt}$,
with $x_{ij0}=\eta_{ij}+\nu_{ij0}$ and $\nu_{ijt}\sim\mathcal{N}(0,1/2)$.
Results are shown for two different assumptions about $V(y_{ijt})$.
The ``Gaussian'' DGP (panel A) assumes $V(\omega_{ijt})=\lambda_{ijt}^{-2}$.
The ``Poisson'' DGP (panel B) assumes $V(\omega_{ijt})=\lambda_{ijt}^{-1}$.$\text{SE/SD}$
refers to the ratio of the average standard error of of $\widehat{\beta}$
relative to the standard deviation of $\widehat{\beta}$ across simulations.
Coverage probability refers to the probability $\beta^{0}$ is covered
in the 95\% confidence interval for $\widehat{\beta}_{NT}$. ``Analytical''
and ``Jackknife'' respectively indicate Analytical and Jackknife
bias-corrected FE-PPML estimates. ``FE-PPML'' indicates uncorrected
estimates. SEs allow for within-$ij$ clustering.}}\tabularnewline
\end{tabular}}\label{table1}
\end{table}

\begin{table}
\centering{}\caption{Finite-sample Properties of Three-way FE-PPML Estimates}
\scalebox{.70}{%
\begin{tabular}{clllcccccccccccc}
\hline 
 &  &  &  & N=20 &  &  &  & N=50 &  &  &  & N=100 &  &  & \tabularnewline
\cline{9-11} \cline{13-16} 
 &  &  & \hspace{0.75em} & T=2 & T=5 & T=10 & \hspace{0.75em} & T=2 & T=5 & T=10 & \hspace{0.75em} & T=2 & T=5 & T=10 & \tabularnewline
\hline 
\multicolumn{16}{l}{\vspace{-0.6cm}
}\tabularnewline
\multicolumn{14}{l}{\textbf{A. Log-homoskedastic DGP (``DGP III'')}} &  & \tabularnewline
\multicolumn{14}{l}{\emph{Average bias ($\times100$)}} &  & \tabularnewline
 & \multicolumn{2}{l}{FE-PPML} &  & 0.368  & -0.086  & -0.248  &  & 0.143  & -0.092  & -0.119  &  & 0.083  & -0.051  & -0.088  & \tabularnewline
 & \multicolumn{2}{l}{Analytical} &  & -0.103  & -0.108  & -0.048  &  & -0.065  & -0.073  & -0.004  &  & -0.012  & -0.024  & -0.016  & \tabularnewline
 & \multicolumn{2}{l}{Jackknife} &  & -0.189  & -0.286  & -0.302  &  & -0.064  & -0.077  & -0.038  &  & -0.012  & -0.021  & -0.023  & \tabularnewline
\multicolumn{16}{l}{\vspace{-0.5cm}
}\tabularnewline
\multicolumn{3}{l}{\emph{Bias / SE ratio}} &  &  & \tabularnewline
 & \multicolumn{2}{l}{FE-PPML} &  & 0.036  & -0.017  & -0.074  &  & 0.032  & -0.042  & -0.081  &  & 0.036  & -0.044  & -0.114  & \tabularnewline
 & \multicolumn{2}{l}{Analytical} &  & -0.010  & -0.021  & -0.014  &  & -0.015  & -0.033  & -0.003  &  & -0.005  & -0.021  & -0.021  & \tabularnewline
 & \multicolumn{2}{l}{Jackknife} &  & -0.019  & -0.057  & -0.090  &  & -0.014  & -0.035  & -0.026  &  & -0.005  & -0.018  & -0.030  & \tabularnewline
\multicolumn{16}{l}{\vspace{-0.5cm}
}\tabularnewline
\multicolumn{3}{l}{\emph{SE / SD ratio}} &  &  &  &  & \tabularnewline
 & \multicolumn{2}{l}{FE-PPML} &  & 0.853  & 0.838  & 0.848  &  & 0.920  & 0.910  & 0.923  &  & 0.949  & 0.943  & 0.956  & \tabularnewline
 & \multicolumn{2}{l}{Analytical} &  & 0.801  & 0.787  & 0.799  &  & 0.881  & 0.874  & 0.885  &  & 0.922  & 0.918  & 0.930  & \tabularnewline
 & \multicolumn{2}{l}{Jackknife} &  & 0.727  & 0.726  & 0.744  &  & 0.853  & 0.850  & 0.863  &  & 0.908  & 0.902  & 0.920  & \tabularnewline
\multicolumn{16}{l}{\vspace{-0.5cm}
}\tabularnewline
\multicolumn{14}{l}{\emph{Coverage probability (should be $0.95$ for an unbiased estimator)}} &  & \tabularnewline
 & \multicolumn{2}{l}{FE-PPML} &  & 0.903  & 0.898  & 0.899  &  & 0.933  & 0.927  & 0.923  &  & 0.937  & 0.936  & 0.940  & \tabularnewline
 & \multicolumn{2}{l}{Analytical} &  & 0.884  & 0.874  & 0.882  &  & 0.919  & 0.917  & 0.914  &  & 0.931  & 0.929  & 0.934  & \tabularnewline
 & \multicolumn{2}{l}{Jackknife} &  & 0.846  & 0.843  & 0.852  &  & 0.912  & 0.907  & 0.902  &  & 0.924  & 0.923  & 0.935  & \tabularnewline
\multicolumn{16}{l}{\vspace{-0.45cm}
}\tabularnewline
\multicolumn{16}{l}{\textbf{B. Quadratic DGP (``DGP IV'')}}\tabularnewline
\multicolumn{16}{l}{\emph{Average bias ($\times100$)}}\tabularnewline
 & \multicolumn{2}{l}{FE-PPML} &  & -6.136  & -5.658  & -5.098  &  & -3.560  & -3.348  & -2.935  &  & -2.207  & -2.114  & -1.878  & \tabularnewline
 & \multicolumn{2}{l}{Analytical} &  & -4.592  & -3.983  & -3.415  &  & -2.088  & -1.907  & -1.588  &  & -1.021  & -0.990  & -0.857  & \tabularnewline
 & \multicolumn{2}{l}{Jackknife} &  & -3.752  & -3.517  & -3.116  &  & -1.794  & -1.674  & -1.414  &  & -0.922  & -0.893  & -0.774  & \tabularnewline
\multicolumn{16}{l}{\vspace{-0.5cm}
}\tabularnewline
\multicolumn{3}{l}{\emph{Bias / SE ratio}} &  &  & \tabularnewline
 & \multicolumn{2}{l}{FE-PPML} &  & -0.510  & -0.898  & -1.171  &  & -0.640  & -1.114  & -1.384  &  & -0.718  & -1.242  & -1.544  & \tabularnewline
 & \multicolumn{2}{l}{Analytical} &  & -0.382  & -0.632  & -0.784  &  & -0.375  & -0.634  & -0.749  &  & -0.332  & -0.582  & -0.704  & \tabularnewline
 & \multicolumn{2}{l}{Jackknife} &  & -0.312  & -0.558  & -0.716  &  & -0.323  & -0.557  & -0.667  &  & -0.300  & -0.525  & -0.636  & \tabularnewline
\multicolumn{16}{l}{\vspace{-0.5cm}
}\tabularnewline
\multicolumn{3}{l}{\emph{SE / SD ratio}} &  &  & \tabularnewline
 & \multicolumn{2}{l}{FE-PPML} &  & 0.799  & 0.759  & 0.749  &  & 0.847  & 0.809  & 0.809  &  & 0.877  & 0.844  & 0.844  & \tabularnewline
 & \multicolumn{2}{l}{Analytical} &  & 0.734  & 0.693  & 0.682  &  & 0.781  & 0.738  & 0.734  &  & 0.811  & 0.773  & 0.770  & \tabularnewline
 & \multicolumn{2}{l}{Jackknife} &  & 0.662  & 0.635  & 0.633  &  & 0.747  & 0.710  & 0.707  &  & 0.795  & 0.756  & 0.755  & \tabularnewline
\multicolumn{16}{l}{\vspace{-0.5cm}
}\tabularnewline
\multicolumn{14}{l}{\emph{Coverage probability (should be $0.95$ for an unbiased estimator)}} &  & \tabularnewline
 & \multicolumn{2}{l}{FE-PPML} &  & 0.849  & 0.760  & 0.700  &  & 0.844  & 0.734  & 0.673  &  & 0.848  & 0.705  & 0.619  & \tabularnewline
 & \multicolumn{2}{l}{Analytical} &  & 0.830  & 0.775  & 0.749  &  & 0.849  & 0.799  & 0.780  &  & 0.871  & 0.823  & 0.801  & \tabularnewline
 & \multicolumn{2}{l}{Jackknife} &  & 0.795  & 0.749  & 0.729  &  & 0.840  & 0.795  & 0.786  &  & 0.867  & 0.818  & 0.803  & \tabularnewline
\multicolumn{16}{l}{\vspace{-0.6cm}
}\tabularnewline
\hline 
\hline 
\multicolumn{16}{>{\centering}p{19cm}}{\raggedright{}\textbf{\footnotesize{}Notes}{\footnotesize{}: Results
computed using 5,000 repetitions. The model being estimated is $y_{ijt}=\lambda_{ijt}\omega_{ijt}$,
where $\lambda_{ijt}=\exp(\alpha_{it}+\gamma_{jt}+\eta_{ij}+\beta x_{ijt})$.
The data is generated using $\alpha_{it}\sim\mathcal{N}(0,1/16)$,
$\gamma_{jt}\sim\mathcal{N}(0,1/16)$, $\eta_{ij}\sim\mathcal{N}(0,1/16)$
and $\beta=1$. $x_{ijt}=x_{ijt-1}/2+\alpha_{it}+\gamma_{jt}+\eta_{ij}+\nu_{ijt}$,
with $x_{ij0}=\eta_{ij}+\nu_{ij0}$ and $\nu_{ijt}\sim\mathcal{N}(0,1/2)$.
Results are shown for two different assumptions about $V(y_{ijt})$.
The ``Log-homoskedastic'' DGP (panel A) assumes $V(y_{ijt})=\lambda_{ijt}^{2}$.
The ``Quadratic'' DGP (Panel D) assumes $\omega_{ijt}$ is log-normal
with variance equal to $\lambda_{ijt}^{-1}+\exp(2x_{ijt})$. $\text{SE/SD}$
refers to the ratio of the average standard error of of $\widehat{\beta}$
relative to the standard deviation of $\widehat{\beta}$ across simulations.
Coverage probability refers to the probability $\beta^{0}$ is covered
in the 95\% confidence interval for $\widehat{\beta}_{NT}$. ``Analytical''
and ``Jackknife'' respectively indicate Analytical and Jackknife
bias-corrected FE-PPML estimates. ``FE-PPML'' indicates uncorrected
estimates. SEs allow for within-$ij$ clustering.}}\tabularnewline
\end{tabular}}\label{table2}
\end{table}

\begin{table}
\centering{}\caption{Improving Coverage in Three-way FE-PPML Gravity Estimates}
\scalebox{.65}{%
\begin{tabular}{clllcccccccccccc}
\hline 
 &  &  &  & N=20 &  &  &  & N=50 &  &  &  & N=100 &  &  & \tabularnewline
\cline{9-11} \cline{13-16} 
 &  &  & \hspace{0.75em} & T=2 & T=5 & T=10 & \hspace{0.75em} & T=2 & T=5 & T=10 & \hspace{0.75em} & T=2 & T=5 & T=10 & \tabularnewline
\hline 
\multicolumn{16}{l}{\vspace{-0.6cm}
}\tabularnewline
\multicolumn{14}{l}{\textbf{A. Gaussian DGP (``DGP I'')}} &  & \tabularnewline
\multicolumn{16}{l}{\emph{SE / SD ratio with corrected SEs}}\tabularnewline
 & \multicolumn{2}{l}{FE-PPML} &  & 0.936  & 0.930  & 0.946  &  & 0.953  & 0.957  & 0.971  &  & 0.965  & 0.972  & 0.988  & \tabularnewline
 & \multicolumn{2}{l}{Analytical} &  & 0.887  & 0.894  & 0.921  &  & 0.908  & 0.927  & 0.956  &  & 0.929  & 0.952  & 0.978  & \tabularnewline
 & \multicolumn{2}{l}{Jackknife} &  & 0.803  & 0.827  & 0.854  &  & 0.879  & 0.902  & 0.935  &  & 0.917  & 0.941  & 0.968  & \tabularnewline
\multicolumn{16}{l}{\vspace{-0.5cm}
}\tabularnewline
\multicolumn{14}{l}{\emph{Coverage probability with corrected SEs (should be $0.95$ for
an unbiased estimator)}} &  & \tabularnewline
 & \multicolumn{2}{l}{(uncorrected)} &  & 0.836  & 0.804  & 0.823  &  & 0.856  & 0.846  & 0.870  &  & 0.874  & 0.870  & 0.889  & \tabularnewline
 & \multicolumn{2}{l}{FE-PPML} &  & 0.877  & 0.846  & 0.856  &  & 0.881  & 0.866  & 0.885  &  & 0.889  & 0.877  & 0.895  & \tabularnewline
 & \multicolumn{2}{l}{Analytical} &  & 0.909  & 0.912  & 0.918  &  & 0.925  & 0.930  & 0.941  &  & 0.931  & 0.940  & 0.948  & \tabularnewline
 & \multicolumn{2}{l}{Jackknife} &  & 0.884  & 0.901  & 0.908  &  & 0.920  & 0.923  & 0.938  &  & 0.926  & 0.935  & 0.944  & \tabularnewline
\multicolumn{16}{l}{\vspace{-0.45cm}
}\tabularnewline
\multicolumn{16}{l}{\textbf{B. Poisson DGP (``DGP II'')}}\tabularnewline
\multicolumn{16}{l}{\emph{SE / SD ratio with corrected SEs}}\tabularnewline
 & \multicolumn{2}{l}{FE-PPML} &  & 0.966  & 0.956  & 0.974  &  & 0.979  & 0.980  & 0.993  &  & 0.983  & 0.986  & 1.001  & \tabularnewline
 & \multicolumn{2}{l}{Analytical} &  & 0.920  & 0.922  & 0.950  &  & 0.949  & 0.963  & 0.983  &  & 0.964  & 0.978  & 0.996  & \tabularnewline
 & \multicolumn{2}{l}{Jackknife} &  & 0.834  & 0.851  & 0.879  &  & 0.920  & 0.936  & 0.960  &  & 0.951  & 0.962  & 0.985  & \tabularnewline
\multicolumn{16}{l}{\vspace{-0.5cm}
}\tabularnewline
\multicolumn{14}{l}{\emph{Coverage probability with corrected SEs (should be $0.95$ for
an unbiased estimator)}} &  & \tabularnewline
 & \multicolumn{2}{l}{(uncorrected)} &  & 0.887  & 0.880  & 0.892  &  & 0.912  & 0.905  & 0.919  &  & 0.918  & 0.919  & 0.925  & \tabularnewline
 & \multicolumn{2}{l}{FE-PPML} &  & 0.923  & 0.915  & 0.916  &  & 0.927  & 0.921  & 0.930  &  & 0.925  & 0.927  & 0.931  & \tabularnewline
 & \multicolumn{2}{l}{Analytical} &  & 0.923  & 0.929  & 0.930  &  & 0.938  & 0.942  & 0.949  &  & 0.942  & 0.945  & 0.952  & \tabularnewline
 & \multicolumn{2}{l}{Jackknife} &  & 0.900  & 0.903  & 0.915  &  & 0.932  & 0.935  & 0.942  &  & 0.936  & 0.941  & 0.949  & \tabularnewline
\multicolumn{16}{l}{\vspace{-0.45cm}
}\tabularnewline
\multicolumn{16}{l}{\textbf{C. Log-homoskedastic DGP (``DGP III'')}}\tabularnewline
\multicolumn{16}{l}{\emph{SE / SD ratio with corrected SEs}}\tabularnewline
 & \multicolumn{2}{l}{FE-PPML} &  & 0.966  & 0.943  & 0.954  &  & 0.979  & 0.967  & 0.979  &  & 0.984  & 0.976  & 0.989  & \tabularnewline
 & \multicolumn{2}{l}{Analytical} &  & 0.906  & 0.886  & 0.899  &  & 0.938  & 0.928  & 0.940  &  & 0.956  & 0.951  & 0.962  & \tabularnewline
 & \multicolumn{2}{l}{Jackknife} &  & 0.823  & 0.817  & 0.837  &  & 0.908  & 0.902  & 0.916  &  & 0.942  & 0.934  & 0.952  & \tabularnewline
\multicolumn{16}{l}{\vspace{-0.5cm}
}\tabularnewline
\multicolumn{16}{l}{\emph{Coverage probability with corrected SEs (should be $0.95$ for
an unbiased estimator)}}\tabularnewline
 & \multicolumn{2}{l}{(uncorrected)} &  & 0.903  & 0.898  & 0.899  &  & 0.933  & 0.927  & 0.923  &  & 0.937  & 0.936  & 0.940  & \tabularnewline
 & \multicolumn{2}{l}{FE-PPML} &  & 0.938  & 0.934  & 0.932  &  & 0.947  & 0.942  & 0.940  &  & 0.945  & 0.943  & 0.949  & \tabularnewline
 & \multicolumn{2}{l}{Analytical} &  & 0.922  & 0.914  & 0.915  &  & 0.938  & 0.932  & 0.929  &  & 0.940  & 0.939  & 0.943  & \tabularnewline
 & \multicolumn{2}{l}{Jackknife} &  & 0.892  & 0.889  & 0.899  &  & 0.930  & 0.925  & 0.924  &  & 0.935  & 0.934  & 0.942  & \tabularnewline
\multicolumn{16}{l}{\vspace{-0.45cm}
}\tabularnewline
\multicolumn{16}{l}{\textbf{D. Quadratic DGP (``DGP IV'')}}\tabularnewline
\multicolumn{16}{l}{\emph{SE / SD ratio with corrected SEs}}\tabularnewline
 & \multicolumn{2}{l}{FE-PPML} &  & 0.952  & 0.861  & 0.929  &  & 0.947  & 0.948  & 0.949  &  & 0.970  & 0.924  & 0.882  & \tabularnewline
 & \multicolumn{2}{l}{Analytical} &  & 0.877  & 0.793  & 0.845  &  & 0.873  & 0.865  & 0.865  &  & 0.897  & 0.853  & 0.810  & \tabularnewline
 & \multicolumn{2}{l}{Jackknife} &  & 0.781  & 0.729  & 0.786  &  & 0.833  & 0.824  & 0.837  &  & 0.889  & 0.828  & 0.789  & \tabularnewline
\multicolumn{16}{l}{\vspace{-0.5cm}
}\tabularnewline
\multicolumn{16}{l}{\emph{Coverage probability with corrected SEs (should be $0.95$ for
an unbiased estimator)}}\tabularnewline
 & \multicolumn{2}{l}{(uncorrected)} &  & 0.849  & 0.760  & 0.700  &  & 0.844  & 0.734  & 0.673  &  & 0.848  & 0.705  & 0.619  & \tabularnewline
 & \multicolumn{2}{l}{FE-PPML} &  & 0.900  & 0.820  & 0.808  &  & 0.894  & 0.806  & 0.762  &  & 0.892  & 0.752  & 0.652  & \tabularnewline
 & \multicolumn{2}{l}{Analytical} &  & 0.894  & 0.830  & 0.838  &  & 0.908  & 0.864  & 0.862  &  & 0.918  & 0.856  & 0.818  & \tabularnewline
 & \multicolumn{2}{l}{Jackknife} &  & 0.844  & 0.808  & 0.814  &  & 0.894  & 0.860  & 0.866  &  & 0.908  & 0.848  & 0.822  & \tabularnewline
\multicolumn{16}{l}{\vspace{-0.6cm}
}\tabularnewline
\hline 
\hline 
\multicolumn{16}{>{\centering}p{19.5cm}}{\raggedright{}\textbf{\footnotesize{}Notes}{\footnotesize{}: Results
computed using 5,000 repetitions. The data is generated in the same manner as Tables \ref{table1} and \ref{table2} $\text{SE/SD}$
refers to the ratio of the average standard error of $\widehat{\beta}$
relative to the standard deviation of $\widehat{\beta}$ across simulations.
Coverage probability refers to the probability $\beta^{0}$ is covered
in the 95\% confidence interval for $\widehat{\beta}$. ``Analytical''
and ``Jackknife'' respectively indicate Analytical and Jackknife
bias-corrected FE-PPML estimates. ``FE-PPML'' estimates use corrections for the SEs only.
SEs allow for within-$ij$ clustering. The corrected SEs
correct for first-order finite sample bias in the estimated variance.}}\tabularnewline
\end{tabular}}\label{table3}
\end{table}

\begin{table}[htbp]
\caption{Bias Correction Results Using BACI Trade Data ($N=167$) }
\scalebox{.81}{%
\begin{tabular}{lllllllll}
\hline 
 &  &  & \multicolumn{2}{>{\centering}p{2.5cm}}{Original estimates} &  & \multicolumn{3}{c}{Bias-corrected estimates}\tabularnewline
\cline{4-5} \cline{7-9} 
Industry & Code &  & $\widehat{\beta}$ & SE &  & Analytical & Jackknife & SE\tabularnewline
\hline 
Agriculture  & 1 &  & 0.097  & (0.046) &  & 0.107  & 0.109  & (0.051)\tabularnewline
Forestry & 2 &  & -0.203  & (0.125) &  & -0.197  & -0.185  & (0.156)\tabularnewline
Fishing & 5 &  & 0.127  & (0.141) &  & 0.140  & 0.189  & (0.164)\tabularnewline
Coal & 10 &  & 0.025  & (0.131) &  & 0.011  & -0.079  & (0.163)\tabularnewline
Metal Ores & 13 &  & 0.040  & (0.100) &  & 0.032  & -0.020  & (0.124)\tabularnewline
Other Mining \& Quarrying n.e.c.  & 14 &  & 0.048  & (0.097) &  & 0.079  & 0.098  & (0.108)\tabularnewline
Food \& Beverages  & 15 &  & 0.020  & (0.043) &  & 0.027  & 0.029  & (0.048)\tabularnewline
Tobacco & 16 &  & 0.535  & (0.139) &  & 0.525  & 0.598  & (0.163)\tabularnewline
Textiles & 17 &  & 0.229  & (0.045) &  & 0.227  & 0.238  & (0.055)\tabularnewline
Apparel & 18 &  & 0.092  & (0.092) &  & 0.094  & 0.132  & (0.120)\tabularnewline
Leather Products & 19 &  & 0.222  & (0.067) &  & 0.218  & 0.242  & (0.079)\tabularnewline
Wood \& Cork Products & 20 &  & 0.077  & (0.109) &  & 0.098  & 0.102  & (0.127)\tabularnewline
Paper \& Paper Products & 21 &  & -0.002  & (0.062) &  & -0.004  & -0.018  & (0.071)\tabularnewline
Printed \& Recorded Media & 22 &  & -0.114  & (0.065) &  & -0.144  & -0.177  & (0.076)\tabularnewline
Coke \& Refined Petroleum  & 23 &  & 0.254  & (0.076) &  & 0.290  & 0.340  & (0.090)\tabularnewline
Chemicals \& Chemical Products  & 24 &  & 0.072  & (0.035) &  & 0.072  & 0.077  & (0.040)\tabularnewline
Rubber \& Plastic Products & 25 &  & 0.141  & (0.030) &  & 0.146  & 0.154  & (0.035)\tabularnewline
Non-metallic Mineral Products & 26 &  & 0.218  & (0.049) &  & 0.225  & 0.232  & (0.058)\tabularnewline
Basic Metal Products & 27 &  & 0.267  & (0.102) &  & 0.272  & 0.302  & (0.115)\tabularnewline
Fabricated Metal Products (excl. Machinery) & 28 &  & 0.196  & (0.036) &  & 0.207  & 0.226  & (0.041)\tabularnewline
Machinery \& Equipment n.e.c. & 29  &  & 0.049  & (0.036) &  & 0.052  & 0.055  & (0.041)\tabularnewline
Office, Accounting, and Computer Equipment & 30  &  & -0.036  & (0.062) &  & -0.044  & -0.038  & (0.074)\tabularnewline
Electrical Equipment & 31  &  & 0.214  & (0.045) &  & 0.226  & 0.240  & (0.052)\tabularnewline
Communications Equipment & 32  &  & -0.127  & (0.067) &  & -0.143  & -0.174  & (0.081)\tabularnewline
Medical \& Scientific Equipment & 33  &  & 0.062  & (0.039) &  & 0.069  & 0.087  & (0.044)\tabularnewline
Motor Vehicles, Trailers \& Semi-trailers & 34  &  & 0.158  & (0.064) &  & 0.170  & 0.196  & (0.077)\tabularnewline
Other Transport Equipment & 35  &  & 0.207  & (0.124) &  & 0.230  & 0.251  & (0.137)\tabularnewline
Furniture \& Other Manufacturing n.e.c. & 36  &  & 0.225  & (0.073) &  & 0.225  & 0.228  & (0.082)\tabularnewline
\hline 
Total & All &  & 0.082  & (0.027) &  & 0.086  & 0.088  & (0.030)\tabularnewline
\hline 
\hline 
\multicolumn{9}{>{\raggedright}p{20cm}}{\textbf{\footnotesize{}Notes}{\footnotesize{}: These results are computed
using ISIC Rev. 3 industry-level trade data for trade between 167
countries during years 1995, 2000, 2005, 2010, \& 2015. The original
data is from BACI. The model being estimated is $y_{ijt}=\exp(\alpha_{it}+\gamma_{jt}+\eta_{ij}+\beta FTA_{ijt})\omega_{ijt}$,
where $y_{ijt}$ is the trade volume and $FTA_{ijt}$ is a dummy for
the presence of an FTA. $\alpha_{it}$, $\gamma_{jt}$, \& $\eta_{ij}$
respectively denote exporter-time, importer-time, \& exporter-importer
fixed effects. We estimate each industry separately. The jackknife
corrections use the average of 200 randomly-assigned split-panel partitions. SEs are clustered by exporter-importer.}}\tabularnewline
\end{tabular}}\label{table4}
\end{table}

\renewcommand{\baselinestretch}{1.12}
\begin{table}[h]
\caption{Further Applications}
\scalebox{.6}{ 
\begin{tabular}{>{\raggedright}p{5cm}l>{\raggedright}p{3.25cm}>{\raggedright}p{3.25cm}>{\raggedright}p{9.5cm}}
\hline 
Article & Covariate & original PPML coefficient and standard error & bias-corrected estimate & Notes\tabularnewline
\hline 
\multirow{2}{5cm}{\citet{baier2018heterogeneous}} & EIA & 0.062 (0.039) & 0.073 (0.044){*} & \multirow{10}{8.5cm}{EIA is a dummy for the presence of an Economic Integration Agreement.
The interaction terms are log distance and dummies for adjacency,
common legal system, common religion, common language, and colonial
history. Our estimation differs from BBF's in that we use PPML, whereas
BBF use OLS with first-differences and pair-specific time trends.
The specification we replicate is their Table 4, column 7. 183 countries,
1965-2010 every 5 years ($N=183,$ $T=10$).}\tabularnewline
 &  &  &  & \tabularnewline
 & EIA $\times$ log DIST & -0.102 (0.031){*}{*}{*} & -0.094 (0.035){*}{*}{*} & \tabularnewline
 &  &  &  & \tabularnewline
 & EIA $\times$ CONTIG & -0.250 (0.096){*}{*}{*} & -0.266 (0.140){*} & \tabularnewline
 &  &  &  & \tabularnewline
 & EIA $\times$ LEGAL & 0.160 (0.078){*}{*} & 0.189 (0.111){*} & \tabularnewline
 &  &  &  & \tabularnewline
 & EIA $\times$ RELG & 0.418 (0.125){*}{*}{*} & 0.456 (0.166){*}{*}{*} & \tabularnewline
 &  &  &  & \tabularnewline
 & EIA $\times$ LANG & -0.414 (0.103){*}{*}{*} & -0.457 (0.147){*}{*}{*} & \tabularnewline
 &  &  &  & \tabularnewline
 & EIA $\times$ CLNY & -0.082 (0.116) & -0.051 (0.147) & \tabularnewline
 &  &  &  & \tabularnewline
\hline 
\multirow{2}{5cm}{\citet{baier_economic_2014}} & Total NRPTA Effect & 0.020 (0.091) & 0.035 (0.110) & \multirow{9}{8.5cm}{NRPTA is a non-reciprocal trade preference. PTA is a reciprocal preferential
trade agreement. FTA is a free trade agreement with additional provisions
beyond tariff reductions. CUCMECU refers to customs unions, monetary
unions, and economic unions. ``Total'' effects are inclusive of
5 year lags. Our estimation differs from BBF's in that we use PPML,
whereas BBF use OLS with first-differences and pair-specific time
trends. The specification we replicate is their Table 1, Column 2a.
149 countries, 1965-2000 ($N=149,$ $T=8$).}\tabularnewline
 &  &  &  & \tabularnewline
 & Total PTA Effect & 0.149 (0.079){*} & 0.160 (0.089){*} & \tabularnewline
 &  &  &  & \tabularnewline
 & Total FTA Effect & 0.365 (0.045){*}{*}{*} & 0.352 (0.061){*}{*}{*} & \tabularnewline
 &  &  &  & \tabularnewline
 & Total CUCMECU Effect & 0.670 (0.060){*}{*}{*} & 0.669 (0.070){*}{*}{*} & \tabularnewline
 &  &  &  & \tabularnewline
 &  &  &  & \tabularnewline
 &  &  &  & \tabularnewline
 &  &  &  & \tabularnewline
 &  &  &  & \tabularnewline
 &  &  &  & \tabularnewline
\hline 
\citet{baier2019widely} & Total FTA Effect & 0.292 (0.051){*}{*}{*} & 0.299 ({0.074}){*}{*}{*} & \multirow{3}{8.5cm}{Reported coefficient is for the ``Total FTA effect'' inclusive of
5-year lags. The specification we estimate is their equation (4).
Our estimate differs slightly because we use asymmetric pair fixed
effects instead of symmetric pair fixed effects. Our standard errors
differ because we cluster by pair. 69 countries, 1986-2006 ($N=69$,
$T=21$).}\tabularnewline
 &  &  &  & \tabularnewline
 &  &  &  & \tabularnewline
 &  &  &  & \tabularnewline
 &  &  &  & \tabularnewline
 &  &  &  & \tabularnewline
 &  &  &  & \tabularnewline
\hline 
\citet{bergstrand_economic_2015} & Total EIA Effect & {0.521} (0.060){*}{*}{*} & {0.546 (0.098)}{*}{*}{*} & Reported coefficient is for the ``Total EIA effect'' inclusive of
4- and 8-year lags for the same specification as in their Table 2,
column 4. Baseline estimates differ slightly because BLY originally
used symmetric pair fixed effects whereas we use asymmetric pair fixed
effects. 41 countries, 1990-2002, every 4 years ($N=41$, $T=4$).\tabularnewline
\hline 
\multirow{2}{5cm}{\citet{larch2019currency}} & Euro & 0.030 (0.042) & 0.021 (0.046) & \multirow{3}{8.5cm}{The specification we replicate is their Table 1, column 5, which in
turn replicates the main specification from \citet{glick_currency_2016}.
Our standard errors differ because we cluster by pair. 213 countries,
1948-2013 ($N=213$, $T=66$). }\tabularnewline
 &  &  &  & \tabularnewline
 & Other CUs & 0.700 (0.107){*}{*}{*} & 0.708 (0.142){*}{*}{*} & \tabularnewline
 &  &  &  & \tabularnewline
 & RTA & 0.169 (0.040){*}{*}{*} & 0.159 (0.050){*}{*}{*} & \tabularnewline
 &  &  &  & \tabularnewline
 & Curr. Colony & 0.545 (0.220){*}{*} & 0.577 (0.262){*}{*} & \tabularnewline
 &  &  &  & \tabularnewline
\hline 
\citet{rose2019soft} & log approval rating & 0.053 (0.022){*}{*}{*} & 0.058 (0.028){*}{*} & \multirow{2}{8.5cm}{These reported coefficient is from Table 4 of Rose (2019). Log approval
rating is a measure of the importing country's approval of the exporting
country's leader. There are 5 exporters, 149 importers, and 12 years
(2006-2017).}\tabularnewline
 &  &  &  & \tabularnewline
 &  &  &  & \tabularnewline
 &  &  &  & \tabularnewline
 &  &  &  & \tabularnewline
 &  &  &  & \tabularnewline
\hline 
\hline 
\multicolumn{5}{>{\raggedright}p{24.25cm}}{This table shows replications of several recent papers that use three-way
gravity models with exporter-time, importer-time, and country-pair
fixed effects. Standard errors shown in parentheses. Bias-corrected
estimates use our analytical corrections for the point estimates and
standard errors. 
{*} \ensuremath{p<0.10} , {*}{*} \ensuremath{p<.05} , {*}{*}{*}
\ensuremath{p<.01}.}\tabularnewline
\end{tabular}} \label{table5}
\end{table}
\renewcommand{\baselinestretch}{1.3}

\clearpage
\appendix
\section{Appendix} 

This Appendix first describes an additional simulation exercise based on the trade data. We then introduce additional notation, definitions, and other technical details that supplement the formal theorems and remarks presented in Section \ref{sec3}. Proofs of our formal results then follow, starting with a poof of Proposition \ref{bias-results}, which characterizes the asymptotic distribution of $\widehat{\beta}$ and its asymptotic bias. {This proof naturally lends itself to further discussion of the ``large $T$'' results from Remark  \ref{Remark:LargeT} as well as the consistency result from Proposition \ref{prop1}, which itself follows as a by-product of Proposition \ref{bias-results}. 
We then demonstrate the uniqueness of this latter result as stated in Proposition \ref{prop2} and highlight the general inconsistency of other three-way gravity estimators. 

{After the proofs, we include further supplementary discussions on
 the downward bias in the estimated standard errors, on allowing for conditional dependence across pairs in the trade data, and on how FE-PPML is affected by IPPs in more general settings beyond the gravity framework.
}

\subsection{Simulation Based on Trade Data}

{As a additional simulation exercise, we revisit the BACI aggregate trade data we used
in Section \ref{emp_app} and ask: if the estimated effect of an FTA and its standard
error were indeed biased to the degrees implied by our bias corrections,
would our corrections be successful at correctly identifying the bias
and improving inference?

To answer this question, we start from the original aggregate trade
data and FTA data but reconstruct the conditional mean $\lambda_{ijt}$
as though $\beta=0.086$. That is, we first adjust the estimated $\widehat{\lambda}_{ijt}$'s
from the original estimation to account for the change in $\beta$
and so that the FOC's for all fixed effects are consistent with $\beta=0.086$.
This gives us new ``true'' values of the conditional mean that we
denote by $\lambda_{ijt}^{(1)}$. The original data is therefore assumed
to have been generated by $y_{ijt}=\lambda_{ijt}^{(1)}\omega_{ijt}$,
where the true disturbance $\omega_{ijt}$ is backed out using $\omega_{ijt}=y_{ijt}/\lambda_{ijt}^{(1)}$.

Next, we choose a DGP for $\omega_{ijt}$ that can reproduce the biases
implied by our corrections. Taking our cues from DGP IV, which we
found earlier to produce a downward bias in $\widehat{\beta}$, we
consider a DGP where the conditional variance of $y_{ijt}$ has the
form $\text{Var}[y_{ijt}|.]=a\lambda_{ijt}+bFTA\lambda_{ijt}^{2}$.
That is, we allow for some overdispersion that depends on the regressor
of interest, as in DGP IV. We choose the two parameters $a$ and $b$
in order to come close to matching the following three values: (i)
the bias in $\widehat{\beta}$, (ii) the standard deviation of $\widehat{\beta}$
(assumed to be $0.03)$, (iii) the bias of the standard error of $\widehat{\beta}$.
To keep things simple, all of our simulations sample new values for
$\omega_{ijt}$ only, holding $\lambda_{ijt}^{(1)}$ fixed. As in
the main text, we use 5,000 replications and assume $\rho=0.3$. For our chosen values of
$a$ and $b$\textemdash $a=200,000$, $b=0.08$\textemdash we obtain
an average $\widehat{\beta}$ of $0.0823$, an average standard error
of $0.0267$, and a standard deviation of $0.0307$. The uncorrected
coverage is $0.9078$.

When our preferred bias corrections are applied, they do not completely
solve the coverage problem but do induce across-the-board improvements.
The average corrected $\widehat{\beta}$ using the analytical bias
correction is $0.0842$ and the corrected standard error is $0.0290$.
Coverage improves as well, but only to $0.9210$. As discussed in
the main text, one important factor that limits the improvement in
coverage is the fact that applying bias corrections to the point estimates
increases their variance. In this case, the standard deviation of
the corrected estimate is $0.0321$. 

Turning to the jackknife bias correction, we find as before that it
does a superior job of bias reduction than the analytical correction,
producing a average corrected estimate of $0.0851$. However, the standard
deviation of the jackknife-corrected estimates is $0.0371$, echoing our
previous finding that the improved bias reduction performance of the
jackknife comes at a steep penalty in terms of increased variance.
As a result, the coverage we obtain when we combine the jackknife
with the corrected standard errors is only $0.8756$.

Interestingly, if we only use the correction to the standard errors, i.e., without applying any 
correction to the point estimates, we obtain a coverage ratio of $0.9312$, which is better than if we also use the analytical correction. 
It nonetheless remains true that using corrections to both the point estimates and standard errors leads to
an improvement in coverage, as we have consistently found throughout our results. In general, these simulations
reinforce our earlier conclusion that bias corrections, though helpful, are not necessarily a panacea to the issues we raise 
in the paper. 

}

{
\subsection{Additional Notation and Definitions}

It is convenient to define the log-likelihood as a function of the index vector $\pi_{ij}$ as follows,
\begin{align*}
\ell_{ij}(\beta,\alpha_{i},\gamma_{j}) & =:\ell_{ij}(\beta,\pi_{ij}), &  & \text{where} & \pi_{ij}=\left(\begin{array}{c}
\pi_{ij1}\\
\vdots\\
\pi_{ijT}
\end{array}\right):=\left(\begin{array}{c}
\alpha_{i1}+\gamma_{j1}\\
\vdots\\
\alpha_{iT}+\gamma_{jT}
\end{array}\right) .
\end{align*}
{In this appendix, we will also be more explicit than in the main text in distinguishing true parameter values
$\beta^0$, $\alpha^0_{it}$, $\gamma_{jt}^0$, and the corresponding $\pi_{ij}^0$ and $\vartheta^0_{ijt}$, 
from their generic equivalents.}
For example, 
the $S_{ij}$, $H_{ij}$ and $G_{ij}$ that were already defined in the main text can more formally be written as
\begin{align*}
     S_{ij} &:=-   \frac{\partial^2 \ell_{ij}(\beta^0,\pi^0_{ij}) } {\partial \pi_{ij}  }    ,
     &
      H_{ij} &:=-   \frac{\partial^2 \ell_{ij}(\beta^0,\pi^0_{ij}) } {\partial \pi_{ij} \, \partial \pi_{ij}' }    ,
\end{align*}
and
\begin{align*}
       G_{ij,tsr} &=  \frac{\partial^3 \ell_{ij}(\beta^0,\pi^0_{ij}) } {\partial \pi_{ijt} \, \partial \pi_{ijs} \, \partial \pi_{ijr}}         
        =  
     \left\{ \begin{array}{ll}
                        -\vartheta^0_{ijt}\left(1-\vartheta^0_{ijt}\right)\left(1-2\vartheta^0_{ijt}\right)\sum_{\tau}y_{ij\tau} & \text{if $t=s=r$,}
                           \\
                        -\vartheta^0_{ijs}\left(1-2\vartheta^0_{ijs}\right)\vartheta^0_{ijt}\sum_{\tau}y_{ij\tau} & \text{if $s=r \neq t$,}
                           \\
                        -\vartheta^0_{ijs}\left(1-2\vartheta^0_{ijs}\right)\vartheta^0_{ijr}\sum_{\tau}y_{ij\tau} & \text{if $t=s \neq r$,}
                           \\
                        -\vartheta^0_{ijt}\left(1-2\vartheta^0_{ijt}\right)\vartheta^0_{ijs}\sum_{\tau}y_{ij\tau} & \text{if $r=t \neq s$,}
                         \\
                       -2\vartheta^0_{ijr}\vartheta^0_{ijs}\vartheta^0_{ijt}\sum_{\tau}y_{ij\tau}   & \text{if $r\neq s \neq t \neq r$.}
              \end{array} 
              \right.
\end{align*}
The  $T \times K$  matrix $\widetilde{x}_{ij} $
that was informally introduced in the main text as a two-way within-transformation of $x_{ij} $,
 can be formally defined by
 $\widetilde{x}_{ij} =x_{ij}-\alpha_{i}^{x}-\gamma_{j}^{x}$,
where $\alpha_{i}^{x}$ and $\gamma_{j}^{x}$ are $T\times K$ matrices that
minimize 
\begin{align}
{
\sum_{i=1}^{N} \sum_{j\in\mathfrak{N}\setminus\{i\}} 
}
 {\rm Tr}\left[\left(x_{ij}-\alpha_{i}^{x}-\gamma_{j}^{x}\right)'\,\bar{H}_{ij}\,\left(x_{ij}-\alpha_{i}^{x}-\gamma_{j}^{x}\right)\right],\label{ProjectionRegressors-1-1}
\end{align}
subject to appropriate normalizations on $\alpha_{i}^{x}$ and $\gamma_{j}^{x}$ (e.g. $\iota_{T}'\alpha_{i}^{x} = \iota_{T}'\gamma_{j}^{x} =0$, where $\iota_{T}=(1,\ldots,1)^{\prime}$ is
a T-vector of ones). Each within-transformed regressor vector $\widetilde{x}_{ij,k}$
can be interpreted as containing the residuals left after partialing
out $x_{ij,k}$ with respect to any $i$- and $j$-specific components
and weighting by $\bar{H}_{ij}$.\footnote{While we present the computation of $\widetilde{x}_{ij}$ as a two-way
within-transformation to preserve the analogy with \citet{fernandez-val_individual_2016},
each individual element $\widetilde{x}_{ijt,k}$ can also be shown
to be equivalent (subject to a normalization) to a three-way within-transformation of $x_{ijt,k}$
with respect to $it$, $jt$, and $ij$ and weighting by $\lambda_{ijt}$.
Readers familiar with \citet{larch2019currency} may find the latter
presentation easier to digest. } 

\subsubsection{Analytical Bias Correction Formulas}

The  analytical bias correction discussed in  Section~\ref{sec3_4} of the main text requires
estimates of the expressions $W_N$, $B_N$, $D_N$ defined in Proposition \ref{bias-results}.
 For this, we first require
plugin objects $\widehat{\widetilde{x}}_{ij}$, $\widehat{S}_{ij}$, $\widehat{H}_{ij}$, $\widehat{\overline{H}}_{ij}$, and $\widehat{G}_{ij}$
---
  these objects are formed in the obvious way by replacing $\lambda_{ijt}$ with $\widehat{\lambda}_{ijt}$ and $\vartheta_{ijt}$ with $\widehat{\vartheta}_{ijt}:=\widehat{\lambda}_{ijt} / \sum_{\tau} \widehat{\lambda}_{ij\tau}$ where needed. 
Then, the $\widehat{B}_N$ and $\widehat{D}_N$ are $K$-vectors with elements given by
\begin{align*}
\widehat{B}_N^{k} & =-\frac{1}{N-1}\sum_{i=1}^{N}\mathrm{Tr}\left[\left(\sum_{j\in\mathfrak{N}\setminus\{i\}}\widehat{\overline{H}}_{ij}\right)^{\dagger}\sum_{j\in\mathfrak{N}\setminus\{i\}}\widehat{H}_{ij}\,\widehat{\widetilde{x}}_{ij,k}\,\widehat{S}_{ij}^{\prime}\right]\\
 & \hspace{-0.5cm}+\frac{1}{2 \left(N-1\right)}\sum_{i=1}^{N}\mathrm{Tr}\left[\left(\sum_{j\in\mathfrak{N}\setminus\{i\}}\widehat{G}_{ij}\,\widehat{\widetilde{x}}_{ij,k}\right)\left(\sum_{j\in\mathfrak{N}\setminus\{i\}}\widehat{\overline{H}}_{ij}\right)^{\dagger}\left[\sum_{j\in\mathfrak{N}\setminus\{i\}}\widehat{S}_{ij}\,\widehat{S}{}_{ij}^{\prime}\right]\left(\sum_{j\in\mathfrak{N}\setminus\{i\}}\widehat{\overline{H}}_{ij}\right)^{\dagger}\right],\\
\widehat{D}_N^{k} & =-\frac{1}{N-1}\sum_{j=1}^{N}\mathrm{Tr}\left[\left(\sum_{i\in\mathfrak{N}\setminus\{j\}}\widehat{\overline{H}}_{ij}\right)^{\dagger}\sum_{i\in\mathfrak{N}\setminus\{j\}}\widehat{H}_{ij}\,\widehat{\widetilde{x}}_{ij,k}\,\widehat{S}_{ij}^{\prime}\right]\\
 & \hspace{-0.5cm}+\frac{1}{2 \left(N-1\right)}\sum_{j=1}^{N}\mathrm{Tr}\left[\left(\sum_{i\in\mathfrak{N}\setminus\{j\}}\widehat{G}_{ij}\,\widehat{\widetilde{x}}_{ij,k}\right)\left(\sum_{i\in\mathfrak{N}\setminus\{j\}}\widehat{\overline{H}}_{ij}\right)^{\dagger}\left[\sum_{i\in\mathfrak{N}\setminus\{j\}}\widehat{S}_{ij}\,\widehat{S}{}_{ij}^{\prime}\right]\left(\sum_{i\in\mathfrak{N}\setminus\{j\}}\widehat{\overline{H}}_{ij}\right)^{\dagger}\right],
\end{align*}
and we have
\begin{align*}
\widehat{W} & =\frac{1}{N\,(N-1)}\sum_{i=1}^{N} \sum_{j\in\mathfrak{N}\setminus\{i\}} \widehat{\widetilde{x}}_{ij}^{\prime}\,\widehat{\overline{H}}_{ij}\,\widehat{\widetilde{x}}_{ij},
\end{align*} 
The replacement of $N$ with $N-1$ in $\widehat{B}_N^{k}$ and $\widehat{D}_N^{k}$ stems from a degrees-of-freedom correction. This correction is needed because creating plug-in values for the $\mathbb{E}\left(S_{ij}^{\prime}H_{ij}\,\big|x_{ij,k}\right)$ and $\mathbb{E}\left(S_{ij}\,S{}_{ij}^{\prime}\big|x_{ij,k}\right)$ objects that appear in Proposition \ref{bias-results} requires computing terms of the form $\mathbb{E}[y_{ijt}^2]$ and $\mathbb{E}[y_{ijs}y_{ijt}]$, as illustrated in {Remark} \ref{T-equals-2}.

\subsubsection{Details on large $T$ bias expansion}
   \label{Remark:RewriteBias}

We also want to explain the result in Remark~\ref{remark-T} of the main text in more detail here by rewriting 
the bias terms $B_{N}$ and $D_{N}$ to illuminate the role of the time dimension. 
	 Using generic definitions for $S_{ij}$, $H_{ij}$, $G_{ij}$, and $\widetilde x_{ij}$ (e.g., $S_{ij}:=\partial\ell_{ij}/\partial\pi_{ij}$,
$H_{ij}:=\partial^{2} \ell_{ij}/\partial\pi_{ij}\partial\pi_{ij}^{\prime}$, etc.), the formulas for the asymptotic distribution in    Proposition~\ref{bias-results} apply generally to M-estimators   based on concave objective functions   $ \ell_{ij}(\beta, \alpha_{it}, \gamma_{jt})$. Unlike in the two-way FE-PPML case, these formulas do not reduce to zero when we further specialize them to the profiled Poisson pseudo-likelihood in \eqref{DefLikelihood}, but
    we still find it instructive to do so (e.g., to discuss the large $T$ limit from Remark~\ref{remark-T}). For that purpose, we define
    the $T \times T$ matrix 
    $M_{ij} = \mathbf{I}_{T} - \vartheta_{ij} \iota_{T}' $.
    Furthermore, let $\Lambda_{ij} $ be the $T \times T$ diagonal matrix with diagonal elements  $ \lambda_{ijt}$,
    and for   $i,j \in \{1,\ldots,N \}$ define the $T \times T$ matrices
    \begin{align*}
    Q_i &=\frac 1 {N-1} \left(  \sum_{j\in\mathfrak{N}\setminus\{i\}}  M_{ij} \,    \Lambda_{ij} \, M_{ij}'  \right)^\dagger
      \left(  \sum_{j\in\mathfrak{N}\setminus\{i\}}  M_{ij} \, \mathbb{E}(y_{ij} y_{ij}') \, M_{ij}' \right)
       \left( \sum_{j\in\mathfrak{N}\setminus\{i\}}  M_{ij} \,    \Lambda_{ij} \, M_{ij}'  \right)^\dagger ,
       \\
    R_{ij} &=   \mathbb{E}(y_{ij} y_{ij}') \,
    M_{ij}' \left(\frac 1 {N-1} \sum_{j' \in\mathfrak{N}\setminus\{i\}}  M_{ij} \,    \Lambda_{ij} \, M_{ij}'  \right)^\dagger \,  \Lambda_{ij} M_{ij}' .
\end{align*}
    The bias term $B_N=(B_{N}^{k})$  in Proposition~\ref{bias-results} can then be expressed as
    \begin{align}
   B_{N}^{k}
   &=   \frac 1 {N (N-1)}  \sum_{i=1}^{N} \sum_{j\in\mathfrak{N}\setminus\{i\}} 
   \left[ - \frac{  \iota_{T}' \, R_{ij} \,  \widetilde x_{ij,k} } 
    { \iota_{T}'  \lambda_{ij}} 
    + 
   \frac{ \lambda_{ij}' \, Q_i \, \Lambda_{ij} \, M_{ij}' \, \widetilde x_{ij,k}  } 
   { \iota_{T}'  \lambda_{ij} }  \label{reexpress-B}
  \right]  ,
\end{align}
and an analogous formula for $D_N$ follows by interchanging $i$ and $j$ appropriately.    
 As long as there is only weak time dependence between observations  the matrix 
objects $R_{ij}$ and $Q_i \Lambda_{ij} M_{ij}'$ above are both of order 1 as $T\rightarrow \infty$, such that both terms in brackets in \eqref{reexpress-B} are likewise of order 1.
This explains the result stated in
Remark \ref{remark-T} of the main text.
}

\subsection{Proof of Proposition~\ref{bias-results}}

\subsubsection*{Known result for two-way fixed effect panel models}

Our proof of Proposition~\ref{bias-results} relies on results from \citet{fernandez-val_individual_2016} 
-- denoted FW in the following. That paper
considers a standard panel setting where individuals $i$ are observed over time periods $t$, and mixing conditions 
(as opposed to conditional independence assumptions) are imposed across time periods. By contrast, we consider a pseudo-panel
setting, where the two panel dimensions are labelled by exporters $i$ and importers $j$, and we impose conditional independence assumptions across both $i$ and $j$ here (see also \citealp{dzemski2018empirical},
who employs those results  in a directed network setting where outcomes are binary,
and \citealp{graham2017econometric}, for the undirected network case.)
Given those differences---and before introducing any further complications---we briefly want to restate the main
result in FW for the two-way pseudo-panel case. 
Outcomes $Y_{ij}$, $i,j = 1,\ldots,N$, conditional on all the strictly exogenous regressors $X=(X_{ij})$,
 fixed effect $N$-vectors $\alpha$ and $\gamma$, and common parameters $\beta$ are assumed to be generated as
\begin{equation*}
Y_{ij}  \mid X, \alpha, \gamma, \beta   \sim f_{Y}(\cdot \mid X_{ij},\alpha_i, \gamma_j, \beta),
\end{equation*}
where the conditional distribution  $ f_{Y}$ is  known, up to the unknown parameters $\alpha_i ,\gamma_j \in \mathbb{R}$
and $\beta \in \mathbb{R}^K$. It is furthermore assumed that $\alpha_i$ and $\gamma_j$  enter the distribution function
only through the single index $\pi_{ij} = \alpha_i + \gamma_j$; that is, the log-likelihood can be defined by
\begin{align*}
   \ell_{ij}(\beta, \, \pi_{ij} ) = \log  f_{Y}(Y_{ij} \mid X_{ij}, \alpha_i, \gamma_j, \beta) .
\end{align*}
The maximum likelihood estimator for $\beta$ is given by
\begin{align*}
  \widehat \beta &= \argmax_{\beta \in \mathbb{R}^K} \max_{\alpha,\gamma \in \mathbb{R}^{N}} \;  {\cal L}(\beta,\alpha,\gamma),
  &
   {\cal L}(\beta,\alpha,\gamma) &= \sum_{i,j}   \ell_{ij}(\beta, \, \alpha_i + \gamma_j ).
\end{align*}
 Also, define the $K$-vector $\Xi_{ij}$ with components, $k=1,\ldots,K$, 
 \begin{align*}
   \Xi_{ij,k} &= \alpha^{\ast}_{i,k} + \gamma^{\ast}_{j,k} ,
   &
  \left(\alpha^{\ast}_{k} , \, \gamma^{\ast}_{k} \right)
   &= \argmin_{\alpha_{i,k},\gamma_{j,k}} \sum_{i,j}
  \mathbb{E}( - \partial_{\pi^2} \ell_{ij} )
        \left(  \frac{\mathbb{E}( \partial_{\beta_k \alpha_i} \ell_{ij} )}
                        {\mathbb{E}( \partial_{\alpha_i^2} \ell_{ij} )}
         - \alpha_{i,k} - \gamma_{j,k} \right)^2,
\end{align*}
where here and in the following all expectations are conditional on regressors $X = (X_{ij})$, and on the parameters $\alpha$, $\gamma$, $\beta$.
For $q \in \{0,1,2\}$, the (within-transformation) differentiation operator  $\mathcal{D}_{\beta \alpha_i^q} = \mathcal{D}_{\beta \gamma_j^q}$ is defined by
\begin{align}
   \mathcal{D}_{\beta \alpha_i^q}  \ell_{ij} 
   & =  \partial_{\beta \alpha_i^q} \ell_{ij} -  \partial_{\alpha_i^{q+1}} \ell_{ij} \, \Xi_{ij},
   &
   \mathcal{D}_{\beta \gamma_j^q} \ell_{ij} 
   & =  \partial_{\beta \gamma_j^q} \ell_{ij} -  \partial_{\gamma_j^{q+1}} \ell_{ij} \, \Xi_{ij}.
   \label{DefProXi}
\end{align}

\setcounter{theorem}{0}  
\setcounter{lemma}{0} 
\begin{theorem} 
   \label{th:FW}
  Assume that
      \begin{itemize}
      \item[(i)] Conditional on $X$, $\alpha^0$, $\gamma^0$, $\beta^0$
      the outcomes $Y_{ij}$ are distributed independently across $i$ and $j$
      with
   \begin{equation*}
Y_{ij}  \mid X, \alpha^0, \gamma^0, \beta^0      \sim \exp[\ell_{ij}(\beta^0,\pi^0_{ij})],
\end{equation*}
where $\pi^0_{ij} = \alpha^0_{i} + \gamma^0_j$.

      \item[(ii)]  The map
      $(\beta,\pi) \mapsto \ell_{ij}(\beta,\pi)$ is
      four times continuously differentiable,  almost surely.
      All partial derivatives of  $\ell_{ij}(\beta,\pi)$
      up to fourth order
      are bounded in absolute value by a function $m(Y_{it}, X_{it})>0$, almost surely,
      uniformly over a convex compact set  ${\cal B} \subset \mathbbm{R}^{\dim \beta+1}$,
      which contains an $\varepsilon$-neighbourhood of $(\beta^0,\pi^0_{ij})$  for all $i,j,N$,
      and some $\varepsilon > 0$. 
      Furthermore, $\max_{i,j} \mathbb{E}[m(Y_{ij}, X_{ij})]^{8+\nu}$
      is uniformly bounded over $N$, almost surely, for some $\nu >0$.

        \item[(iii)]   For all $N$, the function
    $(\beta,\alpha,\gamma) \mapsto \mathcal{L}(\beta,\alpha,\gamma)$
    is almost surely strictly concave over $\mathbb{R}^{K + 2N}$,
    apart from one ``flat direction'' described by the transformation $\alpha_i \mapsto \alpha_i + c$,
    $\gamma_j \mapsto \gamma_j - c$, which leaves $\mathcal{L}(\beta,\alpha,\gamma)$ unchanged 
    for all $c \in \mathbb{R}$.
    Furthermore, there exist constants $b_{\min}$
    and $b_{\max}$ such that for all $(\beta,\pi) \in {\cal B}$,
    $0< b_{\min} \leq - \mathbb{E}\left[ \partial_{\alpha_i^2} \ell_{ij}(\beta,\pi) \right]   \leq b_{\max}$, almost surely,  uniformly over $i,j,N$.

    \end{itemize}  
In addition, assume that the following limits exist
\begin{align*}
   \overline B&= 
      \lim_{N \rightarrow \infty} \left[ - \frac {1} {N}  \sum_{i,j}
            \frac{    
        \mathbb{E} \left(
                \partial_{\alpha_i} \ell_{ij}  \mathcal{D}_{\beta \alpha_i} \ell_{ij}
                 + \frac 1 2  
        \mathcal{D}_{\beta \alpha_i^2} \ell_{ij}     \right)   }
        {  \sum_{j'} \mathbb{E}\left(  \partial_{\alpha_i^2} \ell_{ij'} \right) }  \right]  ,
        \nonumber \\
      \overline D&=     \lim_{N \rightarrow \infty} \left[ -
         \frac {1} {N}  \sum_{i,j}
            \frac{  
        \mathbb{E}\left(
                \partial_{\gamma_j} \ell_{ij} \mathcal{D}_{\beta \gamma_j} \ell_{ij}
              +  \frac 1 2  \mathcal{D}_{\beta \gamma_j^2} \ell_{ij}  \right)    }
        {  \sum_{i'} \mathbb{E}\left(  \partial_{\gamma_j^2} \ell_{i' j} \right) } \right], \\
          \overline W &=    \lim_{N \rightarrow \infty} \left[ - \frac 1 {N^2} \sum_{i,j}
  \mathbb{E} \left(
            \partial_{\beta \beta'} \ell_{ij}
              -  \partial_{\alpha_i^2} \ell_{ij} \Xi_{ij} \Xi'_{ij} \right) \right],
   \end{align*}
where expectations are conditional on $X$, $\alpha$, $\gamma$, $\beta$.
Finally, assume that $\overline W>0$.
  Then, as $N \rightarrow \infty $, we have
   \begin{align*}
      N \left( \widehat \beta - \beta^0 \right)
          \; \to_d \;
       \overline{W}^{-1} \, {\cal N}(   \overline B
                    +  \overline D,
           \;\overline W),
    \end{align*}
\end{theorem}

\noindent {\bf Remarks:}
\begin{itemize}
    \item[(a)] This  is just a reformulation of Theorem 4.1 in FW to the case of
    pseudo-panels, and
    the proof is provided in that paper.
    Since we consider only strictly exogenous  regressors, all the analysis is conditional on $X$;
    and the bias term $\overline B$ simplifies here, since conditional on $X$ (and the other parameters), we assume 
    independence across both $i$ and~$j$. Thus, no Nickell-type bias (\citealp{nickell1981biases,hahn_asymptotically_2002}) appears here, but we still have incidental parameter biases because the model is nonlinear
     (\citealp{neyman_consistent_1948,hahn_jackknife_2004}).

    \item[(b)] In the original version of this theorem, the sums in the definitions
    of  $ {\cal L}(\beta,\alpha,\gamma)$, $ \overline B$, $ \overline D$, and $  \overline W $
    run over all possible pairs $(i,j) \in \{1,\ldots,N\}^2$. However, for the trade application in the current paper we assume we only have 
    observations for $i \neq j$; that is, those sums over $i$ and $j$ only run over the set $\{ (i,j) \in \{1,\ldots,N\}^2 \, : \, i \neq j\}$
    of $N (N-1)$ observed country pairs.  The sum over $j'$ (in $ \overline B$) then also only runs over $j' \neq i$,
    and the sum over $i'$ (in  $\overline D$) only runs over $i' \neq j$. It turns out that those changes make no difference to
    the proof of the theorem, because the proportion of missing  observations for each $i$ and $j$ is 
    asymptotically vanishing. For that reason it also
    does not matter whether we change
    the $1/N^2$ in $  \overline W$ to $1/[N(N-1)]$, or whether we change $ N \left( \widehat \beta - \beta^0 \right)$
    to  $\sqrt{N (N-1)} \left( \widehat \beta - \beta^0 \right)$. The same equivalence holds throughout our own results for applications in which researchers wish to use observations for which $i=j$ (simply replace $N-1$ with $N$ where appropriate.) {It also does not matter for the proof that the number of exporters and importers is the same, since this is already allowed for in FW's existing results. 
If we let $I$ be the number of exporters and $J$ be the number of importers, FW's results apply so long as $I$ and $J$ grow large at the same rate.}
     
    \item[(c)] {More generally, careful examination of these proofs and results reveals that all explicit appearances of $N$ and $N-1$ in the definitions of 
    $\overline W$, $\overline B$, and $\overline D$
    actually play no role in the fully expressed formula for the asymptotic bias, i.e., $N^{-1} \overline W (\overline B + \overline D)$. Thus,
    there is no need to adjust the terms that explicitly depend on $N$ if some of the data are missing. So long as the missing data occur at random, applying the formulas as 
    written should still generally be expected to deliver an asymptotically valid bias correction. A similar observation applies for missing values in the three-way model.
    That said, if the missing values occur in such a way that some of the $\alpha_i$'s or $\gamma_j$'s appear only a small number of times in the data,
    they will tend to be estimated with a larger degree of estimation noise than the other fixed effects, which could affect the 
    performance of bias corrections based on these formulas in practice.}

    \item[(d)] The above theorem assumes that 
    the log-likelihood $ \ell_{ij}(\beta, \, \alpha_i + \gamma_j )$
    for $Y_{ij}  \mid X, \alpha, \gamma, \beta $ is correctly specified. This is an unrealistic assumption
    for the PPML estimators in this paper, where we only want to assume that
    the score of the pseudo-log-likelihood has zero mean at the true parameters, that is, 
    $\mathbb{E}\big[ \partial_\beta \ell_{ij}(\beta^0, \, \alpha^0_i + \gamma^0_j ) \mid X_{ij},\alpha^0_i, \gamma^0_j, \beta^0 \big] = 0$
    and 
    $\mathbb{E}\big[ \partial_{\alpha_i} \ell_{ij}(\beta^0, \, \alpha^0_i + \gamma^0_j ) \mid X_{ij},\alpha^0_i, \gamma^0_j, \beta^0 \big] =0$
    and
     $\mathbb{E}\big[ \partial_{\gamma_j} \ell_{ij}(\beta^0, \, \alpha^0_i + \gamma^0_j ) \mid X_{ij},\alpha^0_i, \gamma^0_j, \beta^0 \big] =0$.
    This extension to ``conditional moment models'' is discussed in Remark 3 of FW.
    The statement of the theorem then needs to be changed as follows:
       \begin{align}
      N \left( \widehat \beta - \beta^0 \right)
          \; \to_d \;
        \overline{W}^{-1} \,  {\cal N}(   \overline B   +    \overline D,
           \;\overline \Omega),
        \label{ResLimitBeta}   
    \end{align}
    where the definition of $  \overline{W}$ is unchanged, but the expression of $  \overline B = \overline B_1 + \overline B_2$,
    $  \overline D = \overline D_1 + \overline D_2 $ and $\overline \Omega$ now read    
    \begin{align}
       \overline B_1 &=     \lim_{N \rightarrow \infty}  \left[   -      \frac {1} {N}  \sum_{i,j}
            \frac{  
        \mathbb{E} \left(
               \partial_{\alpha_i} \ell_{ij} \mathcal{D}_{\beta {\alpha_i}} \ell_{ij} \right) }
        { \sum_{j'} \mathbb{E}\left( \partial_{{\alpha_i}^2} \ell_{ij'} \right) }  \right] ,
    \nonumber  \\  
       \overline B_2 &=     \lim_{N \rightarrow \infty}  \left[     
       \frac 1 2          \frac 1 {N}  \sum_{i}
         \frac{  \left[  \sum_{j}   \mathbb{E}  ( \partial_{{\alpha_i}} \ell_{ij} )^2 \right]
          \sum_{j}  \mathbb{E} ( \mathcal{D}_{\beta {\alpha_i}^2 } \ell_{ij}  )  }
             {\left[  \sum_{j}  \mathbb{E}\left( \partial_{{\alpha_i}^2} \ell_{ij} \right) \right]^2}  \right] \; ,
   \nonumber \\
      \overline D_1 &=      \lim_{N \rightarrow \infty}  \left[ -
         \frac {1} {N}  \sum_{j}
            \frac{  \sum_{i}
        \mathbb{E}\left[
               \partial_{\gamma_j} \ell_{ij}  \mathcal{D}_{\beta {\gamma_j}} \ell_{ij}  \right] }
        { \sum_{i} \mathbb{E}\left( \partial_{{\gamma_j}^2} \ell_{ij} \right) }   \right] ,
    \nonumber  \\  
          \overline D_2 &=      \lim_{N \rightarrow \infty}  \left[   
          \frac 1 2  \,    
        \frac 1 {N}  \sum_{j}
         \frac{   \sum_{i}
              \left[ \mathbb{E}  ( \partial_{{\gamma_j}} \ell_{ij} )^2 \right]
         \sum_{i}  \mathbb{E} (  \mathcal{D}_{\beta {\gamma_j}^2} \ell_{ij}
             ) }
             {\left[  \sum_{i}  \mathbb{E}\left( \partial_{{\gamma_j}^2} \ell_{i j} \right) \right]^2} \right],
   \nonumber \\
   \overline \Omega &= 
           \lim_{N \rightarrow \infty} \left[ \frac 1 {N^2}   \sum_{i,j}
     \mathbb{E} \left[
          \mathcal{D}_{\beta} \ell_{ij} (\mathcal{D}_{\beta} \ell_{ij})'  \right] \right].
          \label{GeneralBias1d}
   \end{align}
    These are the formulas that we have to use as a starting point for the bias results derived in this paper.
  
\end{itemize}
Our task in the following
is to translate and generalize the conditions, statement, and proof of Theorem~\ref{th:FW}, as
extended in \eqref{ResLimitBeta} and \eqref{GeneralBias1d}, to the case of the three-way PPML estimator discussed in the main text.

\subsubsection*{Regularity conditions for Proposition~\ref{bias-results}}

The following regularity conditions are required for the statement of Proposition~\ref{bias-results} to hold.

\newtheorem{MAINassumption}{Assumption}
           \renewcommand{\theMAINassumption}{A}
\begin{MAINassumption}
    \label{ass:MAIN}
    \begin{itemize}
        \item[(i)] Conditional on $x=(x_{ijt})$, $\alpha^0=(\alpha^0_{it})$, $\gamma^0=(\gamma^0_{jt})$, $\eta^0=(\eta^0_{ij})$
        and $\beta^0$,
      the outcomes $y_{ij} = (y_{ij,1},\ldots,y_{ij,T})'$ are distributed independently across $i$ and $j$,
      and the conditional mean of $y_{ijt}$ is given by equation \eqref{MeanY} for all $i$, $j$, $t$.

        \item[(ii)] The range of $x_{ijt}$, $\alpha^0_{it}$ and $\gamma^0_{jt}$ is uniformly bounded,
        and there exists $\nu>0$ such that
$\mathbb{E}(y_{ijt}^{8+\nu}|x_{ijt},\alpha_{it},\gamma_{jt},\eta_{ij})$ is uniformly bounded over $i$, $j$, $t$, $N$.
        
        \item[(iii)] $\lim_{N \rightarrow \infty} \, W_{N}>0$, with  $W_{N}$ defined in Proposition~\ref{bias-results}.    
   \end{itemize}
\end{MAINassumption}
Those assumptions are very similar to those in Theorem~\ref{th:FW} above:
Assumption~\ref{ass:MAIN}(i) is analogous to  condition (i) in the theorem, except that we only impose the conditional
mean of $y_{ijt}$ to be correctly specified, as already discussed in remark (c) above. Notice also that this assumption
requires conditional independence across $i$ and $j$, but we do not have to restrict the dependence of $y_{ijt}$ over $t$
for our results.

We consider the Poisson log-likelihood in this paper, which after profiling out $\eta_{ij}$ gives the pseudo-log-likelihood
function $\ell_{ij}(\beta, \alpha_{it}, \gamma_{jt}) $ defined in equation \eqref{DefLikelihood}. This log-likelihood
is strictly concave and arbitrarily often differentiable in the parameters, so corresponding assumptions 
in Theorem~\ref{th:FW} are automatically satisfied. Assumption~\ref{ass:MAIN}(ii) is therefore already
sufficient for the corresponding assumptions (ii) and (iii) in Theorem~\ref{th:FW}.
Finally, Assumption~\ref{ass:MAIN}(iii) simply corresponds to the condition $\overline W>0$, which is just
an appropriate non-collinearity condition on the regressors $x_{ijt}$.

\subsubsection*{Translation   to our main text notation}

The main difference between Theorem~\ref{th:FW} in the Appendix and Proposition~\ref{bias-results}
in the main text is that Theorem~\ref{th:FW} only covers
 the case where $\pi_{ij}  = \alpha_i + \gamma_j$ is a scalar, while in our model in the main text
$\alpha_i$, $\gamma_j$ and $\pi_{ij}  = \alpha_i + \gamma_j$ are all $T$-vectors. We can impose  additional
normalizations on those $T$-vectors, because the profile likelihood $   \mathcal{L}(\beta, \alpha, \gamma)$
in \eqref{profile} is invariant under parameter transformations 
$\alpha_i \mapsto \alpha_i + c_i \, \iota_{T}$
and $\gamma_j \mapsto \gamma_j + d_j \, \iota_{T}$
for arbitrary $c_i, d_j \in \mathbb{R}$, 
where $\iota_{T}=(1,\ldots,1)^{\prime}$ is the $T$-vector of ones.\footnote{%
Those invariances $\alpha_i \mapsto \alpha_i + c_i \, \iota_{T}$
and $\gamma_j \mapsto \gamma_j + d_j \, \iota_{T}$ correspond to parameter transformations that in the original
model could be absorbed by the parameters $\eta_{ij}$.
}
In the following we choose the normalizations $\iota'_{T} \alpha_i =0$
and $\iota'_{T} \gamma_j =0$, implying
  $\iota'_{T}  \pi_{ij}  =0$ for all $i,j$.
 Accounting for this normalization
we actually only have $(T-1)$ fixed effects $\alpha_i$ and $\gamma_j$ for each $i,j$ here. Theorem~\ref{th:FW}  is
therfore directly applicable 
to the case $T=2$, but for $T>2$ we need to provide an appropriate extension.

The appropriate generalization of the operator  $\mathcal{D}_{\beta \alpha_i^q} = \mathcal{D}_{\beta \gamma_j^q}$ in \eqref{DefProXi}
to the case of vector-valued $\alpha_i$ and $\gamma_j$ was described in Section 4.2 of \cite{ARE}.
Remember the definition of $\ell_{ij}(\beta,\pi_{ij}) = \ell_{ij}(\beta,\alpha_{i},\gamma_{j})$
and $\widetilde{x}_{ij}:=x_{ij}-\alpha_{i}^{x}-\gamma_{j}^{x}$. Then, by reparameterizing the pseudo-log-likelihood
$\ell_{ij}(\beta,\alpha_{i},\gamma_{j})$ as follows
\begin{align}
     \ell^*_{ij}(\beta,\alpha_i,\gamma_j)  
   :=  \ell_{ij}(\beta,\pi_{ij} - \beta' (\alpha_{i}^{x}+\gamma_{j}^{x}) )
   =   \ell_{ij}(\beta,  \alpha_i   - \beta'  \alpha_{i}^{x}, \gamma_j  -    \beta' \gamma_{j}^{x}) 
   \label{DefEllStar}
\end{align}
one achieves that the expected Hessian of ${\cal L}^*(\beta,\alpha,\gamma) =  \sum_{i,j}   \ell^*_{ij}(\beta,\alpha_i,\gamma_j)  $
is block-diagonal, in the sense that  $\mathbb{E} \, \partial_{\beta \alpha_i} {\cal L}^*(\beta_0,\alpha_0,\gamma_0) = 0$
and  $\mathbb{E} \, \partial_{\beta \gamma_j} {\cal L}^*(\beta_0,\alpha_0,\gamma_0) = 0$ --- the definition of 
$\alpha_{i}^{x}$ and $\gamma_{j}^{x}$
by \eqref{ProjectionRegressors-1-1} in the main text
 exactly corresponds to those block-diagonality conditions. With those definitions, we then
have that
\begin{align*}
     \mathcal{D}_{\beta \alpha_i^q} \ell_{ij} =   \partial_{\beta \alpha_i^q} \ell^*_{ij} =  \widetilde{x}_{ij} \, \partial_{\alpha_i^{q+1}} \ell_{ij} .
\end{align*}
In particular, we find that our definitions of 
\begin{align*}
W_{N} & =\frac{1}{N\,(N-1)}\sum_{i=1}^{N} \sum_{j\in\mathfrak{N}\setminus\{i\}} \widetilde{x}_{ij}^{\prime}\,\bar{H}_{ij}\,\widetilde{x}_{ij},\\
\Omega_{N} & =\frac{1}{N\,(N-1)}\sum_{i=1}^{N} \sum_{j\in\mathfrak{N}\setminus\{i\}} \widetilde{x}_{ij}^{\prime}\,\left[{\rm Var}\left(S_{ij}\,\big|\,x_{ij}\right)\right]\,\widetilde{x}_{ij},
\end{align*}
in Proposition~\ref{bias-results}  correspond  to  
$- \frac 1 {N(N-1)} \sum_{i,j} 
  \mathbb{E} \left(
            \partial_{\beta \beta'} \ell_{ij}
              -  \partial_{\alpha_i^2} \ell_{ij} \Xi_{ij} \Xi'_{ij} \right) $
and               
$\frac 1 {N (N-1)}   \sum_{i,j} \allowbreak
     \mathbb{E} \big[
          \mathcal{D}_{\beta} \ell_{ij} (\mathcal{D}_{\beta} \ell_{ij})'  \big]$
in the notation of  Theorem~\ref{th:FW} and equation
\eqref{GeneralBias1d}.
   Thus, the asymptotic variance in \eqref{ResLimitBeta} indeed corresponds to the asymptotic variance formula in 
   Proposition~\ref{bias-results}.

\subsubsection*{Inverse expected incidental parameter Hessian}

The asymptotic bias results that follow require that we first derive some key properties of the expected Hessian with respect to the incidental parameters. 
Remember the definitions of the $2NT$-vector $\phi={\rm vec}(\alpha,\gamma)$ from the main text.
The expected incidental parameter Hessian is the $2NT \times 2NT$ matrix given by
\begin{align*}
    \bar {\cal H} :=  \mathbb{E}\left[  - \partial_{\phi \phi'}   {\cal L}(\beta_0,\phi_0)  \right] 
     =    
\left(\begin{array}{cc}  \bar {\mathcal{H}}_{(\alpha\alpha)} & \bar {\mathcal{H}}_{(\alpha\gamma)}  \\  {[\bar{\mathcal{H}}_{(\alpha\gamma)}]}'  & \bar {\mathcal{H}}_{(\gamma\gamma)}
\end{array}\right)  ,
\end{align*}
 where ${\cal L}(\beta,\phi)  = {\cal L}(\beta,\alpha,\gamma)$   is defined in  \eqref{profile},
 and $\bar {\mathcal{H}}_{(\alpha\alpha)} $, $\bar {\mathcal{H}}_{(\alpha\gamma)} $ 
 and $\bar {\mathcal{H}}_{(\gamma\gamma)}$ are $NT \times NT$ submatrices.
Here and in the following
 all expectations are conditional on all the regressor realizations.
The matrix $\bar {\mathcal{H}}_{(\alpha\alpha)} = \mathbb{E}\left[  - \partial_{\alpha \alpha'}   {\cal L}(\beta_0,\phi_0)  \right] $ is block-diagonal with $N$ non-zero diagonal $T \times T$ blocks
given by $ \mathbb{E}\left[  - \partial_{\alpha_i \alpha_i'}   {\cal L}(\beta_0,\phi_0)  \right]  = \sum_{j\in\mathfrak{N}\setminus\{i\}} \bar H_{ij}$, because 
for $i \neq j$ we have $ \mathbb{E}\left[  - \partial_{\alpha_i \alpha'_j}   {\cal L}(\beta_0,\phi_0)  \right] =0$,
since the parameters $\alpha_i$ and $\alpha_j$  
never enter into the same observation. Analogously, the matrix 
$\bar {\mathcal{H}}_{(\gamma\gamma)} = \mathbb{E}\left[  - \partial_{\gamma \gamma'}   {\cal L}(\beta_0,\phi_0)  \right] $
is block-diagonal with 
$N$ non-zero diagonal $T \times T$ blocks
given by $\sum_{i\in\mathfrak{N}\setminus\{j\}} \bar H_{ij}$. By contrast, the matrix 
$\bar{\mathcal{H}}_{(\alpha\gamma)}$  consistents of  blocks $ \mathbb{E}\left[  - \partial_{\alpha_i \gamma'_j}   {\cal L}(\beta_0,\phi_0)  \right] = \bar H_{ij}$ 
that are  non-zero for $i \neq j$, because any two parameters $\alpha_i$ and $\gamma_j$ jointly enter into $T$ observations.
The  incidental parameter Hessian matrix  $\bar {\cal H}$
therefore has strong diagonal $T \times T$ blocks of order $N$, but also many off-diagonal elements of order one.
 
 The pseudoinverse of  $ \bar {\cal H}$ crucially
 enters in the stochastic expansion for $\widehat \beta$ below. It is therefore necessary to understand the
 asymptotic properties of this pseudoinverse $ \bar {\cal H}^\dagger$.  The following lemma shows that 
 $ \bar {\cal H}^\dagger$ has a structure analogous to $  \bar {\cal H}$, namely, strong diagonal 
  $T \times T$ blocks of order $1/N$, and much smaller off-diagonal elements of order $1/N^2$.
We can write $ \bar {\cal H} = {\mathfrak D} + {\cal G}$, where
   \begin{align*}
       {\mathfrak D}  &:=
\left(\begin{array}{cc}  \bar {\mathcal{H}}_{(\alpha\alpha)} &   0_{NT \times NT}   \\   0_{NT \times NT}  & \bar {\mathcal{H}}_{(\gamma\gamma)}
\end{array}\right) ,
      &
         {\cal G} &:=  \left(\begin{array}{cc}    0_{NT \times NT}  & \bar {\mathcal{H}}_{(\alpha\gamma)}  \\  {[\bar{\mathcal{H}}_{(\alpha\gamma)}]}'  &  0_{NT \times NT} 
\end{array}\right)  .
     \end{align*} 
     The matrix ${\mathfrak D} $ is block-diagonal, and its pseudoinverse   $ {\mathfrak D}^\dagger$
 is therefore also block-diagonal with $T \times T$ blocks on its diagonal given by     
      $\left( \sum_{j\in\mathfrak{N}\setminus\{i\}} \bar H_{ij} \right)^\dagger$, $i=1,\ldots,N$
      and
      $\left( \sum_{i\in\mathfrak{N}\setminus\{j\}} \bar H_{ij} \right)^\dagger$, $j =1, \ldots,N$.
    Thus, $ {\mathfrak D}^\dagger$ has diagonal elements of order $N^{-1}$.    
     For any matrix $A$ we denote by $\|A\|_{\max}$ the maximum over the absolute values of all elements of $A$.

\begin{lemma}
   \label{lemma:Hessian}
   Under Assumption~\ref{ass:MAIN} we have, as $N \rightarrow \infty$,
   \begin{equation*}
   \left\| \bar {\cal H}^\dagger -
 {\mathfrak D}^\dagger
    \right\|_{\max}
        = O_P\left(  N^{-2} \right)  .
   \end{equation*}
\end{lemma}

This result is crucial in order to derive the stochastic expansion of $\widehat \beta$. Indeed, as we will see below, once 
Lemma~\ref{lemma:Hessian} is available, then the proof of Proposition~\ref{bias-results} is a straightforward
extension of the proof of Theorem 4.1 in FW. Lemma~\ref{lemma:Hessian} is analogous to Lemma D.1 
in FW, but our proof strategy for Lemma~\ref{lemma:Hessian} is different here, because
we need to account for the vector-valued nature of $\alpha_i$ and $\gamma_j$, which requires  new arguments.

\begin{proof}[\bf Proof of Lemma~\ref{lemma:Hessian}]
    \# \underline{Expansion of $ \bar {\cal H}^\dagger $ in powers of $   {\cal G} $:}
     The matrix  $\bar {\cal H}$ is (minus) the expected Hessian of
     the profile log-likelihood ${\cal L} = \sum_{i,j} \ell_{ij}$. Because in that objective function we have already profiled out
     the fixed effect parameters $\eta_{ij}$ we have $\bar {\cal H}_{ij} \iota_T = 0$ for all $i,j$,
     where $\iota_T = (1,\ldots,1)'$ is the $T$-vector of ones.     
     This implies that
     \begin{align}
          \bar {\cal H}  \left( \mathbb{I}_{2N} \otimes   \iota_T \right)  = 0 .
          \label{ZeroEV1}
     \end{align}
     The last equation describes $2N$ zero-eigenvectors of $ \bar {\cal H} $ (i.e.~the eigenvalue zero
     of $ \bar {\cal H} $ has multiplicity at least $2N$).
     Because the original log-likelihood function of the Poisson model was strictly concave 
     in the single index $x_{ijt}'\beta+\alpha_{it}+\gamma_{jt}+\eta_{ij}$ 
     it must be the case that any additional zero-eigenvalue of $\bar {\cal H}$
     is due to linear transformations of the parameters
      $\alpha$ and $\gamma$ that leave $ \alpha_{it} + \gamma_{jt}$ unchanged for all $i,j,t$.\footnote{%
     Notice that any collinearity problem in the likelihood involving the regression parameters $\beta$ is ruled out for sufficiently large sample sizes
     by our assumption that
     $\lim_{N \rightarrow \infty} \, W_{N}>0$, which guarantees that the expected Hessian wrt $\beta $ is positive definite
     asymptotically.}
   There is exactly one such
     transformation for every $t \in \{1,\ldots,T\}$, namely the likelihood is invariant under $\alpha_{it} \mapsto  \alpha_{it} + c_t$
     and  $\gamma_{jt} \mapsto  \gamma_{jt} - c_t$ for any $c_t \in \mathbb{R}$. The expected Hessian $\bar {\cal H}$ therefore has
     additional zero-eigenvectors, which are given by the columns of the $2NT \times T$ matrix
     \begin{align}
          v :=        (\iota_N' , - \iota_N')'   \otimes  M_{\iota_T} ,
          \label{DefMatrixV}
     \end{align}
     where $M_{\iota_T} := \mathbbm{I}_{T}  - P_{\iota_T}$
     and $P_{\iota_T} := T^{-1} \iota_T \iota_T'$. In the last display we could have used the identity matrix
     $\mathbbm{I}_{T}$ instead of $M_{\iota_T}$, but we want the columns of $v$ to be orthogonal 
     to the zero-eigenvectors already given by \eqref{ZeroEV1}, which is achieved by using $M_{\iota_T}$. 
     As a consequence of this, we have ${\rm rank}(v)= T-1$; that is, since we already have \eqref{ZeroEV1} we only find
     $T-1$ additional zero-eigenvectors here. Thus, the total number of zero eigenvalues of $\bar {\cal H}$
     (i.e.~the multiplicity of the eigenvalue zero)
    is equal to $2N  + T-1$.   It is easy to verify that indeed
     \begin{align}
          \bar {\cal H}  v  = 0 .
          \label{ZeroEV2}
     \end{align}
    Equations \eqref{ZeroEV1} and \eqref{ZeroEV2} describe all the zero-eigenvectors of $  \bar {\cal H} $.
     The  projector onto the null-space of $  \bar {\cal H} $
      is therefore given by
     \begin{align}
          P_{\rm null} := \mathbb{I}_{2N} \otimes   P_{\iota_T} + P_v  ,
          \label{DefPnull}
     \end{align}     
     where $P_v = v (v'v)^\dagger v'$.
     The Moore-Penrose pseudoinverse of $    \bar {\cal H}  $ therefore satisfies
     \begin{align}
           \bar {\cal H} \,   \bar {\cal H}^\dagger  =  \bar {\cal H}^\dagger  \,   \bar {\cal H} 
            =   \mathbb{I}_{2NT}  -   P_{\rm null}   
            =   M_{(\iota_N' , - \iota_N')'} \otimes  M_{\iota_T} ,
          \label{MoorePenroseDef1}
     \end{align}     
     where the RHS is the projector orthogonal to the null-space of $ \bar {\cal H}$
     (i.e.~the projector onto the span of $ \bar {\cal H}$).  The definition of the Moore-Penrose pseudoinverse 
     guarantees that $ \bar {\cal H}^\dagger$ has the same zero-eigenvectors
     as $\bar {\cal H}$; that is, we also have $\bar {\cal H}^\dagger  v  = 0$
     and $ \bar {\cal H}^\dagger  \left( \mathbb{I}_{2N} \otimes   \iota_T \right)  = 0$. The last equation together with the
     symmetry of  $\bar {\cal H}^\dagger$ implies that
     \begin{align}
         \left( \mathbbm{I}_{2N} \otimes  P_{\iota_T} \right) \bar {\cal H}^\dagger = 0 .
         \label{ProjectionHdagger}
     \end{align}     
     Next,
     similar to the above argument for $ \bar {\cal H}$ we have that  the only 
     zero-eigenvector of the $T \times T$ matrices $ \sum_{j\in\mathfrak{N}\setminus\{i\}} \bar H_{ij}$
     and $\sum_{i\in\mathfrak{N}\setminus\{j\}} \bar H_{ij} $
     is given by $\iota_T$,  and therefore we have
     \begin{align*}
            \left( \sum_{j\in\mathfrak{N}\setminus\{i\}} \bar H_{ij} \right) \,   \left( \sum_{j\in\mathfrak{N}\setminus\{i\}} \bar H_{ij} \right)^\dagger
             &= M_{\iota_T},
             & 
            \left( \sum_{i\in\mathfrak{N}\setminus\{j\}} \bar H_{ij} \right) \,   \left( \sum_{i\in\mathfrak{N}\setminus\{j\}} \bar H_{ij} \right)^\dagger 
                   &= M_{\iota_T} ,
     \end{align*}
     which can equivalently be written as
     \begin{align}
         {\mathfrak D}^\dagger  \,  {\mathfrak D}     = {\mathfrak D}  \, {\mathfrak D}^\dagger  =  \mathbbm{I}_{2N} \otimes  M_{\iota_T} 
        =  \mathbbm{I}_{2NT} - \mathbbm{I}_{2N} \otimes  P_{\iota_T} ,
          \label{MoorePenroseDef2}
     \end{align}     
     where $P_{\iota_T} := T^{-1} \iota_T \iota_T'$.
    Now, using \eqref{MoorePenroseDef1} and $ \bar {\cal H} = {\mathfrak D} + {\cal G}$ we have
     \begin{align*}
             \bar {\cal H}^\dagger  \left( {\mathfrak D} + {\cal G} \right) &=  \mathbbm{I}_{2NT}  -  P_{\rm null}   .
     \end{align*}
    Multiplying this with ${\mathfrak D}^\dagger$ from the right, using \eqref{MoorePenroseDef2}
    and  \eqref{ProjectionHdagger}, and  bringing $ \bar {\cal H}^\dagger  {\cal G} {\mathfrak D}^\dagger $
    to the RHS gives
     \begin{align}
             \bar {\cal H}^\dagger  
                &={\mathfrak D}^\dagger -  P_{\rm null}   {\mathfrak D}^\dagger -   \bar {\cal H}^\dagger  {\cal G} {\mathfrak D}^\dagger .
             \label{ExpansionHessian1}
     \end{align}
     By transposing  this last equation we obtain
     \begin{align}
             \bar {\cal H}^\dagger 
               &={\mathfrak D}^\dagger - {\mathfrak D}^\dagger   P_{\rm null}   -  {\mathfrak D}^\dagger  {\cal G} \bar {\cal H}^\dagger  ,
             \label{ExpansionHessian2}
     \end{align}
     and now plugging \eqref{ExpansionHessian1} into the RHS of \eqref{ExpansionHessian2} gives
     \begin{align*}
           \bar {\cal H}^\dagger   &=  {\mathfrak D}^\dagger   - {\mathfrak D}^\dagger   P_{\rm null}   
           -  {\mathfrak D}^\dagger  {\cal G}  {\mathfrak D}^\dagger 
            + {\mathfrak D}^\dagger  {\cal G}   P_{\rm null}   {\mathfrak D}^\dagger 
             -  {\mathfrak D}^\dagger  {\cal G} \bar   {\cal H}^\dagger  {\cal G} {\mathfrak D}^\dagger
          \nonumber \\
            &=    {\mathfrak D}^\dagger     -  {\mathfrak D}^\dagger  {\cal G}  {\mathfrak D}^\dagger 
             - {\mathfrak D}^\dagger   P_{\rm null}   
              -   P_{\rm null}   {\mathfrak D}^\dagger 
                +  {\mathfrak D}^\dagger  {\cal G} \bar   {\cal H}^\dagger  {\cal G} {\mathfrak D}^\dagger ,  
     \end{align*}
     where in the second step we used that $ {\mathfrak D}^\dagger  {\cal G}   P_{\rm null}    = - P_{\rm null}   $, which follows from 
     $0 =  \bar {\cal H} \,  P_{\rm null}    =   {\mathfrak D}  P_{\rm null}    +   {\cal G}   P_{\rm null}   $
     by left-multiplication with $ {\mathfrak D}^\dagger$ and using that $ {\mathfrak D}^\dagger {\mathfrak D}  P_{\rm null} = 0$.     
     This expansion argument for $  \bar {\cal H}^\dagger $ so far has followed the proof of Theorem 2 
     in \cite{jochmans2019fixed}.   
     We furthermore have here that
     $ {\mathfrak D}^\dagger \left(  \mathbb{I}_{2N} \otimes   P_{\iota_T} \right) = 0$,
     because $\bar H_{ij}  \iota_T = 0$,
     implying that  $ {\mathfrak D}^\dagger   P_{\rm null}    =  {\mathfrak D}^\dagger  P_v$.
   The expansion in the last display therefore becomes
     \begin{align}
           \bar {\cal H}^\dagger - {\mathfrak D}^\dagger    &=       -  {\mathfrak D}^\dagger  {\cal G}  {\mathfrak D}^\dagger 
             - {\mathfrak D}^\dagger    P_v 
              -    P_v  {\cal D}^\dagger 
                +  {\mathfrak D}^\dagger  {\cal G} \bar   {\cal H}^\dagger  {\cal G} {\mathfrak D}^\dagger ,
             \label{ExpansionExpHessian}   
     \end{align}     
     with $2NT \times T$ matrix $v$ defined in \eqref{DefMatrixV}.
This expansion    is the first key step in the proof of the lemma.

\medskip          
   \# \underline{Bound on the spectral norm of $ \bar {\cal H}^\dagger $:}    
     The term $ {\mathfrak D}^\dagger  {\cal G} \bar   {\cal H}^\dagger  {\cal G} {\mathfrak D}^\dagger$ 
     in the expansion \eqref{ExpansionExpHessian} still contains $ \bar {\cal H}^\dagger $ itself. In order to bound contributions
     from this term we therefore need a preliminary bound on the spectral norm of  $ \bar {\cal H}^\dagger $.
     
     The objective function $  \ell_{ij}(\beta, \pi_{ij})  :=  \ell_{ij}(\beta, \alpha_{it}, \gamma_{jt}) $ in  \eqref{DefLikelihood}
     is strictly convex in $\pi_{ij}$, apart from the flat direction given by the invariance $\pi_{ij} \mapsto \pi_{ij} + c_{ij} \, \iota_T$, $c_{it} \in \mathbb{R}$. This strict convexity together with our Assumption~\ref{ass:MAIN}(ii) that all regressors and parameters are uniformly bounded over $i,j,N,T$
     implies that for the $T \times T$ expected Hessian $\bar H_{ij}:= \mathbb{E}\left[ -\partial^{2}\ell_{ij}/\partial\pi_{ij}\partial\pi_{ij}^{\prime}(\beta_0,\alpha_0,\gamma_0) \right]$ there exists a constant $b>0$ that is independent of $i,j,N,T$ such that
     \begin{align*}
            \min_{\left\{ v \in \mathbb{R} \, : \,  \iota_T' v = 0  \right\}}  v' \bar H_{ij} v \geq b> 0 .
     \end{align*}
     The last display states that $ \bar H_{ij}$ is positive definite in all directions orthogonal to $ \iota_T$.
     Again, the lower bound $b>0$ holds uniformly due to Assumption~\ref{ass:MAIN}(ii).
     The last display result can equivalently be written as
     \begin{align}
         \bar H_{ij} \geq  b  \,  M_{\iota_T} ,
         \label{LowerBoundHessian}
     \end{align}
     where $\geq$
     means that the difference between the matrices is positive definite.      
     
     Next, let $e_i= (0,\ldots,0,1,0,\ldots,0)'$ be the $i$'th standard unit vector of dimension $N$.
     For all $i,j \in \mathfrak{N} :=\{1,\ldots,N\}$ we then have
     \begin{align*}
         \partial_{\phi} \pi'_{ij}  = {e_i \choose e_j} \otimes \mathbb{I}_T ,
     \end{align*} 
     which are $2NT \times T$ matrices.
     Because 
     $ \mathcal{L}(\beta, \phi)
      =  \sum_{i=1}^N \sum_{j\in\mathfrak{N}\setminus\{i\}} 
        \ell_{ij}(\beta, \pi_{ij}) $  
        we thus find that
             \begin{align*}
           \bar   {\cal H} &= \mathbb{E}\left[  - \partial_{\phi \phi'}   {\cal L}   \right] 
           =  \sum_{i=1}^N \sum_{j\in\mathfrak{N}\setminus\{i\}}   \left( \partial_{\phi} \pi'_{ij}  \right) \mathbb{E}\left[ -\partial_{\pi_{ij} \pi_{ij}'} \ell_{ij}  \right] \left( \partial_{\phi} \pi'_{ij}  \right)'
         \\  
           &=  \sum_{i=1}^N \sum_{j\in\mathfrak{N}\setminus\{i\}}  \left[  {e_i \choose e_j} \otimes \mathbb{I}_T \right] \bar H_{ij}   \left[  {e_i \choose e_j} \otimes \mathbb{I}_T \right]'
         \\
          &\geq \, b \,   \sum_{i=1}^N \sum_{j\in\mathfrak{N}\setminus\{i\}}   \left[  {e_i \choose e_j} \otimes \mathbb{I}_T \right] M_{\iota_T}   \left[  {e_i \choose e_j} \otimes \mathbb{I}_T \right]'
        \\
          &=  b \,   \left[  \sum_{i=1}^N \sum_{j\in\mathfrak{N}\setminus\{i\}}  {e_i \choose e_j}   {e_i \choose e_j}' \right] \otimes M_{\iota_T} 
        \\
         &= b \, \underbrace{
         \left( \begin{array}{cc}  (N-1) \mathbb{I}_N &  \iota_N \iota_N' - \mathbb{I}_N \\
                              \iota_N \iota_N' - \mathbb{I}_N &  (N-1) \mathbb{I}_N 
          \end{array}   \right) 
          }_{=: Q_{N}}
           \otimes M_{\iota_T} 
     \end{align*}   
     where we also used \eqref{LowerBoundHessian}. It is easy to show that for $N>2$ the $2N \times 2N$ matrix
     $Q_N$ has an eigenvalue zero  with multiplicity one,
     an eigenvalue $N-2$ with multiplicity $N-1$,  an eigenvalue $N$  with multiplicity $N-1$, and an eigenvalue $2(N-1)$  with multiplicity one.
     Thus, the smallest non-zero eigenvalue of $Q_N$ is $(N-2)$. Also, the zero-eigenvector of $Q_N$ is
       given by $v_0 := (\iota_N' , - \iota_N')'$, and therefore we have $Q_{N} \geq (N-2) \, M_{v_0}$,
       where $M_{v_0} = \mathbb{I}_{2N} - (2N)^{-1} v_0 v_0'$
       is the projector orthogonal to $v_0$. We therefore have
       \begin{align*}
              \bar   {\cal H} &\geq  b \,(N-2) \, M_{(\iota_N' , - \iota_N')'} \otimes  M_{\iota_T} 
              \\
              &=   b \,(N-2) \, \left( \mathbb{I}_{2NT}  -   P_{\rm null}  \right) ,
       \end{align*}
       where  $P_{\rm null}$ is the  projector onto the null-space of $  \bar {\cal H} $, as already defined above. From this it follows
       that
       \begin{align*}
             \bar   {\cal H}^\dagger \leq \frac 1 {b \, (N-2)}  \left( \mathbb{I}_{2NT}  -   P_{\rm null}  \right) ,
       \end{align*}
       and therefore for the spectral norm
       \begin{align}
            \left\|  \bar   {\cal H}^\dagger \right\| \leq  \frac 1 {b \, (N-2)}  = O(1/N).
            \label{BoundSpectralHessian}
       \end{align}

\medskip          
   \# \underline{Final bound on   $\left\| \bar {\cal H}^\dagger -
 {\mathfrak D}^\dagger
    \right\|_{\max}$:}    
    Using \eqref{LowerBoundHessian} we find
    \begin{align*}
         \max_{i \in \mathfrak{N}} \left(\frac 1 {N-1} \sum_{j\in\mathfrak{N}\setminus\{i\}}\bar{H}_{ij}\right)^{\dagger} &= O_P(1),
      &
        \max_{j \in \mathfrak{N}} \left(\frac 1 {N-1} \sum_{i\in\mathfrak{N}\setminus\{j\}}\bar{H}_{ij}\right)^{\dagger} &= O_P(1) .
    \end{align*}
    This together with our boundedness Assumption~\ref{ass:MAIN}(ii)
    implies that 
    \begin{align}
           \left\| {\mathfrak D}^\dagger \right\| _{\max} &= O_P(1/N),
         &
          \left\|  {\cal G} \right\|_{\max}    &= O_P(1).
           \label{MaxNormOrders}
    \end{align} 
    The definition of the  $2NT \times T$ matrix  $v$ in \eqref{DefMatrixV} implies that
     \begin{align}
        \left\| P_v \right\|_{\max} 
        &=  \left\|     P_{(\iota_N' , - \iota_N')'}   \otimes  M_{\iota_T}    \right\|_{\max}
       \leq  \left\|     P_{(\iota_N' , - \iota_N')'}     \right\|_{\max}
         =  (2N)^{-1}  \left\|    (\iota_N' , - \iota_N')'      (\iota_N' , - \iota_N')   \right\|_{\max}
  \nonumber  \\
      &=    (2N)^{-1}  = O(1/N) ,
           \label{MaxNormOrder1}
    \end{align}    
  where we also used that $\left\|  M_{\iota_T}    \right\|_{\max} \leq 1$.    
 In the following display, let $e_k= (0,\ldots,0,1,0,\ldots,0)'$ be the $k$'th standard unit vector of dimension $2NT$.
   We find that
    \begin{align}
          \left\|  {\cal G} \bar   {\cal H}^\dagger  {\cal G}  \right\|_{\max} 
       &= \max_{k,\ell \in \{1,\ldots,2NT\}}
      \left|   e_k' \,  {\cal G} \bar   {\cal H}^\dagger  {\cal G} e_\ell \right|
   \nonumber \\  
      &  \leq 
      \left( \max_{k  \in \{1,\ldots,2NT\}}  \left\|    {\cal G}   e_k \right\| \right)
         \left\|   \bar   {\cal H}^\dagger  \right\|
          \left( \max_{\ell  \in \{1,\ldots,2NT\}}  \left\|    {\cal G}   e_\ell \right\| \right)   
   \nonumber \\  
       &=   \left( \max_{k  \in \{1,\ldots,2NT\}}  \left\|    {\cal G}   e_k \right\| \right)^2
         \left\|   \bar   {\cal H}^\dagger  \right\|
    \nonumber \\       
      &\leq 
     \left(    \sqrt{2NT}   \left\|  {\cal G}   \right\|_{\max}   \right)^2
         \left\|   \bar   {\cal H}^\dagger  \right\|
       \nonumber \\          
        & =    O_P(1) ,
          \label{MaxNormOrder2}
    \end{align}    
    where the first line is just the definition of  $  \left\| \cdot \right\|_{\max}$,
    the second step uses definition of the spectral norm $ \left\|   \bar   {\cal H}^\dagger  \right\|$,
    the third step is an obvious rewriting,
    the fourth step uses that the norm of $2NT$-vector $  {\cal G}   e_k$ 
    can at most be $ \sqrt{2NT}$ times the maximal absolute element of the vector,
    and the final step uses that $T$ is fixed in our asymptotic and $ \left\|  {\cal G} \right\|_{\max}     = O_P(1)$
    and also
     \eqref{BoundSpectralHessian}.
    
    Next, for general $2NT \times 2NT$ matrices $A$ and $B$ we  have the bound
       (notice that $\| \cdot \|_{\max}$ is not a matrix norm)
       \begin{align*}
            \| A B \|_{\max} \leq 2NT \, \|A\|_{\max} \, \|B\|_{\max} ,
       \end{align*}
       but
       because   ${\mathfrak D}$ is block-diagonal (with non-zero $T \times T$ blocks on the diagonal)
    we have for any $2NT \times 2NT$ matrix $A$ the much improved bound
    \begin{align*}
        \left\|  {\mathfrak D} A  \right\|_{\max} \leq   T   \left\|  {\mathfrak D} \right\|_{\max} \left\| A  \right\|_{\max} .
    \end{align*}
    Applying those inequalities to the expansion of $ \bar {\cal H}^\dagger  -  {\mathfrak D}^\dagger$ obtained
    from \eqref{ExpansionExpHessian}, and also using \eqref{MaxNormOrders}
    and \eqref{MaxNormOrder1} and \eqref{MaxNormOrder2}, we find that
     \begin{align*}
             \left\| \bar {\cal H}^\dagger -
 {\mathfrak D}^\dagger
    \right\|_{\max}
      & \leq   T^2 \left\| {\mathfrak D}^\dagger \right\|^2_{\max} \left\|  {\cal G} \right\|_{\max}   
       + 2 T  \left\| {\mathfrak D}^\dagger \right\|_{\max} \left\|     P_v  \right\|_{\max} 
           + T^2   \left\| {\mathfrak D}^\dagger \right\|^2_{\max} \left\|  {\cal G} \bar   {\cal H}^\dagger  {\cal G}  \right\|_{\max}    
          \\
         &=   O_P(1/N^2),
     \end{align*}
     as $N \rightarrow \infty$ (remember that $T$ is fixed in our asymptotic.)     This is what we wanted to show.
\end{proof}

\subsubsection*{Proof of Proposition~\ref{bias-results}}
    The pseudo-likelihood function of the Poisson model is strictly concave in the single index.
    Therefore,
    Assumption~\ref{ass:MAIN} together  with Lemma~\ref{lemma:Hessian} 
    guarantee that the conditions of Theorem~B.1 in \cite{fernandez-val_individual_2016} are satisfied
    for the rescaled and penalized objective function\footnote{
    Since we have a concave objective function, we can apply Theorem~B.3 in FW to obtain preliminary convergence results for 
     both $ \widehat \beta$ and $\widehat \phi$. That theorem guarantees that     
     that the consistency condition on $\widehat \phi(\beta)$ in
    Assumption (iii) of Theorem~B.1 in FW is  satisfied under our Assumption~\ref{ass:MAIN},
    and it also guarantees $\left\| \widehat \beta - \beta^0 \right\| = O_P(N^{-1/2})$, which is important
    to apply 
     Corollary B.2 in FW to obtain the expansion result in our equation \eqref{MainExpansionBeta}.
    }
    \begin{align*}
         \frac 1 {\sqrt{N(N-1)}} \,  \mathcal{L}(\beta, \phi)  - \frac 1 2  \phi' \, P_{\rm null} \, \phi   ,
    \end{align*}
    with $ P_{\rm null} $ defined in \eqref{DefPnull}. Here, the penalty term $\phi' \, P_{\rm null} \, \phi $     
    guarantees {\it strict} concavity in $(\beta, \phi) $.
    However, in the following all derivatives of $ \mathcal{L}(\beta, \phi) $ are evaluated at the true parameters,
    and since we impose the normalization $P_{\rm null} \, \phi_0 = 0 $ the only derivative of  $ \mathcal{L}(\beta, \phi) $ where the
    penalty term gives a non-zero contribution is the incidental parameter Hessian matrix 
      $\bar {\cal H} =  \mathbb{E}\left[  - \partial_{\phi \phi'}   {\cal L}(\beta_0,\phi_0)  \right] $ for which the
      penalty term provides exactly the correct regularization. 
      However, instead of that regularization, we can equivalently use the pseudoinverse; namely we have
      \begin{align*}
           \left(  \bar {\cal H}  +  c \, P_{\rm null} \right)^{-1}
           &=  \bar {\cal H}^\dagger + \frac 1 c \, P_{\rm null},
      \end{align*}
      for any $c>0$.
      In all expressions below where $ \bar {\cal H}^\dagger $ appears we could equivalently write 
      $\bar {\cal H}^\dagger + \frac 1 N  P_{\rm null}$, but the additional
       contributions from $ \frac 1 N  P_{\rm null}$ will always vanish
       because the gradient of $  \mathcal{L}(\beta, \phi)$
      with respect to $\phi$ is orthogonal to $P_{\rm null}$.

    By applying Theorem~B.1 and its 
     Corollary B.2 in FW we thus obtain
    \begin{align}
         \sqrt{N(N-1)} (\widehat \beta - \beta^0) = W_N^{-1} U_N+ o_P(1) ,
         \label{MainExpansionBeta}
    \end{align}
    where 
    \begin{align*}
    W_N   &= - \, \frac 1 {N(N-1)} \,
                \left( \partial_{\beta \beta'} \bar {\cal L}
                     + [\partial_{\beta \phi'} \bar {\cal L}] \; \bar {\cal H}^\dagger \;
             [\partial_{\phi \beta'} \bar {\cal L}]  \right)
          \\   
           &=  - \frac 1 {N(N-1)}   \sum_{i=1}^{N} \sum_{j\in\mathfrak{N}\setminus\{i\}}  \partial_{\beta \beta'}  \bar \ell^*_{ij}   
     \end{align*}      
    was already defined in Proposition~\ref{bias-results}, and we have
     $U_N := U_N^{(0)} + U_N^{(1)}$, with
   \begin{align*}
     U_N^{(0)} &=  \frac 1 {\sqrt{N(N-1)}} 
                \left[      \partial_{\beta} {\cal L}
                   + [\partial_{\beta \phi'} \bar {\cal L} ]\, \bar {\cal H}^\dagger  \partial_{\phi} {\cal L} \right]
                 = \frac 1 {\sqrt{N(N-1)}}   \,   \partial_{\beta} {\cal L}^* 
                \\   
                    &= \frac 1 {\sqrt{N(N-1)}}   \sum_{i=1}^{N} \sum_{j\in\mathfrak{N}\setminus\{i\}}   \partial_{\beta} \ell^*_{ij} ,
 \\[5pt]
   \sqrt{N(N-1)}   U_N^{(1)} &=   
      [\partial_{\beta \phi'} {\cal L} - \partial_{\beta \phi'} \bar {\cal L} ] \bar {\cal H}^\dagger \partial_{\phi} {\cal L} 
        -  [\partial_{\beta \phi'} \bar {\cal L}] \,
                      \bar {\cal H}^\dagger \, 
                      \left[  {\cal H}  - \bar {\cal H} \right] \,
                      \bar {\cal H}^\dagger \, \partial_{\phi} {\cal L} 
     \nonumber \\ &\qquad 
    +  
             \frac 1 2 \,    \sum_{g=1}^{\dim \phi}
           \left( \partial_{\beta \phi' \phi_g} \bar {\cal L}
               +  [\partial_{\beta \phi'} \bar {\cal L}] \, \bar {\cal H}^\dagger
            [\partial_{\phi \phi' \phi_g} \bar {\cal L}]  \right)
              [\bar {\cal H}^\dagger\partial_{\phi} {\cal L} ]_g
              \bar {\cal H}^\dagger \partial_{\phi} {\cal L}  
         \\
          &=       [\partial_{\beta \phi'} {\cal L}^* - \partial_{\beta \phi'} \bar {\cal L}^* ] \bar {\cal H}^\dagger \partial_{\phi} {\cal L} 
    +  
            \frac 1 2 \,    \sum_{g=1}^{\dim \phi}
 \partial_{\beta \phi' \phi_g} \bar {\cal L}^*
              [\bar {\cal H}^\dagger\partial_{\phi} {\cal L} ]_g
              \bar {\cal H}^\dagger \partial_{\phi} {\cal L}       .
   \end{align*} 
   Here, $\ell^*_{ij} $ was defined in \eqref{DefEllStar}, 
   all ``bars'' denote expectations conditional on $X$ and $\phi$,
   and all the partial derivatives are evaluated at the true parameters.  
   We also defined
   $ {\cal L}^*(\beta,\phi) :=   \sum_{i=1}^{N} \sum_{j\in\mathfrak{N}\setminus\{i\}}   \ell^*_{ij}(\beta, \alpha_{it}, \gamma_{jt}) $.   
  Remember that we use a different scaling of the (profile) likelihood function than FW; namely
   in \eqref{profile} we define 
   ${\cal L}(\beta,\phi)  =   \sum_{i=1}^{N} \sum_{j\in\mathfrak{N}\setminus\{i\}}   \ell_{ij}(\beta, \alpha_{it}, \gamma_{jt}) $, while in FW this function would have an additional factor $1/\sqrt{N(N-1)}$.
   This explains the additional $\sqrt{N(N-1)}$ factors  in $ W_N $, $  U_N^{(0)} $ and $U_N^{(1)} $ as compared to
   Theorem~B.1 in FW.
   
   The score term $ \partial_{\beta} \ell^*_{ij}  = \widetilde x_{ij}' S_{ij}$ has zero mean
   and finite variance and is independent across $i$ and $j$,
   conditional on $X$ and $\phi$. By  the central limit theorem we thus find
   \begin{align*}
         U_N^{(0)}  &\Rightarrow {\cal N}(0, \Omega_N)  ,
   \end{align*}  
   where
   \begin{align*}   
         \Omega_N 
          &= \frac 1 {N(N-1)}   \sum_{i=1}^{N} \sum_{j\in\mathfrak{N}\setminus\{i\}}
        {\rm Var}\left( \partial_{\beta} \ell^*_{ij} \,\big|\,x_{ij}\right) 
          \\
         &= \frac 1 {N(N-1)}   \sum_{i=1}^{N} \sum_{j\in\mathfrak{N}\setminus\{i\}}
          \widetilde{x}_{ij}^{\prime}\,\left[{\rm Var}\left(S_{ij}\,\big|\,x_{ij}\right)\right]\,\widetilde{x}_{ij} .
   \end{align*}  
   Thus, the term $ U_N^{(0)}$ only contributes variance to the asymptotic distribution of $\widehat \beta $, but
   no asymptotic bias. By contrast, the term $U_N^{(1)} $ only contributes 
   bias to the asymptotic distribution of $\widehat \beta $, but no variance. Namely, one finds that
   \begin{align}
             U_N^{(1)} \, \rightarrow_p \,  B_N + D_N  ,
        \label{BIASformula}
    \end{align} 
    with $B_N$ and $D_N$ as given in the proposition.
     The proof of \eqref{BIASformula}    is exactly analogous to the corresponding discussion
   of those terms in the proof of Theorem 4.1 in FW, which we restated above 
   as Theorem~\ref{th:FW} (remember that for $T=2$ our result here is indeed just a special case of Theorem 4.1 in FW.)
   Therefore, instead of repeating those derivations here, we  provide in the following a slightly less rigorous, but 
   much easier to follow, derivation of those bias terms.
  
\subsubsection*{Derivation of the asymptotic bias in Proposition~\ref{bias-results}}

Remember that the main difference  between Theorem~\ref{th:FW} and our case here
is that for us the incidental parameters $\alpha_i$ and $\gamma_j$ are $T$-vectors, while
in Theorem~\ref{th:FW} the index  $\pi_{ij}  = \alpha_i + \gamma_j$ is just a scalar.
An easy way to generalize the asymptotic bias formulas
 in Theorem~\ref{th:FW} and display \eqref{GeneralBias1d} to vector-valued incidental parameters is to use
a suitable parameterization for the incidental parameters  $\alpha_i$ and $\gamma_j$. The formulas for $ \overline B_1$
and $ \overline D_1 $ can most easily be generalized by parameterizing  the incidental parameters as follows
\begin{align}
      \alpha_i  &=  A_i  \, \widetilde \alpha_i ,
      &
      \gamma_j  &=  C_j \, \widetilde \gamma_j ,
   \label{TransformationAlphaGamma}   
\end{align}
where $\widetilde \alpha_i $ and $\widetilde \gamma_j$ are $T-1$ vectors,
and $A_i$ and $C_j$ are $T \times (T-1)$ matrices that satisfy
\begin{align}
    A_i A_i' &=   \left( \sum_j  \bar H_{ij} \right)^{\dagger} ,
 &
   C_j C_j' &=   \left( \sum_i  \bar H_{ij} \right)^{\dagger}  .
   \label{DefAC1}
\end{align}
Let $\widetilde {\cal L}(\beta, \widetilde \alpha, \widetilde \gamma )
= {\cal L}(\beta,  (A_i  \, \widetilde \alpha_i) , (C_j \, \widetilde \gamma_j))$.
This reparameterization guarantees that
\begin{align}
    \frac{\partial^2   \widetilde {\cal L}(\beta^0, \widetilde \alpha^0, \widetilde \gamma^0 ) }
     {(\partial  \widetilde \alpha_i) (\partial  \widetilde \alpha_i)'}
     &=  A_i'  \left( \sum_j  \bar H_{ij} \right) A_i = \mathbb{I}_{T-1} ,
    \nonumber \\
       \frac{\partial^2   \widetilde {\cal L}(\beta^0, \widetilde \alpha^0, \widetilde \gamma^0 ) }
     {(\partial  \widetilde \gamma_j) (\partial  \widetilde \gamma_j)'}
     &=  C_j'  \left( \sum_i  \bar H_{ij} \right) C_j = \mathbb{I}_{T-1} .
     \label{Normalization1}
\end{align}
That is, the Hessian matrix with respect to the incidental parameters $ \widetilde \alpha_i$ and $\widetilde \gamma_j$
is normalized to be an identity matrix under that normalization. It can be shown that this implies that the incidental parameter 
biases $\overline B_1$  and $\overline D_1$ ``decouple'' across the $T-1$ components of $\widetilde \alpha_i$
and $ \widetilde \gamma_j$; that is, the total contribution to the incidental parameter bias of $\widehat \beta$
just becomes a sum over $T-1$ contributions of the form $\overline B_1$  and $\overline D_1$ in \eqref{GeneralBias1d}.
Thus, for $k \in \{1,\ldots,K\}$ we have
\begin{align*}
   B_{1,k} &=
       \sum_{q=1}^{T-1}  \left[   -      \frac {1} {N}  \sum_{i,j}
            \frac{  
        \mathbb{E} \left(
               \partial_{\tilde \alpha_{i,q}} \ell_{ij} \mathcal{D}_{\beta_k {\tilde  \alpha_{i,q}}} \ell_{ij} \right) }
        { \sum_{j'} \mathbb{E}\left( \partial_{{\tilde  \alpha_{i,q}}^2} \ell_{ij'} \right) } 
     \right] 
    =    
       \sum_{q=1}^{T-1}  \left[   -      \frac {1} {N}  \sum_{i,j}
        \mathbb{E} \left(
               \partial_{\tilde \alpha_{i,q}} \ell_{ij} \mathcal{D}_{\beta_k {\tilde  \alpha_{i,q}}} \ell_{ij} \right)  
     \right]      
  \\
    &=    -   
    \frac {1} {N}  \sum_{i,j}
        \mathbb{E}       \left[   \left(   \partial_{\tilde \alpha_{i}} \ell_{ij}  \right)' \left( \mathcal{D}_{\beta_k {\tilde  \alpha_{i}}} \ell_{ij} \right)      \right]   
     =  -   
    \frac {1} {N}  \sum_{i,j}
        \mathbb{E}       \left[   \left(   \partial_{\alpha_{i}} \ell_{ij}  \right)' 
         A_i A_i' 
        \left( \mathcal{D}_{\beta_k {\alpha_{i}}} \ell_{ij} \right)      \right]      
   \\
     &=    -   
    \frac {1} {N}  \sum_{i,j}
        \mathbb{E}       \left[  S_{ij}' 
      \left( \sum_{j'}  \bar H_{ij'} \right)^{\dagger}
       H_{ij} \, \widetilde x_{ij,k}    \right]      ,
\end{align*}
where in the second step 
we used the fact that $\sum_{j'} \mathbb{E}\left( \partial_{{\tilde  \alpha_{i,q}}^2} \ell_{ij'} \right)=1$ according to \eqref{Normalization1},
in the third step we rewrote the sum over $q \in \{1,\ldots,T-1\}$ in terms of the vector product of the $T-1$ vectors 
$   \partial_{\tilde \alpha_{i}} \ell_{ij}$ and $\mathcal{D}_{\beta_k {\tilde  \alpha_{i}}} \ell_{ij}$, in the 
fourth step we used that $ \alpha_i   =  A_i  \, \widetilde \alpha_i$, and in the final step we used \eqref{DefAC1}
and the definitions of $S_{ij}$, $H_{ij}$ and $\widetilde x_{ij,k}$.
All expectations here are conditional on $X$ (in the main text we always make that conditioning
explicit), and $\bar H_{ij'}$ and $ \widetilde x_{ij,k} $ are non-random conditional on $X$; that is, we can also write this last expression
as 
\begin{align*}
   B_{1,k} &= -   
    \frac {1} {N}  \sum_{i}
 {\rm Tr}\left[    \left( \sum_{j'}  \bar H_{ij'} \right)^{\dagger}
     \sum_j   \mathbb{E}       \left(    H_{ij} \, \widetilde x_{ij,k} \, S_{ij}'     \right)
        \right] .
\end{align*}   
Analogously we find
\begin{align*}      
    D_{1,k} &=     -    \,
    \frac {1} {N}  \sum_{i,j}
        \mathbb{E}       \left[  S_{ij}' 
      \left( \sum_{i'}  \bar H_{i'j} \right)^{\dagger}
        H_{ij} \, \widetilde x_{ij,k}     \right] .
\end{align*}     
Next, to generalize the incidental parameter 
biases $\overline B_2$  and $\overline D_2$ in \eqref{GeneralBias1d} to vector-values $\alpha_i$ and $\gamma_j$ 
we again make a transformation \eqref{TransformationAlphaGamma}, but this time we choose
\begin{align}
    A_i A_i' &=   \left( \sum_j  \bar H_{ij} \right)^{\dagger}
     \left[ \sum_j \mathbb{E}\left(S_{ij}\,S'_{ij}\big|x_{ij}\right)     \right]
     \left( \sum_j  \bar H_{ij} \right)^{\dagger} .
  \nonumber \\  
   C_j C_j' &=   \left( \sum_i  \bar H_{ij} \right)^{\dagger}
     \left[ \sum_i \mathbb{E}\left(S_{ij}\,S'_{ij}\big|x_{ij}\right)     \right]
     \left( \sum_i  \bar H_{ij} \right)^{\dagger} .
  \label{DefAC2}   
\end{align}
Notice that for a correctly specified likelihood we have the Bartlett identities $  \bar H_{ij} =   \mathbb{E}\left(S_{ij}\,S'_{ij}\big|x_{ij}\right) $,
implying that \eqref{DefAC1} and \eqref{DefAC2} are identical for correctly specified likelihoods.
In general, however, the transformation now is different. Instead of normalizing the Hessian matrices
to be identities, as in \eqref{Normalization1}, the new transformation defined by \eqref{DefAC2} guarantees that
\begin{align}
    {\rm AsyVar}\left(  \widehat {\widetilde \alpha}_i \right)
    &=  
    \left[ \frac{\partial^2   \widetilde {\cal L}(\beta^0, \widetilde \alpha^0, \widetilde \gamma^0 ) }
     {(\partial  \widetilde \alpha_i) (\partial  \widetilde \alpha_i)'} \right]^\dagger
     {\rm Var}\left[   \frac{\partial   \widetilde {\cal L}(\beta^0, \widetilde \alpha^0, \widetilde \gamma^0 ) } 
     {\partial  \widetilde \alpha_i }  \, \Bigg| \, X \right]
         \left[ \frac{\partial^2   \widetilde {\cal L}(\beta^0, \widetilde \alpha^0, \widetilde \gamma^0 ) }
     {(\partial  \widetilde \alpha_i) (\partial  \widetilde \alpha_i)'} \right]^\dagger
     = \mathbb{I}_{T-1},
  \nonumber  \\  
    {\rm AsyVar}\left(  \widehat {\widetilde \gamma}_j \right)
    &=  
    \left[ \frac{\partial^2   \widetilde {\cal L}(\beta^0, \widetilde \alpha^0, \widetilde \gamma^0 ) }
     {(\partial  \widetilde \gamma_j) (\partial  \widetilde \gamma_j)'} \right]^\dagger
     {\rm Var}\left[   \frac{\partial   \widetilde {\cal L}(\beta^0, \widetilde \alpha^0, \widetilde \gamma^0 ) } 
     {\partial  \widetilde \gamma_j }  \, \Bigg| \, X \right]
         \left[ \frac{\partial^2   \widetilde {\cal L}(\beta^0, \widetilde \alpha^0, \widetilde \gamma^0 ) }
     {(\partial  \widetilde \gamma_j) (\partial  \widetilde \gamma_j)'} \right]^\dagger
     = \mathbb{I}_{T-1}.
        \label{Normalization2}  
\end{align}
Again, it can be shown that with this normalization the incidental parameter 
bias contributions $\overline B_2$  and $\overline D_2$ ``decouple''; that is,
each component of  $ \widehat {\widetilde \alpha}_i $ contributes an incidental parameter bias of the form $   \overline B_2 $
in \eqref{GeneralBias1d} to $\widehat \beta$,
and each component of $ \widehat {\widetilde \gamma}_i $ contributes an incidental parameter bias of the form $   \overline D_2 $
in \eqref{GeneralBias1d} to $\widehat \beta$. 
The total contribution thus reads, for $k \in \{1,\ldots,K\}$,
\begin{align*}
 B_{2,k} &=
       \sum_{q=1}^{T-1}  \left[    \frac 1 2          \frac 1 {N}  \sum_{i}
         \frac{  \left[  \sum_{j}   \mathbb{E}  ( \partial_{{\tilde  \alpha_{i,q}}} \ell_{ij} )^2 \right]
          \sum_{j}  \mathbb{E} ( \mathcal{D}_{\beta_k {\tilde  \alpha_{i,q}}^2 } \ell_{ij}  )  }
             {\left[  \sum_{j}  \mathbb{E}\left( \partial_{{\tilde \alpha_{i,q}}^2} \ell_{ij} \right) \right]^2}  \right] 
      \\     
      &=   \sum_{q=1}^{T-1}    \frac 1 2          \frac 1 {N}  \sum_{i,j}
     \mathbb{E} ( \mathcal{D}_{\beta_k {\tilde  \alpha_{i,q}}^2 } \ell_{ij}  ) 
     =      \frac 1 2          \frac 1 {N}  \sum_{i,j}
 {\rm Tr}\left[    \mathbb{E} ( \mathcal{D}_{\beta_k \, \tilde  \alpha_i \tilde  \alpha_i' } \ell_{ij}  )  \right]
  \\
   &=      \frac 1 2          \frac 1 {N}  \sum_{i,j}
 {\rm Tr}\left[  A_i'  \,  \mathbb{E} ( \mathcal{D}_{\beta_k \,   \alpha_i   \alpha_i' } \ell_{ij}  ) \,  A_i \right] 
 \\
 &= \frac{1}{2\,N}\sum_{i}\mathrm{Tr}\left[
 { \left(\sum_{j} \bar G_{ij}\,\widetilde{x}_{ij,k}\right)}
 \left(\sum_{j}\bar{H}_{ij}\right)^{\dagger}\left[\sum_{j}\mathbb{E}\left(S_{ij}\,S{}_{ij}^{\prime}\big|x_{ij,k}\right)\right]\left(\sum_{j}\bar{H}_{ij}\right)^{\dagger}\right],
\end{align*}
where in the second step we used that
  $ \left[ \sum_{j}   \mathbb{E}  ( \partial_{{\tilde  \alpha_{i,q}}} \ell_{ij} )^2  \right] / 
             \left[  \sum_{j}  \mathbb{E}\left( \partial_{{\tilde \alpha_{i,q}}^2} \ell_{ij} \right) \right]^2   =1$
             according to \eqref{Normalization2}, in the third step we rewrote the sum over $q \in \{1,\ldots,T-1\}$
             as a trace over the $(T-1) \times (T-1)$ matrix of third-order partial derivatives 
               $\mathbb{E} ( \mathcal{D}_{\beta_k \, \tilde  \alpha_i \tilde  \alpha_i' } \ell_{ij}  )$,
in the 
fourth step we used that $ \alpha_i   =  A_i  \, \widetilde \alpha_i$,
and in the final step we used  the   cyclicity of the trace and
\eqref{DefAC2}
and the definitions of $\bar G_{ij}$, $\widetilde x_{ij,k}$, and the tensor-vector product $\bar G_{ij} \widetilde x_{ij,k}$ (which, recall, is a $T\times T$ matrix).

Analogously we find
\begin{align*}
D_{2,k} & =
     \sum_{q=1}^{T-1}  \left[    \frac 1 2          \frac 1 {N}  \sum_{j}
         \frac{  \left[  \sum_{i}   \mathbb{E}  ( \partial_{{\tilde  \gamma_{j,q}}} \ell_{ij} )^2 \right]
          \sum_{i}  \mathbb{E} ( \mathcal{D}_{\beta_k {\tilde  \gamma_{j,q}}^2 } \ell_{ij}  )  }
             {\left[  \sum_{i}  \mathbb{E}\left( \partial_{{\tilde \gamma_{j,q}}^2} \ell_{ij} \right) \right]^2}  \right] 
  \\
  &= \frac{1}{2\,N}\sum_{j} \mathrm{Tr}\left[\left(\sum_{i} \bar G_{ij}\,\widetilde{x}_{ij,k}\right)\left(\sum_{i}\bar{H}_{ij}\right)^{\dagger}\left[\sum_{i}\mathbb{E}\left(S_{ij}\,S'_{ij}\big|x_{ij,k}\right)\right]\left(\sum_{i}\bar{H}_{ij}\right)^{\dagger}\right].
\end{align*}
We have thus translated all the formulas in Theorem~\ref{th:FW}
and in display \eqref{GeneralBias1d} to the case of vector-valued $\alpha_i$ and $\gamma_j$
to find exactly the expression for the asymptotic biases
$B_N^k = B_{1,k} + B_{2,k} $ and $D_N^k = D_{1,k} + D_{2,k} $
  in Proposition~\ref{bias-results}.

\subsubsection*{Rewriting the bias expressions as in Appendix~\ref{Remark:RewriteBias}}
 
In the following, we unpack the formulas provided in Appendix~\ref{Remark:RewriteBias} in order to provide additional detail on why the leading bias term is of order $1/(NT)$ as both $N$ and $T$ $\rightarrow \infty$ simultaneously. Remember that
$
\mathbb{E}(y_{ijt}|x_{ijt},\alpha_{it},\gamma_{ij})   =\lambda_{ijt}  :=\exp(x_{ijt}'\beta+\alpha_{it}+\gamma_{ij})$
and
$ \vartheta_{ijt}  := \frac{\lambda_{ijt}} {\sum_{\tau}\lambda_{ij\tau}} $,
and denote the corresponding $T$-vectors by $y_{ij}$, $\lambda_{ij}$ and $\vartheta_{ij}$.
  It is convenient to define the $T \times T$ matrices
\begin{align*}
    \Lambda_{ij}  := {\rm diag} \left( \lambda_{ij} \right) ,
\end{align*}
and
\begin{align*}
    M_{ij} := \mathbf{I}_{T} - \frac{\lambda_{ij} \, \iota_{T}' } {\iota_{T}'  \lambda_{ij}} =  \mathbf{I}_{T} - \vartheta_{ij} \iota_{T}'  ,
\end{align*}
which is the unique idempotent $T \times T$ matrix (i.e.\ $M_{ij} M_{ij}= M_{ij}$) that satisfies ${\rm rank}(M_{ij} ) = T-1$, 
$M_{ij}  \lambda_{ij}  = 0$, and  $\iota_{T}'  M_{ij}    = 0$. Notice also that
$\lambda_{ij}  =  \Lambda_{ij} \iota_{T}$, and therefore $M_{ij}    \Lambda_{ij} =  \Lambda_{ij}  M_{ij}'$.
We then have
\begin{align*}
    S_{ij} &= M_{ij}' y_{ij} ,
    \\
    \bar H_{ij} &=  M_{ij} \,    \Lambda_{ij} \, M_{ij}'  = M_{ij} \,    \Lambda_{ij} = \Lambda_{ij}  M_{ij}' 
      =    \Lambda_{ij}  -  \frac{ \lambda_{ij} \lambda_{ij}'  } {\iota_{T}'  \lambda_{ij}} ,
    \\
    H_{ij} &=  \bar H_{ij}  \left( \frac { \iota_{T}'  y_{ij} }  {\iota_{T}'  \lambda_{ij}} \right) ,
\end{align*}
and
\begin{align*}
     \bar G_{ij,tsr} = - \sum_{u=1}^T \, \lambda_{ij,u} \, M_{ij,tu} \, M_{ij,su}  \, M_{ij,ru} ,
\end{align*}
where $t,s,r \in \{1,\ldots,T\}$.

Next, define $\widetilde x^*_{ij,k}:=M_{ij}' \widetilde x_{ij,k}$. 
Noting that
$ \lambda_{ij}'  \widetilde x^*_{ij,k} = 0$, we find
\begin{align*}
   W_{N,k\ell} &=\frac{1}{N\,(N-1)} \sum_{i,j} \, \widetilde{x}_{ij,k}^{* \prime}\,\Lambda_{ij}\,\widetilde{x}_{ij,\ell}^*
   \\
   &= \frac{1}{N\,(N-1)} \sum_{i,j,t} \,  \lambda_{ijt}\,  \widetilde{x}_{ijt,k}^* \, \widetilde{x}_{ijt,\ell}^* .
\end{align*}
This shows that $ W_{N}$ has an additional sum over $t$, so $ W_{N}$  increases linearly in $T$,
and $W_N^{-1} = O(T^{-1})$, for $T \rightarrow \infty$.

Now, also define
$D_{ij,k} :=  \diag \left[ \left(  \lambda_{ijt} \, \widetilde{x}^*_{ijt,k}  \right)_{t=1,\ldots,T} \right]$,
which is the diagonal $T \times T$ matrix with diagonal entries $ \lambda_{ijt} \, \widetilde{x}^*_{ijt,k}  $.
The first-order conditions of the optimization problem that defines $\widetilde x_{ij,k} $ read
\begin{align*}
    \sum_i  \bar H_{ij} \, \widetilde x_{ij,k} &=0 ,
    &
    \sum_j  \bar H_{ij} \, \widetilde x_{ij,k} &=0 ,
\end{align*}
or equivalently
\begin{align*}
    \sum_i  \Lambda_{ij} \, \widetilde x_{ij,k}^* &=0 ,
    &
    \sum_j  \Lambda_{ij} \, \widetilde x_{ij,k}^* &=0 ,
\end{align*}
which can also be written as
\begin{align}
    \sum_i D_{ij,k} &=0 ,
    &
    \sum_j    D_{ij,k}  &=0 .
   \label{FOC} 
\end{align}
These FOC's are only important to simplify the term $B_{2,k} $ in what follows.
We have
\begin{align*} 
    B_{1,k}  &= - \frac 1 N \sum_{i,j}  \frac{ \mathbb{E}\left[ \left(   \iota_{T}'  y_{ij} \right)   S'_{ij} \right] }  {\iota_{T}'  \lambda_{ij}} 
    \left( \sum_{j'} \bar H_{ij'} \right)^\dagger \Lambda_{ij} \, \widetilde x_{ij,k}^* 
    \\
    &=- \frac 1 {N (N-1)} \sum_{i,j} \frac{    \iota_{T}'    }  {\iota_{T}'  \lambda_{ij}}  {\rm Var}(y_{ij})
    M_{ij}' \left(\frac 1 {N} \sum_{j'} \bar H_{ij'} \right)^\dagger   \Lambda_{ij}  M_{ij}' \, \widetilde x_{ij,k}  ,
  \\    
   B_{2,k}   &= - \, \frac 1 {2 \, N} \sum_i {\rm Tr}\left\{ 
  \left[ \sum_j
   M_{ij} \,
    D_{ij,k} \, M_{ij}' 
     \right] \left( \sum_{j} \bar H_{ij} \right)^\dagger  \left[ \sum_j M_{ij}    {\rm Var}(y_{ij}) M_{ij}' \right]  \left( \sum_{j} \bar H_{ij} \right)^\dagger
    \right\}
   \\
   &=    \frac 1 {N (N-1)}  \sum_{i,j} \left\{
   \frac{ \lambda_{ij}' \, Q_i \, \Lambda_{ij} \, \widetilde x_{ij,k}^*   } {\iota_{T}'  \lambda_{ij} }
 -  
  \frac{ \left( \lambda_{ij}' \ \widetilde x_{ij,k}^* \right) \left( \lambda_{ij}'   Q_i \lambda_{ij} \right) } {\left( \iota_{T}'  \lambda_{ij} \right)^2 }  \right\} \\
   &=   \frac 1 {N (N-1)}  \sum_{i,j}
   \frac{ \lambda_{ij}' \, Q_i \, \Lambda_{ij} M_{ij}'\, \widetilde x_{ij,k}   } {\iota_{T}'  \lambda_{ij} }
,
\end{align*}
where, in the second-to-last step, we used the definition of $M_{ij}$, \eqref{FOC}, that $\iota_{T}' D_{ij,k} \iota_{T} =  \lambda_{ij}' \ \widetilde x_{ij,k}^*$, and that $D_{ij,k} \iota_{T} = \Lambda_{ij} \, \widetilde x_{ij,k}^*$; and in the last step, we used that $\Lambda_{ij} \, \widetilde x_{ij,k}^*  =\Lambda_{ij} \, M_{ij}' \widetilde x_{ij,k}$
and $\lambda_{ij}' \ \widetilde x_{ij,k}^* =0$. We also used the definition of $Q_i$ given in Appendix~\ref{Remark:RewriteBias}. We then have for $B_N^k = B_{1,k} + B_{2,k} $  that
\begin{align*}
  B_{N}^{k}
   &=  - \frac 1 {N (N-1)} \sum_{i,j} \frac{\frac 1 T \,  \iota_{T}' \, R_{ij}  \,  \widetilde x_{ij,k}} 
    {\frac 1 T \, \iota_{T}'  \lambda_{ij}} 
   +  \frac 1 {N (N-1)} \sum_{i,j} 
   \frac{\frac 1 T  \, \lambda_{ij}' \, Q_i \, \Lambda_{ij} M_{ij}' \, \widetilde x_{ij,k}  } 
  {\frac 1 T \, \iota_{T}'  \lambda_{ij} }
 ,
\end{align*}
where we have now also used the definition of $R_{ij}$ from Appendix~\ref{Remark:RewriteBias} in order to simplify $B_{1,k}$.
Under appropriate regularity conditions, the $T \times T$ matrices $Q_i$ and $R_{ij}$ each maintain diagonal elements of order one and off-diagonal elements of order  $1/T^2$ through their dependence on ${\rm Var}(y_{ij})$.
Therefore, all the numerators
and denominators in the last expression for $B_{N}^{k}$ remain of order one as $T \rightarrow \infty$,  such that $B_{N}^{k} = O(1)$ as $T \rightarrow \infty$, with an analogous result also following for $D_{N}^{k}$. Recalling that $W_N$ increases linearly with $T$, we thus conclude 
that the bias term
\begin{align*}
 \frac{W_{N}^{-1}(B_{N}+D_{N})}{N-1}  ,
\end{align*}
is of order $1/(NT)$ as both $N$ and $T$ grow large.

\subsubsection*{Comment on Proposition \ref{prop1}}

We note that the consistency result from Proposition \ref{prop1} also follows from the above proof of Proposition~\ref{bias-results}.

\begin{remark}If the asymptotic bias in $\widehat{\beta}$ is characterized by Proposition~\ref{bias-results}, then $\widehat{\beta}$ is consistently estimated as $N\rightarrow\infty$. \label{consistency-remark}\end{remark}

{As we have noted in the text, for this consistency result to hold, we need for the score of the profile log-likelihood $\ell_{ij} (\beta, \alpha_{it}, \gamma_{jt})$ from \eqref{DefLikelihood} to be unbiased when evaluated at the true parameters $(\beta^0,\alpha^0,\gamma^0)$. 
 In particular, we need for there to be no incidental parameter bias term of order $1/T$ associated with the pair fixed effect $\eta_{ij}$. As the following proof and subsequent discussion clarify, the FE-PPML estimator is quite special in this regard.}

\subsection{Proof of Proposition \ref{prop2}}

To prove  Proposition \ref{prop2}, it will first be useful to prove the following lemma:

\begin{lemma} Assume a ``one way'' panel data model with $\lambda_{it}=\exp(x_{it}'\beta+\alpha_{i})$
and consider the class of FE-PML panel estimators
with FOC's given by 
\begin{align*}
\widehat{\beta}\!\!:\,\sum_{i=1}^{N}\sum_{t=1}^{T}\,\!x_{it}\!\left(y_{it}-\widehat{\lambda}_{it}\right)\!g(\widehat{\lambda}_{it}) & =0, & \widehat{\alpha}_{i}\!\!:\,\sum_{t=1}^{T}\left(y_{it}-\widehat{\lambda}_{it}\right)\!g(\widehat{\lambda}_{it}) & =0,
\end{align*}
where $i=1,\ldots,N$, $t=1,...,T,$ and $g(\widehat{\lambda}_{it})$
is an arbitrary positive function of $\widehat{\lambda}_{it}$. If
$T$ is {fixed}, $\widehat{\beta}$ is only consistent 
under general assumptions about ${\rm Var}(y|x,\alpha)$ if $g(\lambda)$ is constant over the range
of $\lambda$'s that are realized in the data-generating process.\label{lemma1}\end{lemma}

Put simply, if Lemma \ref{lemma1} holds, then no other FE-PML estimator
of the form described in Proposition \ref{prop2} aside from FE-PPML can be consistent
under general assumptions about the conditional variance ${\rm Var}(y|x,\alpha,\gamma,\eta)$.
We have already shown that the three-way FE-PPML estimator is generally
consistent regardless of the conditional variance. Thus, if we can
prove Lemma \ref{lemma1}, Proposition \ref{prop2} follows directly.

\begin{proof}[\bf Proof of Lemma~\ref{lemma1}]
Our strategy here will be to
adopt a specific parameterization for the conditional variance ${\rm Var}(y|x,\alpha)$
and then examine the conditions under which $\widehat{\beta}$ is
sensitive to small changes in the conditional variance. If $\widehat{\beta}$
depends on ${\rm Var}(y|x,\alpha)$ even for large $N$, then it is not possible
for $\widehat{\beta}$ to be consistent under general assumptions
about ${\rm Var}(y|x,\alpha)$.

To proceed, let the true data generating process be given by
\begin{align*}
y_{it} & =\lambda_{it}\omega_{it},
\end{align*}
where $\lambda_{it}$ is the true conditional mean and
\begin{align}
\omega_{it} & :=\exp\left[-\frac{1}{2}\ln\left(1+\lambda_{it}^{\rho}\right)+\sqrt{\ln\left(1+\lambda_{it}^{\rho}\right)}z_{it}\right] \label{eq:omega}
\end{align}
with $z_{it}$ a randomly-generated variable distributed $\mathcal{N}(0,1)$.
$\omega_{it}$ is therefore a heteroskedastic multiplicative disturbance
that follows a log-normal distribution with $\mathbb{E}[\omega_{it}]=1$
and ${\rm Var}(\omega_{it})=\lambda_{it}^{\rho}$.
The conditional mean of $y_{it}$ is in turn given by $\mathbb{E}[y_{it}|x,\alpha]=\lambda_{it}$
and the conditional variance is given by ${\rm Var}(y_{it}|x,\alpha)={\rm Var}(y_{it}|\lambda_{it})=\lambda_{it}^{2}{\rm Var}(\omega_{it})=\lambda_{it}^{\rho+2}$.
Our focus is the exponent $\rho$, which governs the nature of the heteroskedasticity
and can be any real number. With this in mind, it is useful to document
the following results,
\begin{align}
\mathbb{E}\left[\frac{\partial\omega_{it}}{\partial\rho}\right] & =\frac{\partial\mathbb{E}\left[\omega_{it}\right]}{\partial\rho}=0\label{eq:Edw}\\
\mathbb{E}\left[\frac{\partial\left(\omega_{it}^{2}\right)}{\partial\rho}\right] & =\mathbb{E}\left[2\omega_{it}\frac{\partial\omega_{it}}{\partial\rho}\right]=\frac{\partial\mathbb{E}\left(\omega_{it}^{2}\right)}{\partial\rho}\nonumber \\
 & =\frac{\partial V\left[\omega_{it}\right]}{\partial\rho}=\lambda_{it}^{\rho}\ln\lambda_{it}\neq0.\label{eq:Edw2}
\end{align}
Put another way, the expected value of the change in $\omega_{it}$
with respect to $\rho$ must always be zero because $\mathbb{E}[\omega_{it}]=1$
regardless of $\rho$. Similarly, the expected change in the second
moment of $\omega_{it}$ must be $\lambda_{it}^{\rho}\ln\lambda_{it}$
because this gives the change in the variance of $\omega_{it}$.\footnote{Note here that $\frac{\partial(\omega_{it}^{2})}{\partial\rho}=2\omega_{it}\frac{\partial\omega_{it}}{\partial\rho}$. } 

To facilitate the rest of the proof, we invoke the following conceit:		
the random disturbance term $z_{it}$, once drawn from $\mathcal{N}(0,1)$,		
is \emph{known} and \emph{fixed}, such that each $\omega_{it}$ may		
be treated as a known transformation of the underlying value for $z_{it}$		
given by \eqref{eq:omega}. Among other things, this means we can		
always treat the partial derivatives $\frac{\partial\omega_{it}}{\partial\rho}$		
and $\frac{\partial y_{it}}{\partial\rho}=\lambda_{it}\frac{\partial\omega_{it}}{\partial\rho}$ as		
well-defined; similarly, we can treat the estimated parameters $\widehat{\beta}$		
and $\widehat{\alpha}_{i}$ as deterministic functions of the variance parameter $\rho$ with well-defined total		
derivatives $\frac{d\widehat{\beta}}{d\rho}$		
and $\frac{d\widehat{\alpha}_{i}}{d\rho}$. That is,		
for a given draw of $z_{it}$'s, we can perturb how the corresponding $\omega_{it}$'s are generated and		
consider comparative statics for how estimates are affected. If $\widehat{\beta}$		
is consistent regardless of the variance assumption used to generate		
$\omega_{it}$, then small changes in $\rho$ should have no effect		
on $\widehat{\beta}$ asymptotically. Thus, our goal in the following is to determine if there are any		
estimators in this class other than FE-PPML under which $\lim_{N\rightarrow\infty}\frac{d\widehat{\beta}}{d\rho}=0$		
in this experiment.

The next step is to totally differentiate the FOC's for $\widehat{\beta}$
and $\widehat{\alpha}_{i}$ with respect to a change in $\rho$. Let
$\mathcal{L}$ denote the pseudo-likelihood function to be maximized.\footnote{The implied pseudo-likelihood function is given here by $\mathcal{L}:=\sum_{i=1}^{N}\!\sum_{t=1}^{T}\!y_{it}\int\!\frac{g(\lambda_{it})}{\lambda_{it}}d\lambda_{it}-\sum_{i=1}^{N}\!\sum_{t=1}^{T}\!\int\!g(\lambda_{it})d\lambda_{it}$.}
For notational convenience, we can express the scores for $\widehat{\beta}$
and $\widehat{\alpha}_{i}$ as $\mathcal{L}_{\beta}$ and $\mathcal{L}_{\alpha_{i}}$,
such that their FOCs can respectively be written as $\mathcal{L}_{\beta}=0$
and $\mathcal{L}_{\alpha_{i}}=0$. Differentiating the FOC for $\widehat{\beta}$,
we obtain
\begin{align}
\frac{d\widehat{\beta}}{d\rho} & =-\mathcal{L}_{\beta\beta}^{-1}\mathcal{L}_{\beta\rho}-\mathcal{L}_{\beta\beta}^{-1}\sum_{i}\mathcal{L}_{\beta\alpha_{i}}\frac{d\widehat{\alpha}_{i}}{d\rho},\label{eq:dB_drho}
\end{align}
where $\mathcal{L}_{\beta\beta}$ is the matrix obtained from partially differentiating
the score for $\widehat{\beta}$ with respect to $\widehat{\beta}$,
$\mathcal{L}_{\beta\rho}$ (a vector) is the partial derivative of $\mathcal{L}_{\beta}$
with respect to $\rho$, and $\text{\ensuremath{\mathcal{L}_{\beta\alpha_{i}}}}$
(also a vector) is its partial derivative with respect to $\widehat{\alpha}_{i}$.
Applying a similar set of operations to the FOC for $\widehat{\alpha}_{i}$
then gives
\begin{align}
\frac{d\widehat{\alpha}_{i}}{d\rho} & =-\mathcal{L}_{\alpha_{i}\alpha_{i}}^{-1}\mathcal{L}_{\alpha_{i}\rho}-\mathcal{L}_{\alpha_{i}\alpha_{i}}^{-1}\mathcal{L}_{\beta\alpha_{i}}^{\prime}\frac{d\widehat{\beta}}{d\rho},\label{eq:da_drho-1}
\end{align}
where $\mathcal{L}_{\alpha_{i}\alpha_{i}}$ and $\mathcal{L}_{\alpha_{i}\rho}$ are
 scalars that respectively contain the partial derivatives of $\mathcal{L}_{\alpha_{i}}$
with respect to $\widehat{\alpha}_{i}$ and $\rho$. Plugging \eqref{eq:da_drho-1}
into \eqref{eq:dB_drho}, we have
\begin{align}
\frac{d\widehat{\beta}}{d\rho} & =-\mathcal{L}_{\beta\beta}^{-1}\mathcal{L}_{\beta\rho}+\mathcal{L}_{\beta\beta}^{-1}\sum_{i=1}^{N}\mathcal{L}_{\alpha_{i}\alpha_{i}}^{-1}\mathcal{L}_{\beta\alpha_{i}}\mathcal{L}_{\alpha_{i}\rho}+\mathcal{L}_{\beta\beta}^{-1}\sum_{i=1}^{N}\mathcal{L}_{\alpha_{i}\alpha_{i}}^{-1}\mathcal{L}_{\beta\alpha_{i}}\mathcal{L}_{\beta\alpha_{i}}^{\prime}\frac{d\widehat{\beta}}{d\rho}\nonumber \\
 & =-\left(\mathbf{I}-\mathcal{L}_{\beta\beta}^{-1}\sum_{i=1}^{N}\mathcal{L}_{\alpha_{i}\alpha_{i}}^{-1}\mathcal{L}_{\beta\alpha_{i}}\mathcal{L}_{\beta\alpha_{i}}^{\prime}\right)^{-1}\mathcal{L}_{\beta\beta}^{-1}\mathcal{L}_{\beta\rho} \\ & +\left(\mathbf{I}-\mathcal{L}_{\beta\beta}^{-1}\sum_{i=1}^{N}\mathcal{L}_{\alpha_{i}\alpha_{i}}^{-1}\mathcal{L}_{\beta\alpha_{i}}\mathcal{L}_{\beta\alpha_{i}}^{\prime}\right)^{-1}\mathcal{L}_{\beta\beta}^{-1}\sum_{i=1}^{N}\mathcal{L}_{\alpha_{i}\alpha_{i}}^{-1}\mathcal{L}_{\beta\alpha_{i}}\mathcal{L}_{\alpha_{i}\rho},\label{eq:dB_drho-1}
\end{align}
where $\mathbf{I}$ is an identity matrix whose dimensions equal the
size of $\beta$. 

Let $\mathbf{P}$ henceforth denote the combined matrix object $\mathbf{I}-\mathcal{L}_{\beta\beta}^{-1}\sum_{i}\mathcal{L}_{\alpha_{i}\alpha_{i}}^{-1}\mathcal{L}_{\beta\alpha_{i}}\mathcal{L}_{\beta\alpha_{i}}^{\prime}$.
It is straightforward to show that that first term in \eqref{eq:dB_drho-1},
$-\mathbf{P}^{-1}\mathcal{L}_{\beta\beta}^{-1}\mathcal{L}_{\beta\rho}$, converges
in probability to a zero vector when $N\rightarrow\infty$. To see
this, note first that $\mathbf{P}$ and $\mathcal{L}_{\beta\beta}$ must
be non-singular and finite for $\widehat{\beta}$ to be at a maximum
point of $\mathcal{L}$ and for $\frac{d\widehat{\beta}}{d\rho}$ to exist.
Furthermore, $\lim_{N\rightarrow\infty}NT\mathcal{L}_{\beta\beta}^{-1}=-\mathbb{E}[x_{it} \widehat{\lambda}_{it} g(\widehat{\lambda}_{it})x^\prime_{it}]^{-1}$
must also be non-singular and finite. Slutsky's theorem then implies
$\lim_{N\rightarrow\infty}-\mathbf{P}^{-1}\mathcal{L}_{\beta\beta}^{-1}\mathcal{L}_{\beta\rho}\rightarrow_{p}0$
if $\lim_{N\rightarrow\infty}N^{-1}T^{-1}\mathcal{L}_{\beta\rho}\rightarrow_{p}0$.
Examining the vector $\mathcal{L}_{\beta\rho}$ more closely, we have
\[
\mathcal{L}_{\beta\rho}=\sum_{i=1}^{N}\sum_{t=1}^{T}\,\!x_{it}\!\frac{\partial y_{it}}{\partial\rho}g(\widehat{\lambda}_{it})=\sum_{i=1}^{N}\sum_{t=1}^{T}\,\!x_{it}\lambda_{it}\frac{\partial\omega_{it}}{\partial\rho}g(\widehat{\lambda}_{it}).
\]
$\lim_{N\rightarrow\infty}N^{-1}T^{-1}\mathcal{L}_{\beta\rho}\rightarrow_{p}0$
then follows via standard arguments because $\mathbb{E}\left[\frac{\partial\omega_{it}}{\partial\rho}\right]=0$
(by \eqref{eq:Edw}). We may therefore focus our attention on the
second term on the RHS in \eqref{eq:dB_drho-1}, $\mathbf{P}^{-1}\mathcal{L}_{\beta\beta}^{-1}\sum_{i}\mathcal{L}_{\alpha_{i}\alpha_{i}}^{-1}\mathcal{L}_{\beta\alpha_{i}}\mathcal{L}_{\alpha_{i}\rho}$.
Noting that $\mathcal{L}_{\alpha_{i}\alpha_{i}}^{-1}$must be $<0$, in this
case we consider the conditions under which $\lim_{N\rightarrow\infty}N^{-1}T^{-1}\sum_{i}\mathcal{L}_{\alpha_{i}\alpha_{i}}^{-1}\mathcal{L}_{\beta\alpha_{i}}\mathcal{L}_{\alpha_{i}\rho}$
similarly converges in probability to zero. The summation in this
latter term may be expressed as
\begin{align*}
\sum_{i=1}^{N}\mathcal{L}_{\alpha_{i}\alpha_{i}}^{-1}\mathcal{L}_{\beta\alpha_{i}}\mathcal{L}_{\alpha_{i}\rho} & =\sum_{i=1}^{N}\mathcal{L}_{\alpha_{i}\alpha_{i}}^{-1}\left[\sum_{t=1}^{T}\,\!x_{it}\!\left(y_{it}-\widehat{\lambda}_{it}\right)\!g^{\prime}(\widehat{\lambda}_{it})\widehat{\lambda}_{it}-\sum_{t=1}^{T}\,\!x_{it}\widehat{\lambda}_{it}g(\widehat{\lambda}_{it})\right]\sum_{t=1}^{T}\frac{\partial y_{it}}{\partial\rho}\!g(\widehat{\lambda}_{it}).
\end{align*}
Re-arranging this expression, we have that 
\begin{align}
\sum_{i=1}^{N}\mathcal{L}_{\alpha_{i}\alpha_{i}}^{-1}\mathcal{L}_{\beta\alpha_{i}}\mathcal{L}_{\alpha_{i}\rho} & =\sum_{i=1}^{N}\sum_{t=1}^{T}\sum_{s=1}^{T}\mathcal{L}_{\alpha_{i}\alpha_{i}}^{-1}\!x_{it}y_{it}g^{\prime}(\widehat{\lambda}_{it})\widehat{\lambda}_{it}g(\widehat{\lambda}_{is})\frac{\partial y_{is}}{\partial\rho}\nonumber \\
 & -\sum_{i=1}^{N}\sum_{t=1}^{T}\sum_{s=1}^{T}\mathcal{L}_{\alpha_{i}\alpha_{i}}^{-1}\!x_{it}\!\left(\widehat{\lambda}_{it} g^{\prime}(\widehat{\lambda}_{it})+g(\widehat{\lambda}_{it})\right)\widehat{\lambda}_{it}g(\widehat{\lambda}_{is})\frac{\partial y_{is}}{\partial\rho}.\label{eq:expanded_larho}
\end{align}
Focusing first on the second of the two summation terms in \eqref{eq:expanded_larho},
we again apply $y_{it}=\lambda_{it}\omega_{it}$, $\frac{\partial y_{is}}{\partial\rho}=\lambda_{it}\frac{\partial\omega_{is}}{\partial\rho},$
and $\mathbb{E}\left[\frac{\partial\omega_{it}}{\partial\rho}\right]=0$.
We have that 
\begin{align*}
\lim_{N\rightarrow\infty}\frac{1}{NT}\sum_{i=1}^{N}\sum_{t=1}^{T}\sum_{s=1}^{T}\mathcal{L}_{\alpha_{i}\alpha_{i}}^{-1}\!x_{it}\!\left(\widehat{\lambda}_{it} g^{\prime}(\widehat{\lambda}_{it})+g(\widehat{\lambda}_{it})\right)\widehat{\lambda}_{it}g(\widehat{\lambda}_{is})\lambda_{is}\frac{\partial\omega_{is}}{\partial\rho} & \rightarrow_{p}0.
\end{align*}
This follows for the same reason $\lim_{N\rightarrow\infty}N^{-1}T^{-1}\mathcal{L}_{\beta\rho}\rightarrow_{p}0$
above. The first summation term in \eqref{eq:expanded_larho} obviously
$\rightarrow_{p}0$ as well if the estimator is FE-PPML, in which
case $g^{\prime}(\widehat{\lambda}_{it})=0$. To complete the proof,
we just need to show that this term does not reduce to $0$ if $g^{\prime}(\widehat{\lambda}_{it})\neq0$.
A final step gives us
\begin{align*}
\lim_{N\rightarrow\infty}\!\frac{1}{NT}\!\sum_{i=1}^{N}\!\sum_{t=1}^{T}\!\sum_{s=1}^{T}\mathcal{L}_{\alpha_{i}\alpha_{i}}^{-1}\!x_{it}y_{it}g^{\prime}(\widehat{\lambda}_{it})\widehat{\lambda}_{it}g(\widehat{\lambda}_{is})\frac{\partial y_{is}}{\partial\rho} & =\lim_{N\rightarrow\infty}\!\frac{1}{NT}\!\sum_{i=1}^{N}\!\sum_{t=1}^{T}\!\mathcal{L}_{\alpha_{i}\alpha_{i}}^{-1}x_{it}g^{\prime}(\widehat{\lambda}_{it})\widehat{\lambda}_{it}g(\widehat{\lambda}_{it})y_{it}\frac{\partial y_{it}}{\partial\rho}\\
 & =\lim_{N\rightarrow\infty}\!\frac{1}{NT}\!\sum_{i=1}^{N}\!\sum_{t=1}^{T}\!\mathcal{L}_{\alpha_{i}\alpha_{i}}^{-1}\!x_{it}g^{\prime}(\widehat{\lambda}_{it})\widehat{\lambda}_{it}g(\widehat{\lambda}_{it})\lambda_{it}^{2}\omega_{it}\frac{\partial\omega_{it}}{\partial\rho} \\ & \neq 0.
\end{align*}
To elaborate, the terms where $s\neq t$ vanish as $N\rightarrow\infty$
because disturbances are assumed to be independently distributed ($\mathbb{E}[\omega_{it}\frac{\partial\omega_{is}}{\partial\rho}]=0$
if $s\neq t$.)\footnote{Note that under FE-PPML, where $g^{\prime}(\widehat{\lambda}_{it})=0$,
the estimator is consistent even if disturbances are correlated. This
is yet another reason why FE-PPML is an especially robust estimator. } The remaining details follow from \eqref{eq:Edw2}.\footnote{Notice that if $T\rightarrow\infty$ also, we have that $\lim_{T\rightarrow\infty}T\mathcal{L}_{\alpha_{i}\alpha_{i}}^{-1}=-\mathbb{E}\left[\widehat{\lambda}_{it} g(\widehat{\lambda}_{it}) \right]^{-1}$
must be finite. We would therefore have
\begin{align*}
\lim_{N,T\rightarrow\infty}\frac{1}{NT}\sum_{i=1}^{N}\sum_{t=1}^{T}\left[T\mathcal{L}_{\alpha_{i}\alpha_{i}}^{-1}\right]\!x_{it}\!g^{\prime}(\widehat{\lambda}_{it})\widehat{\lambda}_{it}\!g(\widehat{\lambda}_{it})\!\lambda_{it}^{2}\!\left[T^{-1}\omega_{it}\frac{\partial\omega_{it}}{\partial\rho}\right] & =0,
\end{align*}
ensuring that $\widehat{\beta}$ does not depend on $\rho$ for the
large $N,$ large $T$ case. This follows because $\lim_{T\rightarrow\infty}T^{-1}V\left[\omega_{it}\right]=0\implies\lim_{T\rightarrow\infty}T^{-1}\mathbb{E}\left[\omega_{it}\frac{\partial\omega_{it}}{\partial\rho}\right]=0$.} We have now shown $\lim_{N\rightarrow\infty}\frac{d\widehat{\beta}}{d\rho}=0$
if and only if $g^{\prime}(\widehat{\lambda}_{it})=0$. In other words,
the estimator must be FE-PPML, which assumes $g(\widehat{\lambda}_{it})$
is a constant. For other FE-PML estimators, even if $\widehat{\beta}$
is consistent for a particular $\rho$, it cannot be consistent for
all $\rho$ because $\widehat{\beta}$ does not converge to the same
value for $N\rightarrow\infty$ when we vary $\rho$. As we discuss
below, this is what happens for FE-Gamma PML (where $g(\widehat{\lambda}_{it})=\widehat{\lambda}_{it}^{-1}$)
and some other similar {estimators}. 
\end{proof}

To be clear, the robustness of the FE-PPML estimator to misspecification
is a known result established by \citet{wooldridge_distribution-free_1999}.
However, to our knowledge, it has not previously been shown that FE-PPML
is the only estimator in the class we consider that has this property.\footnote{Alternatively, it is possible to extend the above result to an even
more general class of estimators by considering estimators that depend
on $g(\widehat{\alpha}_{i})$ rather than $g(\widehat{\lambda}_{it})$.
The same type of proof may be used to show that $\widehat{\beta}$
depends on the variance assumption if $g^{\prime}(\widehat{\alpha}_{i})\neq0$.
Furthermore, the estimator can be shown to be consistent if $g^{\prime}(\widehat{\alpha}_{i})=0$. } At the same time, it is worth clarifying that FE-PPML is not the
only estimator that is capable of producing consistent estimates of
three-way gravity models. Rather, it is the only estimator in the
class we consider that only requires correct specification of the
conditional mean and for the covariates to be conditionally exogenous
in order to be consistent. The following discussion describes some
known cases in which other estimators will be consistent.

\subsection{Results for Other Three-way Estimators}

Depending on the distribution of the data, there may be some other
consistent estimator available aside from FE-PPML. In particular,
if $g(\widehat{\lambda}_{ijt})$ is of the form $g(\widehat{\lambda}_{ijt})=\widehat{\lambda}_{ijt}^{q}$,
with $q$ an arbitrary real number, the FOC for $\widehat{\eta}_{ij}$
has a solution of the form $\widehat{\eta}_{ij}=[\sum_{t=1}^{T}\widehat{\mu}_{ijt}^{q+1}]^{-1}\sum_{t=1}^{T}y_{ijt}\widehat{\mu}_{ijt}^{q}$.
It is therefore possible to ``profile out'' $\widehat{\eta}_{ij}$
from the FOC for $\widehat{\beta}$, just as in the FE-PPML case.
As such, it is possible for the estimator to be consistently estimated,
but only if the conditional variance is correctly specified (more precisely, we must have ${\rm Var}(y|x,\alpha,\gamma,\eta)\propto\widehat{\lambda}_{it}^{1-q},$
the equivalent of $\rho=-1-q$.) In this case, the estimator is not
only consistent, but should be more efficient as well.

An interesting example to consider in the gravity context is the Gamma
PML (GPML) estimator, which imposes $g(\widehat{\lambda}_{ijt})=\widehat{\lambda}_{ijt}^{-1}$.
Generally speaking, GPML is 
considered the primary alternative
to PPML and OLS as an estimator for use with gravity equations (see
\citealp{head_gravity_2014,bosquet2015really}.) However, to our knowledge,
no references to date on gravity estimation make it clear that, unlike
in a two-way setting, the three-way FE-GPML estimator is only consistent
when the conditional variance is correctly specified.\footnote{As discussed in \citet{greene_fixed_2004}, the fixed effects Gamma
{model} is generally known not to suffer from an incidental parameter
problem, similar to FE-Poisson. However, the result stated in \citet{greene_fixed_2004}
is for the Gamma MLE estimator, which restricts the conditional variance
to be equal to the square of the conditional mean. The FE-Gamma PML
estimator is  
consistent under the slightly more general assumption
that the conditional variance is proportional to the square of the
conditional mean.} Thus, it is possible that researchers could mistakenly infer that
the appeal of FE-GPML as an alternative to FE-PPML in the two-way gravity
setting carries over to the three-way setting.\footnote{For example, \citet{head_gravity_2014}, arguably the leading reference
to date on gravity estimation, suggest comparing PPML estimates with
GPML estimates to determine if the RHS of the model is potentially
misspecified. Such a comparison is not straightforward in a three-way
setting because the GPML estimator is likely to be inconsistent. Their
other suggestion to compare GPML and OLS estimates still seems sensible,
however. As we show below, both estimators give similar results when
the Gamma variance assumption is satisfied and give different results
otherwise.} This is especially a concern now that recent computational advances
have made estimation of FE-GLM models significantly more feasible.

\renewcommand{\baselinestretch}{1.05}
\begin{figure}[h]
\centering{}\includegraphics[scale=1]{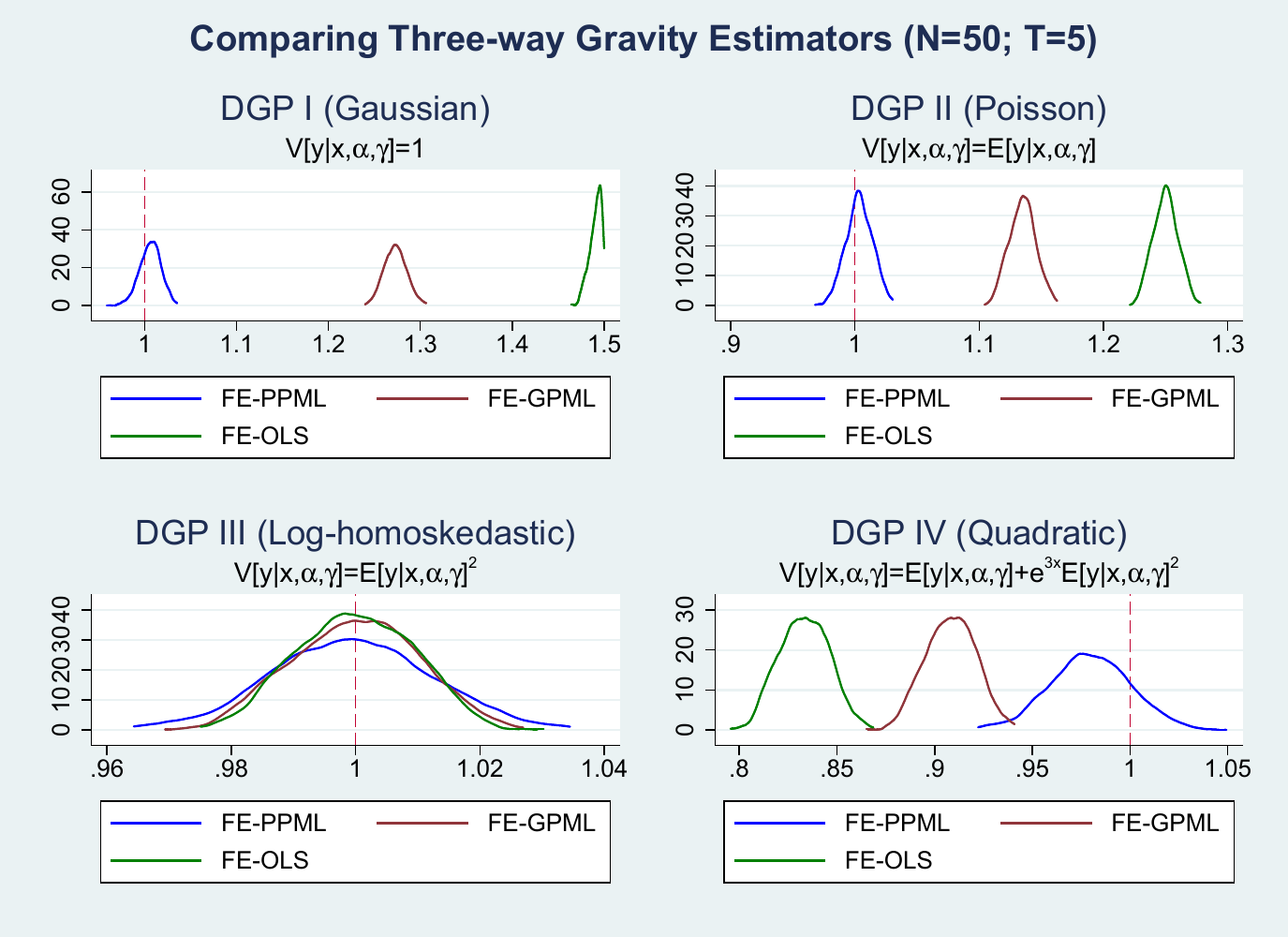}\caption{\footnotesize{Kernel density plots of three-way gravity model estimates using different
FE estimators, based on 500 replications. The model being estimated
is $y_{ijt}=\exp[\alpha_{it}+\gamma_{jt}+\eta_{ij}+x_{ijt}\beta]\omega_{ijt}$,
where the distribution of $\omega_{ijt}$ depends on the DGP and the
true value of $\beta$ is 1 (indicated by the vertical dotted lines).
The size of the $i$ and $j$ dimensions is given by $N=50$ and the
$t$ dimension has size $T=5$. See text for further details.}\label{other-three-way}}
\end{figure}
\renewcommand{\baselinestretch}{1.3}

To illuminate the unique IPP-robustness properties of FE-PPML in the
three-way context, Fig. \ref{other-three-way} shows a comparison
of simulation results for FE-PPML versus log-OLS and Gamma PML.\footnote{We were able to compute three-way FE-Gamma PML estimates using a modified
version of the HDFE-IRLS algorithm used in \citet{ppmlhdfe}. To our
knowledge, these are the first results presented anywhere documenting
the inconsistency of the three-way Gamma PML estimator. } The displayed kernel densities are computed using 500 replications
of a three-way panel structure with $N=50$ and $T=5$.\footnote{Simulations with larger $N$ are more narrowly distributed, but otherwise
are very similar.} 
The $i$ and $j$ dimensions of the panel both have size $N=50$ and
the size of the time dimension is $T=5$. The fixed effects are generated
according to the same procedures described in the text and we again
model four different scenarios for the distribution of the error term
(Gaussian, Poisson, Log-homoskedastic, and Quadratic).

As we would expect based on Proposition \ref{prop2}, FE-PPML is relatively
unbiased across all four different assumptions considered for the
distribution of the error term. The general inconsistency of the three-way
OLS estimator\textemdash which is only unbiased for DGP III where the
error term is log-homoskedastic\textemdash is also as expected. However,
the reasons behind the bias in the OLS estimate are well-documented
(see \citealp{santos_silva_log_2006}) and do not have to do with
the incidental parameters included in the model. The three-way FE-GPML estimator
is also consistent under DGP III because it assumes the error term
has a variance equal to the square of the conditional mean. Both OLS
and GPML are also more efficient than PPML in this case. However,
as the other three panels show, when this variance assumption is relaxed,
three-way FE-GPML clearly suffers from an IPP, exhibiting
an average bias equal to roughly half that of OLS in all three cases.

We have also performed some simulations with three-way FE-Gaussian
PML, which imposes $g(\widehat{\lambda}_{ijt})=\widehat{\lambda}_{ijt}$.
We do not show results for this other estimator because the HDFE-IRLS
algorithm we used to produce the FE-PPML and FE-Gamma PML estimates
frequently did not converge for the FE-Gaussian PML estimator. However,
the results we did obtain were in line with our results for FE-GPML
and with our discussion of Proposition \ref{prop2} above: the FE-Gaussian
PML estimates were consistent when the DGP for $\omega_{ijt}$ was itself
Gaussian (as in DGP I), but were inconsistent otherwise.

{

\subsection{Allowing for Conditional Dependence across Pairs}

The bias expansion in Proposition \ref{bias-results} allows for errors to be clustered within each pair $(i,j)$, but assumes conditional
independence of $y_{ij}$ and $y_{i'j'}$ for all $(i,j)\neq(i',j')$. This assumption is consistent with the standard practice in the literature of assuming that
errors are clustered within pairs when computing standard errors (see \citealp{yotov_advanced_2016}.) However, it is important to clarify that the results in Proposition \ref{bias-results} may change
when other assumptions are used. For example, if we want to allow $y_{ij}$ and $y_{ji}$ (i.e., both directions of trade) to be correlated, then
the bias results would not actually change, but we would need to modify
the definition of $\Omega_{N}$ to allow for the additional clustering; namely, we would need
\begin{align}
\Omega_{N} & =\frac{1}{N\,(N-1)}\sum_{i=1}^{N-1}\sum_{j=i+1}^{N}{\rm Var}\left(\widetilde{x}_{ij}'S_{ij}+\widetilde{x}_{ji}'S_{ji}\,\big|\,x\right)\nonumber \\
 & =\frac{1}{N\,(N-1)}\sum_{i=1}^{N-1}\sum_{j=i+1}^{N}\bigg\{\widetilde{x}_{ij}'\,\left[{\rm Var}\left(S_{ij}\,\big|\,x_{ij}\right)\right]\,\widetilde{x}_{ij}+\widetilde{x}_{ji}'\,\left[{\rm Var}\left(S_{ji}\,\big|\,x_{ji}\right)\right]\,\widetilde{x}_{ji}\nonumber \\
 & \qquad\qquad\qquad\qquad\qquad+\widetilde{x}_{ij}'\,\left[{\rm Cov}\left(S_{ij},S_{ji}\,\big|\,x_{ij}\right)\right]\,\widetilde{x}_{ji}+\widetilde{x}_{ji}'\,\left[{\rm Cov}\left(S_{ji},S_{ij}\,\big|\,x_{ji}\right)\right]\,\widetilde{x}_{ij}\bigg\}.\label{GeneralizedVariance-1}
\end{align}
Notice, however, this is just one possibility. Similar adjustments could be made to allow
for clustering by exporter or importer, for example, or even for multi-way
clustering \`a la \citet{cameron_robust_2011}. In these cases, the bias
would also need to be modified; specifically, one would have to modify the portions of $D_{N}^{k}$ that
$B_{N}^{k}$ that depend on the variance of $S_{ij}$ to allow for correlations across $i$ and/or $j$. 
}

\subsection{Showing Bias in the Cluster-robust Sandwich Estimator}

For convenience, let
$\mathbf{x}_{ij}:=(x_{ij},\,d_{ij})$ be the matrix of covariates
associated with pair $ij$, inclusive of the $it$- and $jt$-specific
dummy variables needed to estimate $\alpha_{i}$ and $\gamma_{j}$.
Similarly, let $b:=(\beta^{\prime},\,\phi^{\prime})^{\prime}$ be the vector of coefficients
to be estimated and let $\widehat{b}$ be the vector of coefficient
estimates. Note that we can write a first-order approximation for $\widehat{S}_{ij}$ as
\begin{align*}
\widehat{S}_{ij} & \approx S_{ij}-\bar{H}_{ij}\mathbf{x}_{ij}(\widehat{b}-b),
\end{align*}
which is consistent with the approximation provided in \eqref{eq:SS-bias}.
We can then replace $\widehat{b}-b$
with the standard first-order expansion $\widehat{b}-b\approx-\mathcal{\bar{L}}_{bb}^{-1}\mathcal{L}_{b}^{0}$, where $\mathcal{L}=\sum_{i,j} \ell_{ij}$ is the profile likelihood.
This expansion in turn can be written out as %
\begin{align*}
\widehat{b}-b & \approx -\mathcal{\bar{L}}_{bb}^{-1}\left[\sum_{m,n}\mathbf{x}_{mn}^{\prime}  S_{mn}\right].
\end{align*}
Now we turn our attention to the outer product $\widehat{S}_{ij}\widehat{S}_{ij}^{\prime}$:
\begin{align*}
\widehat{S}_{ij}\widehat{S}_{ij}^{\prime} & \approx S_{ij}S_{ij}^{\prime}+\bar{H}_{ij}\mathbf{x}_{ij}(\widehat{b}-b)^{2}\mathbf{x}_{ij}^{\prime}\bar{H}_{ij}-2\bar{H}_{ij}\left[\mathbf{x}_{ij}(\widehat{b}-b)\right]S_{ij}^{\prime}\\
 & =S_{ij}S_{ij}^{\prime}+\bar{H}_{ij}\mathbf{x}_{ij}(\widehat{b}-b)^{2}\mathbf{x}_{ij}^{\prime}\bar{H}_{ij}+2\bar{H}_{ij}\mathbf{x}_{ij}\mathcal{\bar{L}}_{bb}^{-1}\left[\sum_{m,n}\mathbf{x}_{mn}^{\prime}  S_{mn} \right]S_{ij}^{\prime}
\end{align*}
Because we assume we are in the special case where FE-PPML is correctly
specified, we have that $\mathbb{E}[(\widehat{b}-b)^{2}]=-\kappa \mathcal{\bar{L}}_{bb}^{-1}$, where $\mathcal{\bar{L}}_{bb}:=\mathbb{E}[\mathcal{L}_{bb}]$.
We also have that $\mathbb{E}[S_{ij}S_{ij}^{\prime}]=\kappa \bar{H}_{ij}$.
Therefore, after applying expectations where appropriate%
, we have that 
\begin{align*}
\mathbb{E}[\widehat{S}_{ij}\widehat{S}_{ij}^{\prime}] & \approx S_{ij}S_{ij}^{\prime}+\kappa \bar{H}_{ij}\mathbf{x}_{ij}\mathcal{\bar{L}}_{bb}^{-1}\mathbf{x}_{ij}^{\prime}\bar{H}_{ij},
\end{align*}
which can be seen as extending \citet{kauermann2001note}'s results
to the case of a panel data pseudo-likelihood model with within-panel
clustering. We are not done, however, as we have not yet isolated
the influence of the incidental parameters. To complete the derivation of the bias, 
we must more carefully consider the full inverse Hessian term $\mathcal{\bar{L}}_{bb}^{-1}$.
Using standard matrix algebra, this inverse can be written as:
\begin{align*}
\mathcal{\bar{L}}_{bb}^{-1}= & \left( \! \! \begin{array}{cc}
\left(\mathcal{\bar{L}}_{\beta\beta}-\mathcal{\bar{L}}_{\phi\beta}^{\prime}\bar{\mathcal{L}}_{\phi\phi}^{-1}\mathcal{\bar{L}}_{\phi\beta}\right)^{-1} & -\left(\bar{\mathcal{L}}_{\beta\beta}-\mathcal{\bar{L}}_{\phi\beta}^{\prime}\bar{\mathcal{L}}_{\phi\phi}^{-1}\bar{\mathcal{L}}_{\phi\beta}\right)^{-1}\mathcal{\bar{L}}_{\phi\beta}^{\prime}\bar{\mathcal{L}}_{\phi\phi}^{-1}\\
-\bar{\mathcal{L}}_{\phi\phi}^{-1}\mathcal{\bar{L}}_{\phi\beta}\left(\bar{\mathcal{L}}_{\beta\beta}-\bar{\mathcal{L}}_{\phi\beta}^{\prime}\mathcal{\bar{L}}_{\phi\phi}^{-1}\bar{\mathcal{L}}_{\phi\beta}\right)^{-1} & \mathcal{\bar{L}}_{\phi\phi}^{-1} \! +\mathcal{\bar{L}}_{\phi\phi}^{-1}\mathcal{\bar{L}}_{\phi\beta}\!\left(\bar{\mathcal{L}}_{\beta\beta}-\mathcal{\bar{L}}_{\phi\beta}^{\prime}\mathcal{\bar{L}}_{\phi\phi}^{-1}\mathcal{\bar{L}}_{\phi\beta}\right)^{-1}\!\mathcal{\bar{L}}_{\phi\beta}^{\prime}\bar{\mathcal{L}}_{\phi\phi}^{-1}
\end{array} \! \! \right),
\end{align*}
where we have used $\mathcal{\bar{L}}_{\phi\phi}$ in place of $\mathcal{\bar{H}}$ in order to add clarity.
Making use of some already-established definitions, we have that the
top-left term $(\mathcal{\bar{L}}_{\beta\beta}-\mathcal{\bar{L}}_{\phi\beta}^{*\prime}\bar{\mathcal{L}}_{\phi\phi}^{-1}\mathcal{\bar{L}}_{\phi\beta})^{-1}=-[N(N-1)]^{-1}W_{N}^{-1}$
and, similarly, that $\mathcal{\bar{L}}_{\phi\phi}^{-1}=-[N(N-1)]^{-1}W_{N}^{(\phi)-1}$.
If we again consider $\mathbb{E}[\widehat{S}_{ij}\widehat{S}_{ij}^{\prime}]$,
we can now write
\begin{align*}
\mathbb{E}[\widehat{S}_{ij}\widehat{S}_{ij}^{\prime}-S_{ij}S_{ij}^{\prime}] & \approx-\frac{\kappa}{N(N-1)}\bar{H}_{ij}(x_{ij}\,d_{ij}) \; \times \\ &\left( \! \!  \begin{array}{cc}
W_{N}^{-1} & -W_{N}^{-1}\mathcal{\bar{L}}_{\phi\beta}^{\prime}\bar{\mathcal{L}}_{\phi\phi}^{-1}\\
-\bar{\mathcal{L}}_{\phi\phi}^{-1}\mathcal{\bar{L}}_{\phi\beta}W_{N}^{-1} & W_{N}^{(\phi)-1} \! \! +\mathcal{\bar{L}}_{\phi\phi}^{-1}\mathcal{\bar{L}}_{\phi\beta}^{*}W_{N}^{-1}\mathcal{\bar{L}}_{\phi\beta}^{\prime}\bar{\mathcal{L}}_{\phi\phi}^{*-1}
\end{array} \! \!  \right) \!  (x_{ij}\,d_{ij})^{\prime}\bar{H}_{ij}\\
 & =-\frac{\kappa}{N(N-1)}\bar{H}_{ij}\left\{ x_{ij}W_{N}^{-1}x_{ij}^{\prime}-x_{ij}W_{N}^{-1}\mathcal{\bar{L}}_{\phi\beta}^{\prime}\bar{\mathcal{L}}_{\phi\phi}^{-1}d_{ij}^{\prime}-d_{ij}\bar{\mathcal{L}}_{\phi\phi}^{-1}\mathcal{\bar{L}}_{\phi\beta}^{*}W_{N}^{-1}x_{ij}^{\prime}\right.\\
 & \left.+d_{ij}\mathcal{\bar{L}}_{\phi\phi}^{-1}\mathcal{\bar{L}}_{\phi\beta}W_{N}^{-1}\mathcal{\bar{L}}_{\phi\beta}^{\prime}\bar{\mathcal{L}}_{\phi\phi}^{-1}d_{ij}^{\prime}+d_{ij}W_{N}^{(\phi)-1}d_{ij}^{\prime}\right\} \bar{H}_{ij},
\end{align*}
which simplifies to the expression shown in \eqref{eq:SS-bias}.

\bigskip{}

\textbf{
Results for the two-way model}. Though we have focused on the downward bias of the sandwich estimator for the three-way gravity model,
it is also known
to be biased for the standard two-way gravity model without pair fixed
effects (\citealp{egger_glm_2015};
\citealp{jochmans_two-way_2016}; \citealp{pfaffermayr2019gravity}).
As it turns out, the analytics for the two-way and three-way models
are very similar here, and we can easily adapt our results
 to the simpler two-way setting. The main change we would need to make
is to replace $H_{ij}$ everywhere it appears with $\Lambda_{ij}$,
including in the definitions of $\widetilde{x}_{ij}$, $W_{N}$, and
$W_{N}^{(\phi)}$. The rest of the derivations then follow in the
same manner as for the three-way model. The resulting correction has been included
in our \href{https://github.com/tomzylkin/ppml_fe_bias}{\tt{ppml\_fe\_bias}} Stata package for users working with two-way gravity models. 
A version of this correction has been studied alongside other methods in a recent paper by
\citet{pfaffermayr2021confidence}.

{
\subsection{More Discussion of IPPs in FE-PPML Models}
 
In this part of the Appendix, we wish to give a more expansive discussion of when IPPs may arise in case of an FE-PPML estimator. We have already reviewed the two-way and three-way gravity models in the main text; thus, here, we will focus first on the classic ``one-way'' FE panel setting where no IPP occurs. Doing so will allow us to draw a contrast with other, more complex models where IPPs could be a problem. As part of this discussion, we give some examples of panel models where FE-PPML is actually inconsistent, unlike the models covered in the main text.

\subsubsection*{The Classic (One-way) Setting}

Consider a static panel data model
with individuals $i=1,\ldots,N$, time periods $t=1,\ldots,T$, outcomes
$y_{it}$, and strictly exogenous regressors $x_{it}$ satisfying
\begin{align}
\mathbb{E}(y_{it}|x_{it},\alpha_{i}) & =\lambda_{it}:=\exp(x_{it}'\beta+\alpha_{i}).\label{eq:1}
\end{align}
The FE-PPML estimator maximizes $\sum_{i,t}\left(y_{it}\log\lambda_{it}+\lambda_{it}\right)$
over $\beta$ and $\alpha$. The corresponding FOC's may be written
as 
\begin{align}
\sum_{i=1}^{N}\sum_{t=1}^{T}\,x_{it}\,\left(y_{it}-\widehat{\lambda}_{it}\right) & =0, & \forall i:\;\;\sum_{t=1}^{T}\left(y_{it}-\widehat{\lambda}_{it}\right) & =0,\label{eq:2}
\end{align}
where $\widehat{\lambda}_{it}:=\exp(x_{it}'\widehat{\beta}+\widehat{\alpha}_{i})$.
Solving for $\widehat{\alpha}_{i}$ and plugging the expression back
into the FOC for $\widehat{\beta}$ we find 
\begin{align}
\sum_{i=1}^{N}\sum_{t=1}^{T}\,x_{it}\,\left[y_{it}-\frac{\exp(x_{it}'\widehat{\beta})}{\sum_{\tau=1}^{T}\exp(x_{i\tau}'\widehat{\beta})}\sum_{\tau=1}^{T}y_{i\tau}\right] & =0,\label{eq:3}
\end{align}%
{%
which, as long as \eqref{eq:1} holds, are valid (sample) moments to estimate $\beta$.
Thus, under standard regularity conditions,
we have that $\sqrt{N}(\widehat{\beta}-\beta^{0})\rightarrow_{d}{\cal N}(0,V)$ as $N\rightarrow\infty$,} 
where $V$ is the asymptotic variance.
The FE-PPML estimator therefore does not suffer from an IPP: even though $\widehat{\alpha}_{i}$ is {an inconsistent} estimate
of $\alpha_{i}$, the FE-PPML score for $\beta$ {has zero mean when evaluated 
at the true parameter $\beta^0$}, and $\widehat{\beta}$ therefore converges in  
probability to $\beta^{0}$ without any
asymptotic bias. This is a well known result that can also be obtained in the Poisson-MLE case by conditioning on $\sum_{t}y_{it}$; see \citet{cameron_count_2013}. For our purposes, it gives us
a benchmark against which other, more complex models may be compared.

\subsubsection*{Examples where FE-PPML is Inconsistent}

In the above ``classic'' setting, every observation is affected by
exactly one fixed effect. In current applied work, it
is common to specify models with what we will call ``overlapping''
fixed effects, where each observation may be affected by more than
one fixed effects. Some standard examples include the gravity model from international trade (as is our focus in the main text) as well as other settings 
where researchers may wish to control for multiple sources of heterogeneity (e.g., firm and employee, teacher and student). 
Thus, it is important to clarify that the presence of overlapping fixed
effects can easily lead to an IPP, even when the underlying estimator
is Poisson or PPML. We give the following simple example: 
\begin{example}Consider a model with three time periods $T=3$ and two fixed effects
$\alpha_{i}$ and $\gamma_{i}$ for each individual: 
\begin{align*}
 & t=1: & \mathbb{E}(y_{i1}|x_{i1},\alpha_{i},\gamma_{i}) & =\lambda_{i1}:=\exp(x_{i1}'\beta+\alpha_{i}),\\
 & t=2: & \mathbb{E}(y_{i2}|x_{i2},\alpha_{i},\gamma_{i}) & =\lambda_{i2}:=\exp(x_{i2}'\beta+\alpha_{i}+\gamma_{i}),\\
 & t=3: & \mathbb{E}(y_{i3}|x_{i3},\alpha_{i},\gamma_{i}) & =\lambda_{i3}:=\exp(x_{i3}'\beta+\gamma_{i}).
\end{align*}
The FE-PPML estimator maximizes $\sum_{i=1}^{N}\sum_{t=1}^{3}\left(y_{it}\log\lambda_{it}+\lambda_{it}\right)$
over $\beta$, $\alpha$ and $\gamma$. $T=3$ is fixed as $N\rightarrow\infty$. 
\label{IPP-example}\end{example}

\begin{example} In addition to $i=1,\ldots,N$ and $t=1,\ldots,T$ we re-introduce another
panel dimension $j=1,\ldots,J$ and consider 
\begin{align*}
\mathbb{E}(y_{ijt}|x_{ijt},\alpha_{it},\gamma_{ij}) & =\lambda_{ijt}:=\exp(x_{ijt}'\beta+\alpha_{it}+\gamma_{ij}),
\end{align*}
where $\alpha_{it}$ is now indexed by both $i$ and $t$ and our second fixed effect is similarly indexed by $i$ and $j$. The FE-PPML estimator in this case maximizes $\sum_{i,j,t}\left(y_{ijt}\log\lambda_{ijt}+\lambda_{ijt}\right)$
over $\beta$, $\alpha$ and $\gamma$. We consider $N\rightarrow\infty$
with both $J$ and $T$ fixed, e.g. $J=T=2$. 
\label{IPP-example2}\end{example}

In these examples, because the fixed effects are overlapping, we have
that $\widehat{\alpha}$ enters into the FOC for $\widehat{\gamma}$,
and vice versa. Therefore, when for a given value $\widehat{\beta}$
we want to solve the FOC for $\widehat{\alpha}$ and $\widehat{\gamma}$
we have to solve a system of equations, and the solutions become much
more complicated functions of the outcome variable than in the one-way model. While having this
type of co-dependence between the FOCs for the various fixed effects
need not necessarily lead to an IPP, as we discuss next, it does create one in models where more than one fixed effect dimension
grows at the same rate as the panel size, as is the case with $\alpha$ and $\gamma$ in both of these examples. 

In gravity settings, by contrast, the crucial distinction is that the dimensions of each fixed effect grow only with the square root of the sample size as the number of countries increases. As mentioned in the text, this ensures that the IPPs associated with each fixed effect ``decouple'' from one another, in the sense described by \citet{fernandez-val_individual_2016}. However, in the inconsistency examples just given, the estimation noise in the estimated $\alpha$ parameters will always depend on the estimation noise in the estimated $\gamma$ parameters, and vice versa, even as $N\rightarrow \infty$. Thus, decoupling cannot occur. For illustration, we have used the model from Example \ref{IPP-example} as our example of inconsistency in our earlier Figure \ref{fig1}. We have also performed simulations for the model in Example \ref{IPP-example2} and found similar results.

\subsubsection*{A Suggested Heuristic for IPPs when the Estimator is FE-PPML}

As has been made clear from our discussion and results, the usual generic bias heuristic shown in
\eqref{BiasGeneral} from \citet{ARE} is generally not appropriate for FE-PPML. Indeed, because FE-PPML can be asymptotically unbiased in special cases, it may not be productive to try to boil down how their biases are likely to behave to a single formula. Instead, we propose the following approach:

\begin{enumerate}
\item If there are no fixed effect dimensions that grow proportionately with the sample, we expect FE-PPML to be unbiased asymptotically.
\item Otherwise, the likely order of the bias can be derived as follows:
\begin{enumerate}
	\item Construct the equivalent {multinomial model} by profiling out the largest fixed effect dimension.
	\item Infer what the order of the asymptotic bias would be for the equivalent {multinomial model} by calculating $p/n$  (i.e., as in \eqref{BiasGeneral}).
\end{enumerate}
\end{enumerate}

For example, in the three-way gravity model, the number of observations is on the order of $N^2 T$ and the number of parameters is on the order of $N^2$ pair fixed effects plus $2NT$ exporter-time and importer-time fixed effects. However, after profiling out the pair fixed effects, we only have the $2NT$ exporter-time and importer-time fixed effects. Thus, we take $p/n$ to be proportional to $1/N$ as $N\rightarrow \infty$, implying an asymptotic bias of order $1/N$. 

For further illustration, consider Examples \ref{IPP-example} and \ref{IPP-example2} above. In these cases, even after profiling out $\alpha$, one still finds that $p$ is proportional to $n$ as $n\rightarrow \infty$, implying inconsistency. As a contrast, consider the two-way gravity model. As we have just discussed, all of the fixed effects grow only with the square root of $n$ in that case, implying it is asymptotically unbiased. 

{\paragraph{``Four-way'' gravity models.} As a more complicated example, consider the following ``four-way'' gravity model:
\begin{equation}
	y_{ijlt} = \exp{\left[\alpha_{ilt} + \alpha_{jlt} + \eta_{ijl} + \zeta_{ijt} + x_{ijlt} \beta \right]} \omega_{ijlt}. \label{fourway}
\end{equation}
This type of model may be used for trade data that is observed separately for different industries or commodities, which here are indexed by $l= 1...L$. $\alpha_{ilt}$, $\alpha_{jlt}$, and $\eta_{ijl}$ respectively are industry-level analogs of $\alpha_{it}$, $\alpha_{jt}$, and $\eta_{ij}$ from the three-way model. Thus, they allow multilateral resistance effects and cross-sectional heterogeneity in trade costs to vary by industry. The fourth fixed effect, $\zeta_{ijt}$, captures general changes in trade across all industries for a given pair. $x_{ijlt}$ is assumed to be an industry-specific policy variable of interest (e.g., tariffs). {We assume that the error term $\omega_{ijlt}$ exhibits correlation over time within the same exporter-importer-industry triplet but is independent across trade partners and across industries within the same exporter-importer pair.}

The four-way model does not conform to the framework from our main analysis, but we can nonetheless use the above heuristic to infer the order of the bias and propose a correction. After profiling out the order-$N^2 L$ exporter-importer-industry fixed effects, the model has on the order of $2NLT$ exporter-industry-time and importer-industry-time fixed effects and $N^2T$ exporter-importer-time fixed effects. The number of observations is on the order of $N^2 L T$. Following our discussion from Section \ref{theIPP}, the bias is thus expected to be of the form
\begin{equation}
	\frac{1}{N} b^{(\alpha)} + \frac{1}{N} b^{(\gamma)} + \frac{1}{L} b^{(\zeta)}. \label{fourway-bias}
\end{equation}
Where $b^{(\alpha)}$, $b^{(\gamma)}$, and $b^{(\zeta)}$ are unknown constants. Two observations stand out. First, for consistency, we require both $N$ and $L$ to be large. For data sets where the number of industries is relatively small, the $1/L$ bias term associated with the $\zeta_{ijt}$ fixed effect is likely to induce substantial bias. We will thus consider the implications for asymptotic bias as $N$ and $L$ grow large at the same rate. Second, the order of the standard error as $N$ and $L$ both $\rightarrow \infty$ while $T$ is fixed is $1/(N \sqrt{L})$. The ratio of the asymptotic bias to the standard error as $N$ and $L$ both $\rightarrow \infty$ is expected to be of the form
 \[
	\frac{\sqrt{L} b^{(\alpha)} + \sqrt{L} b^{(\gamma)} + \frac{N}{\sqrt{L}} b^{(\zeta)}} {c}, \label{fourway-bse}
\]
where $c>0$ is a positive constant. This ratio diverges to infinity as $N,L \rightarrow \infty$, implying that the standard error shrinks to zero faster than the bias does. This is a more severe form of the asymptotic bias problem than the one we found for the three-way model, where the bias and standard error both decreased at the same rate asymptotically. It is therefore advised that a jackknife correction should be used to reduce the bias. This can be done by holding out industries to inflate the $1/L$ bias while simultaneuously holding out countries to inflate the $1/N$ bias. For example, a jackknife sample with half the number of countries and half the number of industries will have an asymptotic bias of order $2/N + 2/L$. 

A closely related model that we can also discuss is the case when the fourth fixed effect, $\zeta_{ijt}$, is not included in the model shown in \eqref{fourway}. An example of when this might be desirable is when the policy variable of interest does not vary across industries (e.g., a trade agreement dummy). In this case, one can infer from the formula in \eqref{fourway-bias} that we would expect an asymptotic bias of order $1/N$, like what we found in the case of the three-way model. Furthermore, if only the number of countries is allowed to become large, while both $L$ and $T$ are held fixed, the standard error is also of order $1/N$, and the behavior of the asymptotic bias is exactly like what we found for the three-way case. However, interestingly, if both $N$ and $L$ $\rightarrow \infty$, we are back in the case where the bias-to-standard error ratio heads to infinity. Thus, the severity of the problem will depend importantly on the number of industries in both of these models.

Finally, note that if $T$ is no longer fixed, so that $N$, $L$, and $T$ all $\rightarrow \infty$ jointly, we expect the special properties of PPML to cause the IPP bias to become more benign. We know from \citet{fernandez-val_individual_2016}'s earlier results for the two-way model and from our own results for the three-way model that the decoupling of the IPPs as all dimensions of the panel become large at the same rate eliminates the asymptotic bias in these settings. Future work can investigate to what extent this holds true for four-way panel models. }

\end{document}